\documentclass[fleqn,usenatbib]{mnras}

\usepackage[T1]{fontenc}

\usepackage{graphicx}	
\usepackage{amsmath}	
\usepackage{amssymb}	
\usepackage{caption}
\usepackage{longtable}
\usepackage{supertabular}
\usepackage{rotating}
\usepackage{multicol}
\usepackage{color}
\usepackage{threeparttable}
\usepackage{newtxtext,newtxmath}

\title[APEX at the QSO MUSEUM]{APEX at the QSO MUSEUM: Molecular gas reservoirs associated with $z\sim$3 quasars and their link to the extended Ly$\alpha$ emission}

\author[N. Mu\~noz-Elgueta et al.]{
N. Mu\~noz-Elgueta$^{1}$,\thanks{E-mail: nahir@mpa-garching.mpg.de }
F. Arrigoni Battaia$^{1}$,
G. Kauffmann$^{1}$,
C. De Breuck$^{2}$,
\newauthor C. Garc\'ia-Vergara$^{3}$,
A. Zanella$^{4}$,
E. P. Farina$^{1}$,
R. Decarli$^{5}$
\\
$^{1}$Max-Planck-Institut fur Astrophysik, Karl-Schwarzschild-Str 1, D-85748 Garching bei M\"unchen, Germany\\
$^{2}$European Southern Observatory, Karl-Schwarzschild-Str 2, D-85748 Garching bei M\"unchen, Germany\\
$^{3}$Leiden Observatory, Leiden University, P.O. Box 9513, 2300 RA Leiden, The Netherlands\\
$^{4}$Istituto Nazionale di Astrofisica, Vicolo dell’Osservatorio 5, I-35122 Padova, Italy\\
$^{5}$INAF—Osservatorio di Astrofisica e Scienza dello Spazio di Bologna, via Gobetti 93/3, I-40129 Bologna, Italy
}

\date{Accepted XXX. Received YYY; in original form ZZZ}

\pubyear{2015}

\begin{document}
\label{firstpage}
\pagerange{\pageref{firstpage}--\pageref{lastpage}}
\maketitle

\begin{abstract}

Cool gas (T$\sim$10$^{4}$~K) traced by hydrogen Ly$\alpha$ emission is now routinely detected around $z\sim3$ quasars, but little is known about their molecular gas reservoirs.
Here, we present an APEX spectroscopic survey of the CO(6-5), CO(7-6) and [C{\sc i}](2-1) emission lines for 9 quasars from the QSO MUSEUM survey which have similar UV luminosities, but very diverse Ly$\alpha$ nebulae.
These observations ($\langle~\rm rms~\rangle=2.6$~mJy in 300~km~s$^{-1}$) detected three CO(6-5) lines with 3.4$\leq I_{\rm CO(6-5)} \leq$5.1~Jy~km~s$^{-1}$, 620$\leq$FWHM$\leq$707~km~s$^{-1}$, and three [C{\sc i}](2-1) lines with 2.3$\leq I_{\rm [C\textsc{i}](2-1)} \leq$15.7~Jy~km~s$^{-1}$, 329$\leq$FWHM$\leq$943~km~s$^{-1}$.
For the CO and [C{\sc i}] detected sources, we constrain the molecular gas reservoirs to be $\rm M_{H_{2}} = (0.4-6.9) \times 10^{11} M_{\odot}$, while the non-detections imply $\rm M_{H_{2}} < 1.1\times 10^{11} M_{\odot}$.
We compare our observations with the extended Ly$\alpha$ properties to understand the link between the cool and the molecular gas phases. We find large velocity shifts between the bulk of Ly$\alpha$ and the molecular gas systemic redshift in five sources (from $\sim$-400 to $\sim+$1200~km~s$^{-1}$).
The sources with the largest shifts have the largest Ly$\alpha$ line widths in the sample, suggesting more turbulent gas conditions and/or large-scale inflows/outflows around these quasars.
We also find that the brightest ($ I_{\rm[C\textsc{i}](2-1)}=15.7\pm3.7~\rm Jy~km~s^{-1}$)
and the widest (FWHM$\sim$900~km~s$^{-1}$) lines are detected for the smallest and dimmest Ly$\alpha$ nebulae.
From this, we speculate that host galaxy obscuration can play an important role in reducing the ionizing and Ly$\alpha$ photons able to escape to halo scales, and/or that these systems are hosted by more massive halos.

\end{abstract}



\section{Introduction}

Super-massive black holes are found at the centre of massive galaxies (e.g., \citealt{1998richstone,2000ferrarese,2000kauffmann,2013kormendy}). They become visible as extremely luminous active galactic nuclei (AGN) through episodes of intense accretion across the history of the
Universe (e.g., \citealt{1963schmidt,2018banados,2020lyke}).
Because of the large budget of rest-mass energy available, these objects can regulate their own growth and the evolution of their host galaxies, even if only a small fraction of their feedback couples efficiently to the surrounding material (e.g., \citealt{1998silk,2005dimatteo,2015steinborn}).

Within AGN, quasars are the most luminous sources where we can see the 
nuclear emission directly (e.g., \citealt{1993antonucci,2000elvis}).
Clustering measurements suggest that quasars  preferentially inhabit dark matter halos with masses $M_{\rm DM}\sim10^{12}-10^{13}$~M$_{\odot}$ (e.g., \citealt{2004porciani,2007shen,2012white,2018timlin} and references therein). This mass range should guarantee that a non-negligible fraction of cool ($T\sim10^4$~K) gas, inflowing from large intergalactic scales at redshifts $z\gtrsim2$ does not shock heat at
the halo boundary, but accretes in cold form (e.g., \citealt{2006dekel}).
Quasars at such epochs are therefore expected to sit in halos with both a cool and a warm/hot gas ($T\sim10^5-10^7$~K) phase (e.g., \citealt{2005keres}). 
Because the quasar number density peaks between $z\sim2$ and $z\sim3$ (e.g., \citealt{2006richards,2020shen}),
these epochs ($\sim10.4-11.6$ ~Gyr ago)
are frequently 
targeted by observations 
to understand how quasars are triggered and which reservoirs, from halo to galaxy scales, sustain their
central engines.

The halo gas, known as circumgalactic medium (CGM, e.g., \citealt{2017tumlinson}), has been studied around quasars mostly targeting the cool phase both in absorption (e.g., \citealt{2006hennawi,2013prochaska,2013farina,2014farina,2018lau}) and in emission (e.g., \citealt{1991heckman,2003bunker,2013hennawi,2019farina,2021fossati}). While the absorption technique usually relies on only one background sightline per foreground halo to provide statistical information on the physical properties of  the CGM of quasars, studies
of the CGM in emission are currently able to map the quasar CGM around individual systems.  At
$z\sim2-3$, the study of projected quasar pairs has led to a number
of new insights: (i) the measurement of the anisotropic clustering of \ion{H}{i} systems around quasars (\citealt{2007hennawi,2019jalan})
suggested that their ionizing radiation escapes anisotropically or intermittently, (ii) the discovery
of large  reservoirs ($>10^{10}$~M$_{\odot}$) of cool and metal-enriched ($Z\gtrsim0.1$~Z$_{\odot}$) halo gas (\citealt{2013prochaska,2014prochaska,2016lau}), and (iii) the study of the kinematics
of the halo, which seems to suggest that the gas is in virial equilibrium with
the dark matter halo, though there is some evidence for outflowing gas (\citealt{2014prochaska,2018lau}).

In recent years, sensitive integral field unit spectrographs like the Multi-Unit Spectroscopic Explorer (MUSE; \citealt{2010bacon}), the Keck Cosmic Web Imager (KCWI; \citealt{2012morrissey}) and the Palomar Cosmic Web Imager (PCWI; \citealt{2010matu}) revolutionized 
the study of CGM gas through emission lines by allowing deeper observations in reasonable amount of time.
The seminal papers by \citet{1988rees} and \citet{2001haiman}, 
predicted that gas surrounding quasars reprocesses the impinging strong UV radiation as Ly$\alpha$ emission. 
Current studies routinely report extended Ly$\alpha$ emission, with $\sim200$ quasars surveyed to date at $2<z<4$ (e.g., \citealt{2015husband,2016fumagalli,2016borisova,2019arrigoni,2019cai,2020osullivan,2021mackenzie,2021fossati}). The bulk of the extended emission traces gas on a few tens of kpc near the quasars, while large-scale structures extending to $R\gtrsim80$~kpc are 
seen at lower surface brightness (SB$_{\rm Ly\alpha}\sim10^{-18}$~erg~s$^{-1}$~cm$^{-2}$~arcsec$^{-2}$). 

These studies reveal few extended structures over hundreds of kpc with
SB$_{\rm Ly\alpha}\gtrsim10^{-17}$~erg~s$^{-1}$~cm$^{-2}$~arcsec$^{-2}$ (\citealt{2018arrigoni}), likely pinpointing very dense environments (\citealt{2015hennawi}, Nowotka et. al subm.). These rare and bright large-scale nebulae are also known as enormous Ly$\alpha$ nebulae (ELAN; \citealt{2017cai}).
The Ly$\alpha$ kinematics in the extended nebulae is consistent with gravitational motions in halos
with masses consistent with typical quasar hosts (e.g., \citealt{2019arrigoni,2020osullivan}),
with a few exceptions with possible quasar winds extending over tens of kpc (\citealt{2020travascio}).

While a large fraction of these Ly$\alpha$ nebulae are likely powered by the quasars, 
the balance between different plausible mechanisms is still debated. Most previous work assumes
that the Ly$\alpha$ emission is due to recombination radiation 
following quasar photoionization (e.g., \citealt{1991heckman,2014cantalupo}).
Resonant scattering of quasar Ly$\alpha$ photons and active companions can, however,
provide a non-negligible contribution  on  scales of tens of kpc near compact sources 
(e.g., \citealt{2014cantalupo,2018husemann,2019arrigoni2}). On top of this, there are large 
uncertainties on the ionizing radiation that impinges on the surrounding gas, because
 quasars are expected to be anisotropic, intermittent  sources with different degrees of obscuration.
These uncertainties hamper the physical interpretation of properties of the
emitting gas (e.g., density $n_{\rm H}$, metallicity; \citealt{2021fossati}). 

There is, however,  evidence in few systems that gas at large projected distances ($\gtrsim50$~kpc) 
is not affected by resonant scattering effects, namely: (i) non-resonant lines follow the kinematics 
of the Ly$\alpha$ emission (e.g., \ion{He}{ii}, \citealt{2017cai}), and (ii) there is no evidence for
double-peaked line profiles at the current resolution of the observational data (e.g., \citealt{2018arrigoni}).
Neglecting resonant scattering, photoionization models match the observed Ly$\alpha$ and low \ion{HeII} emission only if interstellar-medium-like 
densities ($n_{\rm h}\gtrsim 1$~cm$^{-3}$) in small-scale structures ($<20$~pc)  are invoked (\citealt{2014cantalupo,2015hennawi,2015arrigoni,2016borisova}).
This finding suggests the presence of dense CGM gas whose survival and entrainment 
in the warm/hot halo seem plausible from current high resolution ``cloud-crushing'' 
simulations(e.g., \citealt{2018gronke,2020gronke,2021kanjilal}). Note that such processes are
still largely unresolved by current cosmological simulations,
even when attempts are made to resolve the CGM (e.g., \citealt{2019hummels,2019peeples}).  

It is therefore of interest to ascertain observationally the maximum density the cool CGM gas
is able to reach, and whether a fraction of the gas is able to transform into a molecular phase.
The molecular gas around quasars can be  best probed through different tracers
depending on its physical properties (e.g., density, temperature) and 
those of the surrounding environment (e.g., radiation field) (e.g., \citealt{2013carilli}).
Most previous works have focused  on the rotational ($J$) transitions of carbon monoxide $^{12}$C$^{16}$O
(hereafter CO), which is the most abundant molecule after $\rm H_{2}$.  
Low-$J$ CO transitions are 
good tracers of the total cold molecular gas due to their low excitation temperatures.
The CO($J$=1-0) ground transition requires an excitation temperature 
of only $\sim$5.5K (e.g., \citealt{2013bolatto}). Using observations of different 
CO transitions and radiative transfer models (e.g., large velocity gradient, LVG; e.g., \citealt{2007vandertak}),
it is possible to constrain the CO spectral line energy distribution (SLED),
 and probe the excitation conditions and physical properties of the gas,
as the density and kinetic temperature  (e.g., \citealt{2007weiss,2009riechers}). 

The detection of CO emission in high-$z$ quasars greatly advanced thanks
to the Atacama Large Millimeter/Submillimeter Array (ALMA, \citealt{2009wootten}), and the Karl J. Jansky Very Large Array (JVLA, \citealt{2011perley}). It is now possible to probe
quasars at very high redshifts ($z\sim 6-7.5$, e.g., \citealt{2016wang,2017venemans,2018decarli,2019novak}).
The population of z$\sim$3 quasars has also been studied in a number
of previous studies\citep[e.g.,][]{2007weiss,2012schumacher,2013carilli,2020bischetti}. 
From these works, we know that at these redshifts, the low CO transitions (i.e., $J \leq 3$) 
are expected to be faint, and that the redshifted CO lines lie at challenging frequencies for current and past instruments.
Past CO observations of $z\sim 2-3$ quasars found molecular gas masses 
in the range of $\sim 10^{9} - 10^{11}\ M_{\odot}$, similar to those found for quasars at higher redshift (e.g., \citealt{2002barvainis,2003weiss,2004beelen,2011walter,2012schumacher,2019hill,2020bischetti}). 
These molecular reservoirs are
characterized by densities of $\sim 10^{3.0}-10^{4.4} \rm cm^{-3}$ and kinetic temperatures of $\sim$30 - 90~K \citep{2003weiss,2007bweiss,2012schumacher},  
and, when resolved, have an effective radius of $\sim$~0.5 - 2.5  kpc \citep[e.g.,][]{2009riechers,2012schumacher,2021stacey}.

Currently, there is only tentative evidence for extended molecular gas reservoirs around individual quasars,
but only few studies attempted long integrations. \cite{2006riechers} presented CO(1-0) detections
in three quasars at $z\sim 4$. 
Using single component LVG models, they found that all the flux detected 
in CO(1-0) was associated with the molecular gas traced by higher CO transitions. 
An extended component up to 30$\%$ of the total CO(1-0) luminosity was allowed by the observations.
The extended component could have larger mass  
if the CO conversion factor was taken to be higher on larger scales.
\citealt{2019emonts} targeted the CO(1-0) transition from the 
MAMMOTH-I ELAN located at $z\sim 2$  (\citealt{2017cai,2018arrigoni2}) and reported emission
extended over tens of kpc, with roughly $50\%$ of the CO(1-0) emission outside of galaxies. 
Finally, \citet{2021decarli} targeted the CO(3-2) transition for two $z\sim2$ ELANe, the Slug (\citealt{2014cantalupo}) and the Jackpot (\citealt{2015hennawi}). Their NOEMA observations did not unveil any extended molecular reservoir in these objects down to molecular gas surface densities typical of starbursting systems ($\Sigma_{\rm H_{2}}<12-68$~M$_{\odot}$~pc$^{-2}$).

Fine structure lines of atomic carbon, for example [C{\sc i}], are an additional tracer to
probe the cold molecular phase (e.g., \citealt{2004papadopoulos,2018valentino}). Observational studies
in the local Universe have shown that CO and [C{\sc i}] can coexist,
suggesting that both transitions arise from the same regions (e.g., \citealt{1994white,2002ikeda,2002israel}),
though spatial variations could be present (e.g., \citealt{2019salak}). 
Analysis using simultaneously [C{\sc i}] and multi-transition CO 
observations at high redshifts found agreement between the H$_2$ masses determined through the two different tracers (e.g., \citealt{2003weiss,2013alaghbandzadeh}), corroborating  the assumption that [C{\sc i}] and CO 
usually coexist (\citealt{2013carilli}). 
The carbon masses found in the literature for quasars at $z\sim 2-3$ are typically
of the order of $10^{6}-10^{7} M_{\odot}$ and do not differ significantly from those found for quasars at $z>3$  \citep[e.g.,][]{2003weiss,2011walter,2012schumacher,2017venemans,2018banerji,2019yang}.
Molecular masses are then usually obtained by assuming the same  abundance of [C{\sc i}] relative to H$_2$
as found in high-z quasars (e.g., \citealt{2005weiss}).

In this framework, we 
targeted the CO(6-5) ($\rm \nu_{rest}$ = 691.4731 GHz), CO(7-6) ($\rm \nu_{rest}$ = 806.6518 GHz) and [C{\sc i}] $^{3}$P$_{2}$-$^{3}$P$_{1}$ (hereafter [C{\sc i}](2-1), $\rm \nu_{rest}$ = 809.3420 GHz) transitions with the SEPIA180 receiver (\citealt{2018belitsky,2018bbelitsky}) on the Atacama Pathfinder Experiment (APEX) for a sample of nine z$\sim$3 quasars, whose halo gas has been studied in the QSO MUSEUM survey \citep{2019arrigoni}. 
With these observations, we aim to (i) constrain the molecular phase around these massive systems 
and thus start characterizing the multiphase nature of the halo gas, and (ii) investigate the relation
between the molecular gas content and the large-scale cool phase.
The molecular line detections also pin down the systemic redshift of the quasar
very accurately, allowing us to probe the kinematics of the halo gas.

This work is structured as follows. In Section \ref{section:sampleanddata}, we describe our sample, observations and data reduction. In Section \ref{section:obsresults} we present the observed line properties and refine the systemic redshift when possible. In Section \ref{section:allmasses}, we describe the estimation of the molecular gas masses using different methods, and present results for these masses.
In Section \ref{section:comparisonlya}, we compare the derived molecular masses  with the Ly$\alpha$ properties 
of our sources. In Section \ref{section:discussion} we discuss our main results, and explore the link between the 
molecular gas content and the large scale Ly$\alpha$ emission. Finally, Section \ref{section:summary} summarizes our findings. 

Throughout this paper, we adopt the cosmological parameter $H_{\rm 0} = 70\ \rm km\ s^{-1} Mpc^{-1}$, $\Omega_{\rm M} = 0.3$ and $\Omega_{\rm \Lambda} = 0.7$.

\section{Sample and observational data}\label{section:sampleanddata}

\begin{figure}
    \centering
    \includegraphics[width=0.9\columnwidth]{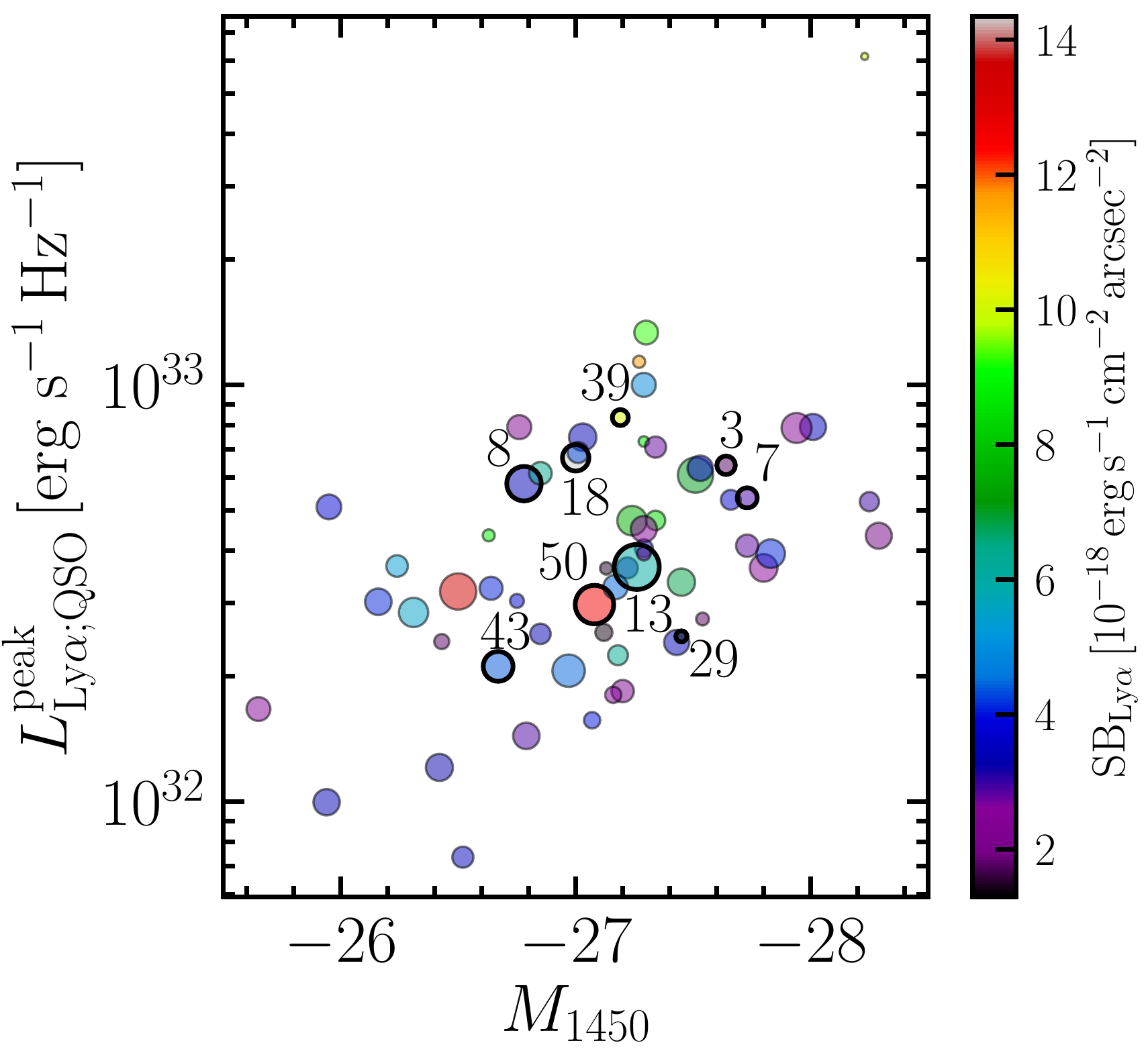}
    \caption{The peak specific luminosity at the Ly$\alpha$ line of the quasar versus  $M_{\rm 1450\AA}$ for the QSO MUSEUM sample (\citealt{2019arrigoni}). The symbols sizes are proportional to the area of the Ly$\alpha$ nebulae, while the colours indicate their average Ly$\alpha$ surface brightness. We highlight the nine sources observed with APEX with black circles and their ID numbers.}
    \label{fig:Fig_sample}
\end{figure}

Our sample is composed of nine quasars at $z\sim$3 selected from the  QSO MUSEUM survey \citep{2019arrigoni}, which targeted  61 quasars with MUSE for the study of their CGM rest-frame UV line emission.
The nine quasars were observed with the  SEPIA180\footnote{The APEX/SEPIA180 dual polarisation 2SB receiver is a pre-production version of the ALMA Band 5 receiver, and covers the frequency range 159-211 GHz.} receiver mounted on the APEX antenna, located in Llano de Chajnantor, Chile.
The targets were selected using the following constraints:
\begin{itemize}
    \item Visibility from the telescope site and the presence of CO rotational transitions, CO(6-5) and CO(7-6), and the [C{\sc i}](2-1) transition within the frequency range covered by the SEPIA180 instrument. 
    \item The expected frequency of the targeted emission lines was required to be located far from the atmospheric 183 GHz water-absorption feature to  best exploit the sensitivity of the SEPIA180 instrument, 
    even under high water vapor conditions. 
    \item Similar absolute magnitudes at rest frame 1450 \AA, 
    ranging between -27.64 and -26.67 mag, with a median of -27.20 (Figure~\ref{fig:Fig_sample}). 
    \item Coverage of a large portion of the physical  parameter space of the QSO MUSEUM survey, namely
    Ly$\alpha$ nebulae with sizes spanning the range  $\sim$29 - 467 arcsec$^{2}$ (or $\sim$1600 - 27000~kpc$^{2}$) and
     surface brightnesses $\sim$1.25$\times$10$^{-18}$-1.43$\times$10$^{-17}$ erg~s$^{-1}$~cm$^{-2}$~arcsec$^{-2}$
     (Figure~\ref{fig:Fig_sample}). 
    \item One of the targets was selected to be radio-loud, reflecting a similar fraction of such objects in the parent quasar
     sample ($\sim$10$\%$, \citealt{2002ivezic}).
\end{itemize}

\begin{figure*}
\includegraphics[scale=0.05]{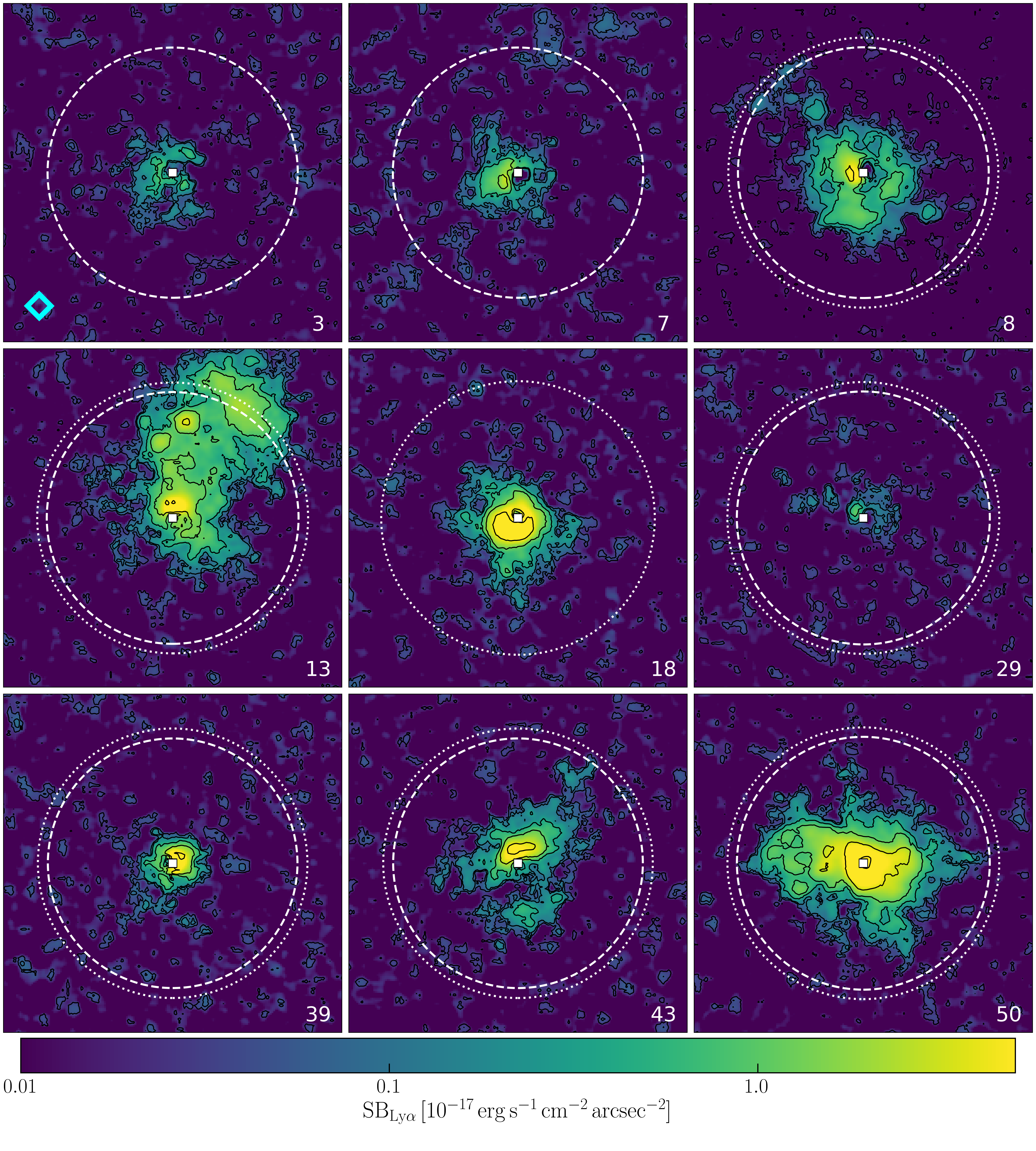}
\caption{$40 \arcsec \times 40 \arcsec (\rm or \sim312\ kpc \times 312\ kpc)$ MUSE Ly$\alpha$ surface brightness maps of the nebulae surrounding the 9 targets of this study (adapted from \citealt{2019arrigoni}). 
The numbers in the bottom right corner are the IDs from the QSO MUSEUM survey (see Table \ref{table:observations}, first column). The white dotted circles represent the APEX/SEPIA180 beam sizes for the CO(6-5) observations ($\sim$32''), while the white dashed circles are the beam sizes for the [C{\sc i}](2-1) - CO(7-6) observations ($\sim$30'').  The cyan diamond indicates that the source is radio-loud.}
 \label{fig:apexbeam}
\end{figure*}

\begin{figure*}
\includegraphics[scale=0.55]{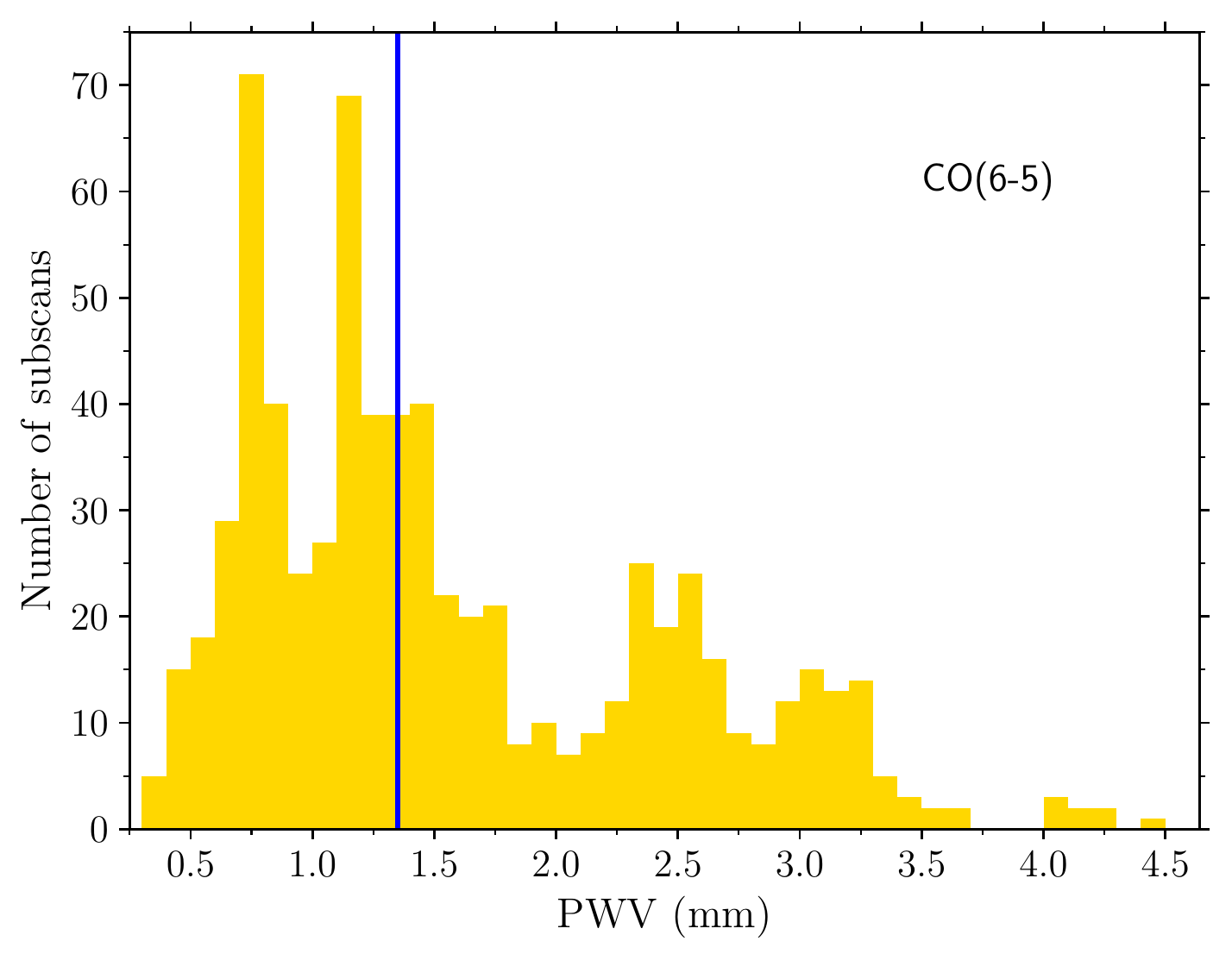}
\includegraphics[scale=0.55]{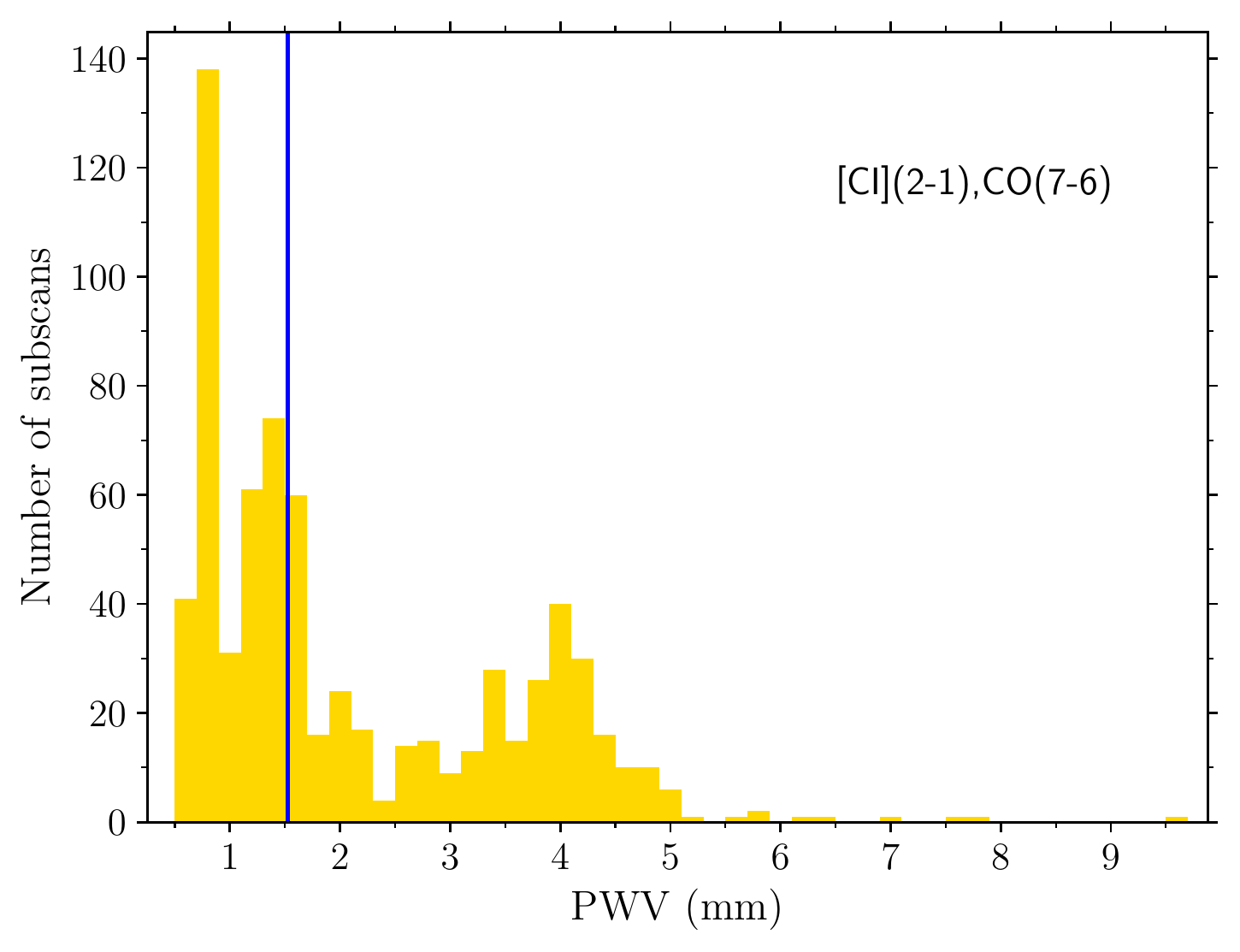}
\caption{Histograms of the PWV values 
during the observations. Left panel: Values for the  ESO programme 0102.A-0394A, the CO(6-5) observations. Right panel: Values for the ESO programme 0103.A-0306A, the [C{\sc i}](2-1) - CO(7-6) observations.  The blue vertical lines indicate the median values.}
 \label{fig:histo_pwv}
\end{figure*}

The CO(6-5) ($\rm \nu_{obs}$\footnote{Observed frequency of the line transition at $z$=3.133, the median redshift of the sample studied
in this paper.} = 167.3 GHz) observations were carried out between October and December of 2018 under the ESO programme 0102.A-0394A (PI: F. Arrigoni Battaia), with a total of $\sim$133 hours of telescope time. The [C{\sc i}](2-1) ($\rm \nu_{obs}$ = 195.8 GHz) and CO(7-6) ($\rm \nu_{obs}$ = 195.2 GHz) observations were performed between May and December of 2019 under the ESO programme 0103.A-0306A (PI: F. Arrigoni Battaia), with a total of $\sim$140 hours of telescope time. The main beam full width half maximum (FWHM) of the SEPIA180 receiver is about $\sim$32'' ($\sim$249~kpc) for the CO(6-5) observations and $\sim$ 30'' ($\sim$234~kpc) for the [C{\sc i}](2-1) - CO(7-6) observations. Figure \ref{fig:apexbeam} shows Ly$\alpha$ images of the nine targets analysed in this work, with superimposed APEX beams shown as 
white circles.  The acquired data will therefore provide an integrated spectrum of the emission within such beams. 
The median value of precipitable water vapor (PWV) was 1.4 and 1.5 mm for CO(6-5) and [C{\sc i}](2-1) - CO(7-6), respectively. The full histograms of the PWV values for the observations are 
shown in Fig.~\ref{fig:histo_pwv}.  A summary of the sample and observational setup is shown in Table~\ref{table:observations}. Due to source visibility and weather constraints, we obtained CO(6-5) data for 7 sources and [C{\sc i}](2-1) - CO(7-6) for 8 sources.

The data reduction was performed using the \textsc{gildas/class}\footnote{\url{http://www.iram.fr/IRAMFR/GILDAS/}} package version 1.1. For each source the data corresponding to a different date were processed separately, before combining them. In this procedure, for every target the noisy edges ($\sim$3\%) of the spectra were trimmed. Then, a velocity window\footnote{The total width of the velocity window was in the range of $\sim$1000 - 1500 km s$^{-1}$ for the cases in which the CO(6-5) emission line was expected, and in the range of $\sim$2000 - 2500 km s$^{-1}$ for the cases in which the [C{\sc i}](2-1) and CO(7-6) emission lines were expected. } was chosen to encompass the expected location of the emission line, according to the redshift of the source. First-degree polynomial baselines were computed neglecting the data within that window, and subtracted from the individual scans. All data for each source were then combined into one final spectrum, after visual inspection of individual scans. These spectra cover an average spectral window of $\sim$6000 km~s$^{-1}$.

To further improve the root mean square (rms) of the final combined spectra, we applied the following procedure. For each target, we computed the rms for each used subscan in order to reject the nosiest data. We computed the median rms of 
the whole dataset (i.e., all dates for each source) and removed the data  farthest away from this median.
New reductions ignoring these subscans were performed following the steps described above,
checking if the final rms of the dataset improved. We found that the removal of the noisiest subscans 
did not improve the final rms, because the
decrease in exposure time  compensates the improvement in rms, so we decided to keep all the data for the final reduction. The final rms for each tuning is reported in Table~\ref{table:observations}.

To illustrate the stability of the SEPIA180 instrument at the targeted frequencies,
in Fig. \ref{fig:appendix_rms} of Appendix \ref{app:stability} we show the rms as a function of the bin size (in km~s$^{-1}$), for the  final combined spectra of each source, starting from the original resolution and up to 600~km~s$^{-1}$. At the bottom of each panel is shown by how much the observed rms deviates from the expected value. At a bin size of 300 km~s$^{-1}$, we found a mild median deviation of 12$\%$ and 14$\%$ for the CO(6-5) and [C{\sc i}](2-1) - CO(7-6) observations, respectively.

As last step, we transformed the intensity units of the spectra, originally in temperature (K), to physical flux units. For this purpose, we assumed the conversion factor $38.4 \pm 2.8$~Jy~K$^{-1}$, calculated for the SEPIA180 receiver \citep{2018belitsky}\footnote{This value is consistent within uncertainties with the SEPIA180 efficiencies computed for the year of our observations (see listed values at \url{http://www.apex-telescope.org/telescope/efficiency/index.php}).}.

\begin{table*}   
\centering
\caption{Sample, observations setup and log.}\label{table:observations}
\resizebox{\textwidth}{!}{
\begin{threeparttable}
 \begin{tabular}{lcccccccccccc}
\hline\hline 
 & & & & & \multicolumn{4}{c}{0102.A-0394A$^{a}$} & \multicolumn{4}{c}{0103.A-0306A$^{b}$}\\ \cline{6-13}
ID$^{c}$ & Quasar         	& RA		& DEC		& z$^{d}_{\rm sys}$ 	&Frequency$^{e}$	& Exp. time$^{f}$	&RMS$^{g}$	&PWV$^{h}$	&Frequency$^{e}$	&Exp. time$^{f}$	&RMS$^{g}$	&PWV$^{h}$	 \\
   &                    & (J2000)	&(J2000)	& 				&(GHz)		&(hr)			&(mK)	&(mm)		&(GHz)		&(hr)			&(mK)	& (mm)       \\ \hline \\
3  & J 0525-233         &05:25:06.500	&-23:38:10.00	&3.110 				&-		&-			&-		&-		&196.000	&4.8			&0.131		&3.6     \\
7  & SDSS J1209+1138    &12:09:18.000	&+11:38:31.00	&3.117 				&-		&-			&-		&-		&195.932	&5.4			&0.086		&1.8     \\
8  & UM683 	        &03:36:26.900	&-20:19:39.00	&3.132 				&167.346	&6.1			&0.034		&1.9		&195.221	&5.8			&0.079		&1.4     \\
13 & PKS-1017+109       &10:20:10.000	&+10:40:02.00	&3.164 				&166.060	&6.2			&0.075 		&1.0		&193.720	&4.7			&0.087		&3.6     \\
18 & SDSS J1557+1540    &15:57:43.300	&+15:40:20.00	&3.265				&162.127	&11.6			&0.047		&1.8		&-		&-			&-		&-       \\
29 & Q-0115-30          &01:17:34.000	&-29:46:29.00	&3.180 				&164.600	&11.5			&0.060		&1.5		&192.000	&13.6			&0.065		&1.7     \\
39 & SDSS J0100+2105    &01:00:27.661	&+21:05:41.57	&3.100 				&168.000	&8.1			&0.071		&1.7		&196.888	&6.0			&0.047 		&2.1     \\
43 & CTSH22.05          &01:48:18.130	&-53:27:02.00	&3.087 				&168.000	&12.7			&0.049		&1.8		&196.800	&15.8			&0.045		&2.3     \\
50 & SDSS J0819+0823    &08:19:40.580	&+08:23:57.98	&3.197 				&164.754	&10.5			&0.057		&1.8		&192.197	&8.9			&0.073		&2.6     \\ \hline
                                                                                	
\end{tabular}                                                                                         
\begin{tablenotes}
\footnotesize
\item[\emph{a}]{ESO programme corresponding to the CO(6-5) observations.}
\item[\emph{b}]{ESO programme corresponding to the [C{\textsc{i}}](2-1) CO(7-6) observations.}
\item[\emph{c}]{Identification number taken from the QSO MUSEUM survey \citep{2019arrigoni}.}
\item[\emph{d}]{Quasar systemic redshift taken from the QSO MUSEUM survey \citep{2019arrigoni}, which has an intrinsic uncertainties of 415 km s$^{-1}$. See Table \ref{table:redshifts} for updated systemic redshifts ($z_{\rm mol,sys}$) calculated from the lines detected in this work.}
\item[\emph{e}]{Tuning frequency used for the observations. }
\item[\emph{f}]{Total ON-OFF exposure time per source. The total telescope time is roughly double this integration time.}
\item[\emph{g}]{RMS per bin of 300 km s$^{-1}$ of the final combined spectrum, in antenna temperature $T^{*}_{A}$ units.}
\item[\emph{h}]{Median PWV between all the observed dates for each source.}
\end{tablenotes}                                  
\end{threeparttable}
}
\end{table*}

\section{Observational results}\label{section:obsresults}
The resulting APEX spectra, reduced and converted to flux density units (mJy), are presented in Figs. \ref{fig:co65spec} and \ref{fig:co76spec} for the CO(6-5) and [C{\sc i}](2-1)-CO(7-6) observations, respectively. 
For all sources, the left panel spectrum has a bin size between 150 and 200 km s$^{-1}$ (depending on the depth of the data), and the right panel spectrum has a bin size of 300 km s$^{-1}$. These two different bin sizes are shown to highlight the reliability of the detections. 
In this work we report as detections the lines that fulfill the following conditions: i) have a peak emission at S/N$>2\sigma$, at bin sizes of 300 km~s$^{-1}$, ii) are also present at the resolution of 150 km~s$^{-1}$ but with lower significance than at 300 km~s$^{-1}$, and consistent integrated fluxes, iii) have an integrated S/N$>3\sigma$. In Appendix~\ref{app:spurious} we show that this detection algorithm is reliable, giving basically a zero-rate of false positive identifications in a negative and jack-knife tests.

Importantly, throughout this work we assume that the detected emission is due to the central quasar hosts, unless specified. We checked this assumption by computing the number of expected line-emitter companions for each transition within the APEX observations, down to the $3\sigma$ limiting luminosities of $L'$=3.13$\times 10^{10}$, 3.02$\times 10^{10}$ and 2.65$\times 10^{10}$~K~km~s$^{-1}$~pc$^{2}$, respectively for CO(6-5), CO(7-6), and [C{\sc i}](2-1). Specifically, we assumed (i) a cylindrical volume defined by the APEX primary beam and the covered velocity range of $\sim6000$~km~s$^{-1}$, (ii) the luminosity function of CO(6-5), CO(7-6) and [C{\sc i}](2-1) emission measured for similar redshifts ($z\sim2$; \citealt{2020decarli}), and (iii) a deterministic bias model for the clustering of sources around quasars (e.g., \citealt{2017garciavergara,2019garciavergara}). 
In this model, we assume a power-law shape for the clustering, with a fixed slope of $\gamma=1.8$, and we use the $z\sim3$ quasar clustering (\citealt{2007shen}) and the clustering of Lyman-break galaxies (LBGs) at $z=4$ \citep{2004ouchi}, which are assumed to have similar clustering as CO and [C{\sc i}] sources.

Following these assumptions, we found that the expected number of companions for the total number of observed fields per line (7, 8, 8) are 0.07, 0.01, 1.39, respectively for CO(6-5), CO(7-6), and [C{\sc i}](2-1). 
We caution that the luminosity functions in \cite{2020decarli} are still associated with large uncertainties, 
and only upper limits are reported at the bright-end sampled by our observations. We have thus extrapolated their measurements up to brighter fluxes, and therefore our estimations represent upper limits for the number of expected companions. Specifically, the worst case is the current [C{\sc i}](2-1) luminosity function, which has a relatively flat shape (see Fig. 7 of \citealt{2020decarli}). In this case, our extrapolation is flat and therefore represents a clear upper limit considering that the luminosity function is expected to steeply decrease for high luminosities.
In summary, the only transition for which we may find a companion is [C{\sc i}](2-1), with a conservative probability $<17$\% for each field.
Follow-up high-resolution observations with interferometers (e.g., ALMA, NOEMA) are required to verify this assumption and assess whether any of the detected emission comes from companions and/or larger scales.

We also computed the probability that one detected line in one field is actually any CO or [C{\textsc{i}}] transition from an interloper galaxy at possible lower redshifts given our tunings, by assuming the luminosity functions in \citet{2019decarli,2020decarli}, and the comoving volume spanned by our observations. Once again, these estimates represent upper limits as our observations sample the bright-end of the luminosity functions. We found that the probability of observing a contaminant is <0.2\% for any CO or [C{\textsc{i}}] line.
Therefore, it is very likely that any detected emission in our observations is associated with the quasar or its environment.

\subsection{Emission line measurements} \label{sec:detections}
We measured the molecular velocity-integrated emission line fluxes  $I_{\rm CO(6-5)}$, $I_{\rm [C{\textsc{i}}](2-1)}$ and I$_{\rm CO(7-6)}$, by fitting Gaussian to the detected lines. 
For those spectra that presented two 
peaks, a double Gaussian fit was applied. The uncertainties of the measured fluxes include 
the aforementioned error on the flux conversion factor. 
The Gaussian fits also provided an estimate of the full width at half maximum (FWHM) of the emission lines and their respective uncertainties. 

When an emission line was not detected,
we derived 3$\sigma$ upper limits using the rms noise within the same velocity range for the emission
lines that were detected for that target, and within a velocity width of 300 km s$^{-1}$ (expected average line width for quasar hosts, e.g., \citealt{2003weiss,2007bweiss,2011walter}) when no line was detected.
In Table \ref{table:fluxes} we tabulate these fluxes (or upper limits), FWHM and their respective uncertainties for each source.

\begin{figure*}
    \centering
    \includegraphics[scale=0.75]{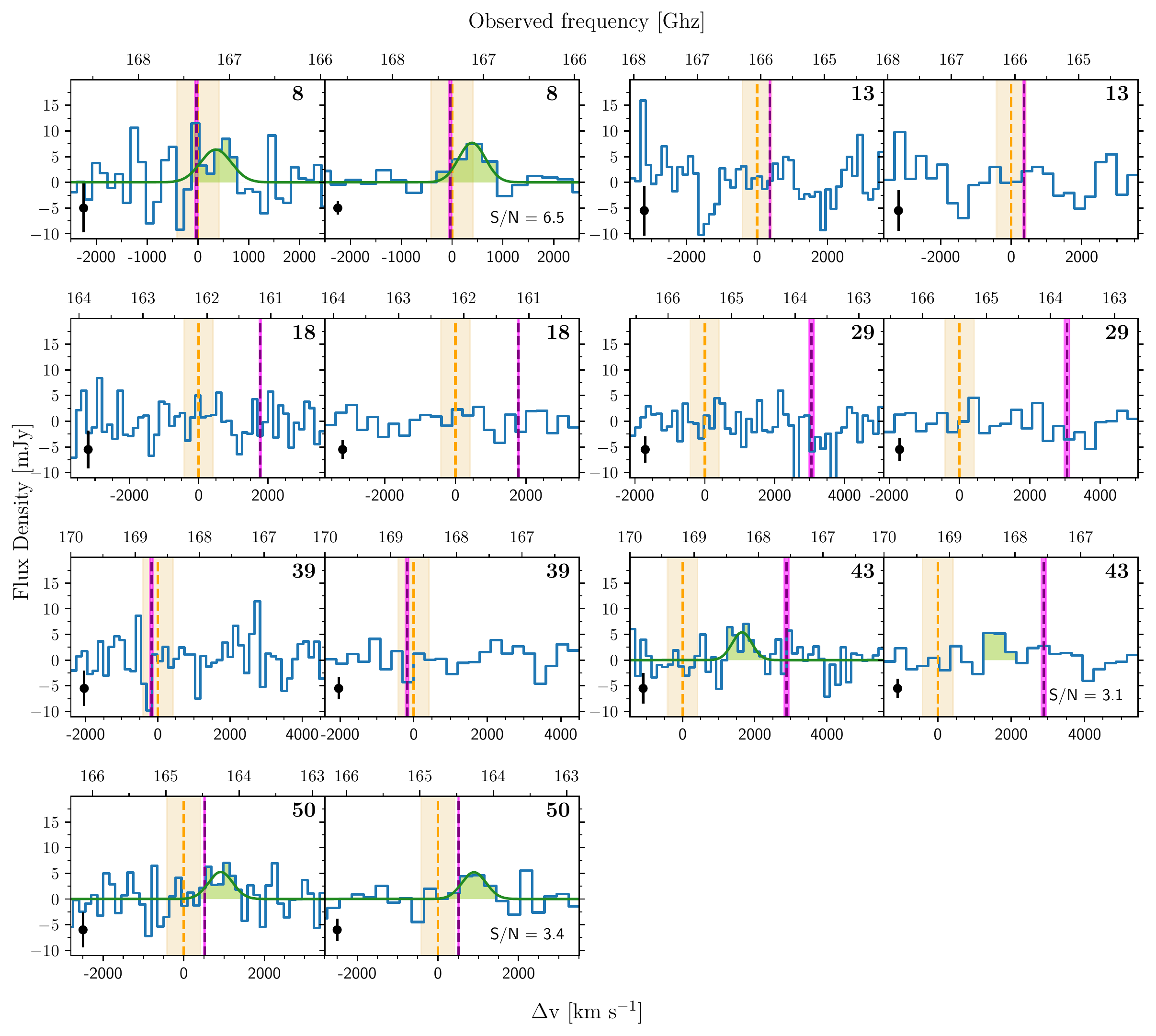}
    \caption{The seven APEX CO(6-5) spectral line observations. 
    Each observed spectrum is shown 
    with bin sizes between 150 and 200 km s$^{-1}$ (left column) and 300 km s$^{-1}$ right (column). 
    In each panel, the vertical black error bar represents the uncertainty (rms) per bin, 
    while the ID number of each quasar is shown in the top right corner. The orange vertical lines at the velocity zero point represent the systemic redshift of the sources before our APEX observations (from C{\sc iv}, see section~\ref{section:redshifts}), and the shaded regions correspond to its uncertainty ($\sim$415 km s$^{-1}$, \citealt{2019arrigoni}). The purple vertical lines represent the nebulae Ly$\alpha$ redshift from MUSE  extracted within the APEX beam (see Fig. \ref{fig:apexbeam}), with their corresponding uncertainty (magenta shaded regions). The green  shaded area indicates the detected CO(6-5) emission line, while the dark green curve represents the Gaussian fit applied to this line, when possible.}
    \label{fig:co65spec}
\end{figure*}

\begin{figure*}
    \centering
    \includegraphics[scale=0.75]{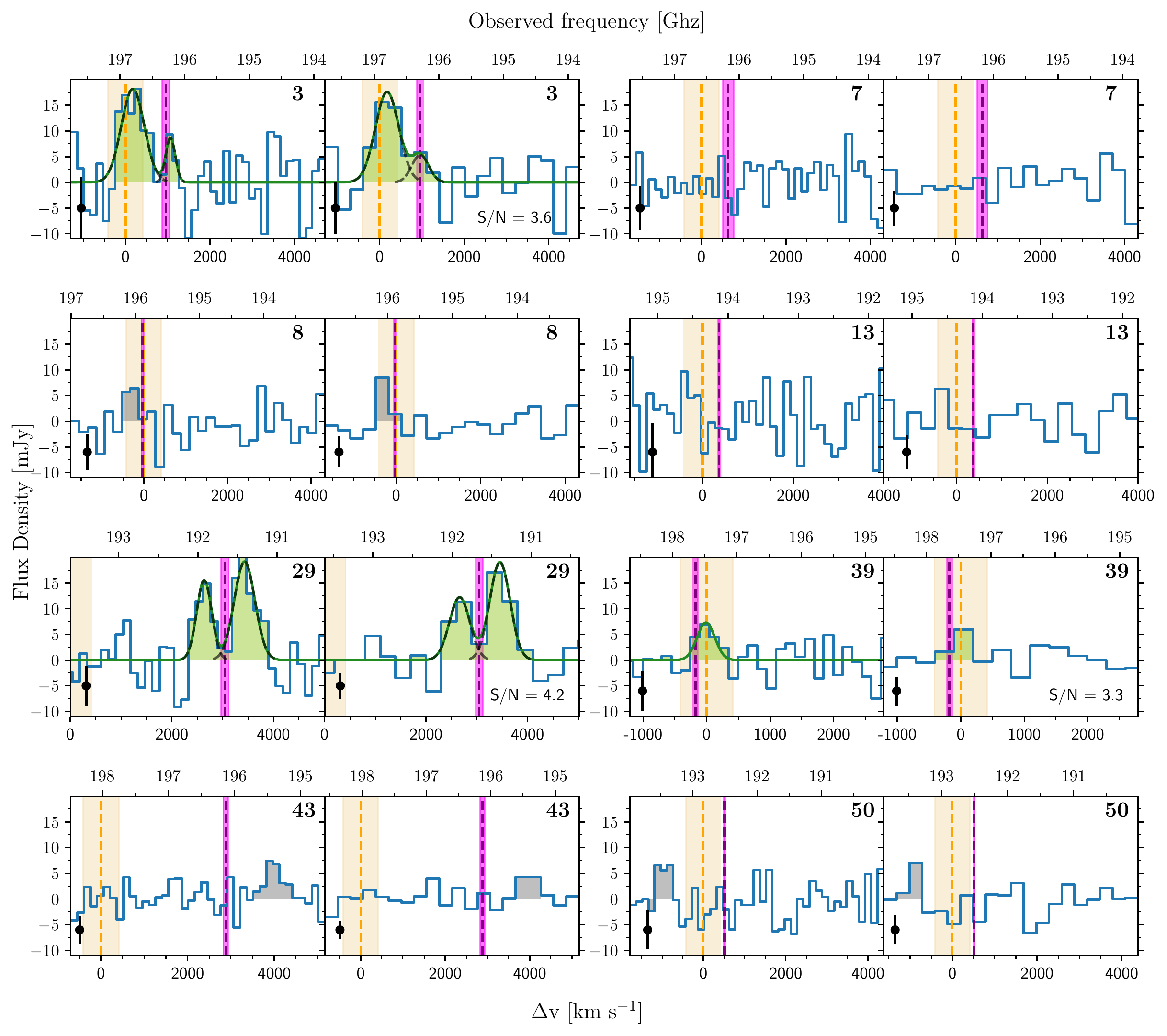}
    \caption{Same as Figure \ref{fig:co65spec}, but for the eight APEX CO(7-6)-[C{\sc i}](2-1) spectral line observations. Here, the zero velocities are placed at the expected [C{\sc i}](2-1) transition, considering the systemic redshift from C{\sc iv} (orange vertical line) with its uncertainty (shaded region) (see section~\ref{section:redshifts}).  The grey shaded areas indicate a likely spurious source (ID8, see section~\ref{sec:detections} and Appendix~\ref{app:spurious}) or tentative features (IDs 43 and 50) 
    that do not belong to the quasar hosts.
    For sources with IDs 3 and 29, we applied a double Gaussian fit, where the black dashed lines represent the single Gaussian components. 
    For ID 3, a second component was added not to overestimate the integrated flux of the [C{\sc i}](2-1) line. The emission lines of quasar with ID 29 are both due to [C{\sc i}](2-1) (see section~\ref{sec:detections}), and its systemic redshift (orange line) lies outside the plotted window.}
    \label{fig:co76spec}
\end{figure*}

For the quasars with IDs  39 ([C{\sc i}](2-1) line) and 43 (CO(6-5) line), it was not possible to obtain a good Gaussian fit at bin sizes of 300 km s$^{-1}$ (see Figs. \ref{fig:co65spec} and \ref{fig:co76spec}). 
Therefore, their integrated fluxes were first estimated by adding the area covered by each bin of 300 km s$^{-1}$ contained within the emission line and verified against a
Gaussian fit at bin sizes between 150 and 200 km s$^{-1}$. Their integrated fluxes computed from both bin sizes are consistent, as per our detection criteria. The FWHMs and fluxes given in Table \ref{table:fluxes} are those estimated with the
Gaussian fit. 
For the quasar with ID 3 (or J 0525-233), all  line properties are listed using bin sizes of 150 km s$^{-1}$, because at this resolution we obtained a better fit (see Fig. \ref{fig:co76spec}) than at bin sizes of 300 km s$^{-1}$.

\begin{table*}   
\caption{CO(6-5), [C{\textsc{i}}](2-1), CO(7-6) fluxes, FWHM, and integrated S/N ratios.} 
\label{table:fluxes}
\resizebox{\textwidth}{!}{
\begin{threeparttable}
 \begin{tabular}{lcccccccccccc}
\hline\hline

ID & Quasar          & I$\rm_{CO(6-5)}$	  &S/N$\rm_{CO(6-5)}$  & FWHM$\rm_{CO(6-5)}$     &I$\rm_{CI(2-1)}$ 	&S/N$\rm_{CI(2-1)}$ & FWHM$\rm_{CI(2-1)}$ &I$\rm_{CO(7-6)}$ 	&S/N$\rm_{CO(7-6)}$ &FWHM$\rm_{CO(7-6)}$\\	
   &                 &  (Jy km s$^{-1}$)  &   			&(km s$^{-1}$)     	& (Jy km s$^{-1}$)      &    		& (km s$^{-1}$)		&(Jy km s$^{-1}$) 	& 		&(km s$^{-1}$)   \\ \hline \\   
3  & J 0525-233      &  -          	  &-   			&-          	 	&12.4$\pm$3.4$^{\dagger}$          	&3.6 		&640$\pm$168$^{\dagger}$		& <6.6 (2.4$\pm$1.9$^{\dagger *}$)     	&- (1.3$^{\dagger *}$) 		&- (265$\pm$217$^{\dagger *}$)     \\    		
7  & SDSS J1209+1138 &  -          	  &-   			&-          	 	&<3.1          	        &-   		&-      		& <3.1  	  	&- $\, \, \, \, \, \, \, \, \, \, \, \, \, \, \, \, $  		&- $\, \, \, \, \, \, \, \, \, \, \, \, \, \, \, \, \, \, \, \, \, \, \, \, \, \, \, \, \, \, $		\\	 	
8  & UM683 	     &  5.1$\pm$0.8       &6.5  		&620$\pm$175    	&<3.9           	&- 		&-			& <3.9  	  	&- $\, \, \, \, \, \, \, \, \, \, \, \, \, \, \, \, $  		&- $\, \, \, \, \, \, \, \, \, \, \, \, \, \, \, \, \, \, \, \, \, \, \, \, \, \, \, \, \, \, $             \\	 	
13 & PKS-1017+109    &  <2.6      & -  		&-  	& <3.0          	&-   		&-      		& <3.0  	  	&- $\, \, \, \, \, \, \, \, \, \, \, \, \, \, \, \, $  		&- $\, \, \, \, \, \, \, \, \, \, \, \, \, \, \, \, \, \, \, \, \, \, \, \, \, \, \, \, \, \, $             \\	 	
18 & SDSS J1557+1540 &  <1.6     	  &-   			&-          	 	& -             	&-   		&-     			&   -     	  	&- $\, \, \, \, \, \, \, \, \, \, \, \, \, \, \, \, $  		&- $\, \, \, \, \, \, \, \, \, \, \, \, \, \, \, \, \, \, \, \, \, \, \, \, \, \, \, \, \, \, $		\\		
29 & Q-0115-30       &  <3.6     	  &-   			&-          	 	&15.7$\pm$3.7$^{\ddagger}$           &4.2 		&943$\pm$249$^{\ddagger}$		&<4.0     	  	&- $\, \, \, \, \, \, \, \, \, \, \, \, \, \, \, \, $  		&- $\, \, \, \, \, \, \, \, \, \, \, \, \, \, \, \, \, \, \, \, \, \, \, \, \, \, \, \, \, \, $		\\	 	
39 & SDSS J0100+2105 &  <2.0     	  &-   			&-          	 	&2.3$\pm$0.7$^{\dagger}$          	&3.3 		&329$\pm$109$^{\dagger}$      		& <1.7  	  	&- $\, \, \, \, \, \, \, \, \, \, \, \, \, \, \, \, $  		&- $\, \, \, \, \, \, \, \, \, \, \, \, \, \, \, \, \, \, \, \, \, \, \, \, \, \, \, \, \, \, $             \\		
43 & CTSH22.05       &  3.4$\pm$1.1$^{\dagger}$    	  &3.1  		&643$\pm$279$^{\dagger}$		    	&<2.3            	&-   		&-			& <2.3  	  	&- $\, \, \, \, \, \, \, \, \, \, \, \, \, \, \, \, $  		&- $\, \, \, \, \, \, \, \, \, \, \, \, \, \, \, \, \, \, \, \, \, \, \, \, \, \, \, \, \, \, $             \\	 	
50 & SDSS J0819+0823 &  3.9$\pm$1.1       &3.4  		&707$\pm$100    	 	&  <3.9       	        &-   		&-      		& <3.9  	  	&- $\, \, \, \, \, \, \, \, \, \, \, \, \, \, \, \, $  		&- $\, \, \, \, \, \, \, \, \, \, \, \, \, \, \, \, \, \, \, \, \, \, \, \, \, \, \, \, \, \, $             \\ \hline	
\end{tabular}                                          
\begin{tablenotes}
\footnotesize
\item Note: The integrated fluxes, S/N and FWHM have been calculated using a bin size of 300 km s$^{-1}$. For the cases with non-detections, 3$\sigma$ upper limits on the fluxes are provided.
\item[\emph{*}]{Feature 
added in the Gaussian fit of quasar with ID 3 (or J 0525-233), to not overestimate the integrated flux of the [C{\sc i}](2-1) line at bin size of 300 km s$^{-1}$ (see Fig. \ref{fig:co76spec}).}
\item[\emph{$\dagger$}]{These values have been calculated using a bin size between 150 and 200 km s$^{-1}$.}
\item[\emph{$\ddagger$}]{These values have been calculated as the sum of the two Gaussian components (see Fig. \ref{fig:co76spec}). For each component, the individual values of I$\rm_{CI(2-1)}$ are 6.1$\pm$2.9 km~s$^{-1}$ and 9.5$\pm$2.3 Jy km~s$^{-1}$, with FWHM$\rm_{CI(2-1)}$ values of 471$\pm$224 km~s$^{-1}$ and 471$\pm$108 km~s$^{-1}$, respectively.}
\end{tablenotes}                                                                                    
\end{threeparttable}
}
\end{table*}

We started the analysis from the CO(6-5) tuning. From the spectra shown in Fig. \ref{fig:co65spec}, we found CO(6-5) emission line detections (represented by a green shaded area) for three sources: quasars with IDs 8, 43, and 50. Two of these detections are closer to the redshift of the Ly$\alpha$ nebula (IDs 43 and 50; purple vertical line) than the uncertain systemic redshift computed from C{\sc iv} (orange vertical line). 
They present fluxes of 3.4$\leq I_{\rm CO(6-5)}$<5.1~Jy~km~s$^{-1}$.  
The other targeted sources,  quasars with IDs 13, 18, 29 and 39, do not present any CO(6-5) emission down to the current rms. We therefore report 3$\sigma$ upper limits for these sources (see Table~\ref{table:fluxes}).

From the spectra in Fig.~\ref{fig:co76spec},
we found [C{\sc i}](2-1) emission line detections in quasars with IDs 3, 29, and 39. They have fluxes of 2.3$\leq I_{\rm [C\textsc{i}](2-1)}\leq $15.7~Jy~km~s$^{-1}$. The quasar with ID~8 (or UM683) has a feature at 196.1 GHz (or -328 km s$^{-1}$, integrated S/N = 3.1, $I_{\rm line}\sim$3 Jy km s$^{-1}$, which could be a spurious line given the results of the jack-knife test for this spectrum (see Appendix \ref{app:spurious}).
For completeness, we also report that the quasars with ID 43 (or CTSH22.05) and ID 50 (or SDSS J0819+0823) have  tentative features at 195.4 GHz (or +3966 km s$^{-1}$, integrated S/N = 2.7, $I_{\rm line}\sim$2.8 Jy km s$^{-1}$) and 193.4 GHz (or -957 km s$^{-1}$, integrated S/N = 2.1, $I_{\rm line}\sim$2.5 Jy km s$^{-1}$), respectively. The redshift of these lines is not consistent with the CO(6-5) detections for these sources, but they might be an associated object. 
We note that sub-millimeter galaxies (SMG) have often being found in the surroundings of quasars (e.g., \citealt{2015silva}), however the CO(6-5) transition (detected for IDs~43 and 50) is expected to be stronger in quasars than in SMG  \citep{2013carilli}. Also, we stress that the detections in the CO(6-5) tuning must come from the quasar, due to the low probability of finding a companion in this transition in our observations (see Section \ref{section:obsresults}). 
The rest of the sources do not have any  detected emission lines at the  rms of the current observations.

It is important to stress that we considered our detected lines as [C{\sc i}](2-1) and not CO(7-6) because they are closer to the expected position of [C{\sc i}](2-1) based on the systemic redshift of the sources (within 1$\sigma$, IDs~3 and 39),
or because if they were CO(7-6) emission, CO(6-5) detections would be also expected unless
unphysical  line ratios are assumed (quasar with ID~29, see next paragraph). We acknowledge that the used systemic redshifts (based on C{\sc iv}) are uncertain (see Section \ref{section:redshifts}) and further data are needed to verify our line identifications.

The spectrum of the quasar with ID 3 (or J 0525-233) shows the [C{\sc i}](2-1) emission line (at 196.8~GHz) 
and a S/N = 1.4 bump (binning of 300~km~s$^{-1}$) at the expected location of  
CO(7-6) (196.2~GHz), which we estimate would contribute $\sim20$\% of the [C{\sc i}](2-1) flux when using the 300~km~s$^{-1}$ binning. In contrast, the quasar with ID 29 (or Q-0115-30)\footnote{See Figure \ref{fig:spectrumID29} for a higher resolution spectrum for this source.} presents two emission lines (at 191.9 and 191.4~GHz) consistent with the redshift of its Ly$\alpha$ nebula. Both these lines likely
correspond to [C{\sc i}](2-1). To confirm this, we first assumed that the line at frequency 191.4 GHz was CO(7-6) and estimated the observed I$\rm_{CO(7-6)}$/I$\rm_{CO(6-5)}$ ratio ($>3.73$). We then compared this observed ratio to theoretical predictions from large velocity gradient (LVG) models \citep{1974goldreich} with different physically plausible kinetic temperatures and densities (see Section \ref{section:massco} for a detailed explanation of these models and input parameters). This test did not allowed us to find any modelled  I$\rm_{CO(7-6)}$/I$\rm_{CO(6-5)}$ ratio as high as that observed, indicating that the line peak at lower frequencies also corresponds to [C{\sc i}](2-1)\footnote{All the LVG models explored in this work have I$\rm_{CO(7-6)}$/I$\rm_{CO(6-5)}$ < 1.34 when considering kinetic temperatures of $20 - 200\ \rm K$, and densities of $10^{3} - 10^{5}\ \rm cm^{-3}$, as frequently used in the literature (e.g., \citealt{2006riechers,2007bweiss}). These LVG models allow a maximum I$\rm_{CO(7-6)}$=3.48~Jy~km~s$^{-1}$~ at a kinetic temperature of $200\ \rm K$, which represent the most extreme SLED found in the literature (\citealt{2007bweiss}). Even in this extreme case, a fraction of $63\%$ of the observed second line peak has to correspond to [C{\sc i}](2-1).}. We stress that this identification needs further confirmation, e.g. by targeting additional transitions from this source. Quasar ID~29, with a total $I_{\rm [C{\textsc{i}}](2-1)}=15.7\pm3.7$~Jy~km~s$^{-1}$, is therefore the brightest [C{\sc i}](2-1) detection in this sample. Furthermore, it is $\sim$3 times brighter than the lensed Cloverleaf quasar ($I_{\rm [C{\textsc{i}}](2-1)}\sim~5.2$~Jy~km~s$^{-1}$,\citealt{2003weiss}), which is as far as we know the brightest [C{\sc i}](2-1) $z\sim 3$ quasar detected in the literature. We further discuss the nature of this double-peak emission of quasar with ID~29 (or Q-0115-30) in Section \ref{dis:q0115}.

Overall, our observations suggest that [C{\sc i}](2-1) is stronger than CO(7-6) in these $z\sim3$ quasars. This is in contrast with other observations of the same transitions in different $z\sim4-7$ quasars (\citealt{2009riechers,2017venemans,2018lu,2019wang,2019yang}) and $z\sim$ 2.5-3 quasars \citep{2003weiss,2011walter,2012schumacher}, likely indicating that the excitation conditions in the molecular gas are different. Also, \cite{2018banerji} observed these transitions in two quasars and their companions at $z\sim2.5$, finding a [C{\sc i}](2-1) line slightly stronger than CO(7-6) in one quasar and one companion. To our knowledge, the studies mentioned above are the only available observations of these lines in the literature. Therefore, if our observations are
confirmed in deeper datasets, current literature could be affected by low number statistics.
We further note that [C{\sc i}](2-1) is stronger than CO(7-6) in high-redshift radio galaxies ($\sim$5 times stronger, e.g., \citealt{2016gullberg}), which can be explained as enhancement of atomic carbon in cosmic ray dominated regions \citep[e.g.,][]{2017bisbas}.

\subsection{Redshifts from molecular lines}\label{section:redshifts}
A precise estimate of the systemic redshifts  of quasars
plays a fundamental role in understanding the physical processes and kinematics of each system. For instance, an accurate systemic redshift would allow us 
to better constrain the cool gas kinematics mapped on large scales by the Ly$\alpha$ emission, and to compare it to cosmological simulations \citep{2019arrigoni}.

The uncertain systemic redshifts for our sample, estimated by \cite{2019arrigoni}, are shown in the fifth column of Table \ref{table:observations}. These redshifts were determined from the peak of the C{\sc iv} line, after correcting from the expected luminosity-dependent blue-shift  \citep{2016shen}, and have an intrinsic uncertainty of 415 km s$^{-1}$. This large uncertainty, comparable to outflow/inflow velocities expected in quasar halos, hampers any kinematical study of these systems.
However, the molecular emission lines  should provide a more robust measure of the quasar's systemic redshift (e.g., \citealt{2017banerji,2020bischetti}).
We therefore derived new systemic redshifts for the objects with detected molecular lines. For the case of ID~29, the redshift was estimated as an average between the centroids of the two peaks.
These new redshifts and their uncertainties are 
listed in Table \ref{table:redshifts}.
They have on average an uncertainty of 74.8 km s$^{-1}$ and a difference of $+$1045 km~s$^{-1}$ (or $\Delta z=+0.014$) with respect to the systemic redshifts from C{\sc iv} (see also Section~\ref{sec:lyacoprofiles}).
Hereafter we will assume these new values as systemic redshifts.

\begin{table*}   
\centering
\caption{Redshifts and central frequencies obtained from the targeted molecular lines, and velocity shifts from the large-scale Ly$\alpha$ emission within the same APEX aperture.} 
\label{table:redshifts}
\resizebox{\textwidth}{!}{
\begin{threeparttable}
 \begin{tabular}{lccccccccccc}
\hline\hline

ID & Quasar          &z$^{a}_{\rm CO(6-5)}$ & $\nu_{\rm CO(6-5)}$ &z$^{b}_{\rm C\textsc{i}(2-1)}$ & $\nu_{\rm C\textsc{i}(2-1)}$ & z$^{c}_{\rm CO(7-6)}$ &  $\nu_{\rm CO(7-6)}$ & z$^{d}_{\rm mol,sys}$ & $\Delta v^{e}_{\rm Ly\alpha Neb-mol}$ & FHWM$_{\rm Ly\alpha}$ \\
   &        &   &(GHz)  &   &(GHz) & &(GHz) & & (km s$^{-1}$) & (km s$^{-1}$) \\ \hline \\
3  & J 0525-233       &-     &- & 3.112$\pm$0.001 &196.81$\pm$0.05 & - (3.111$\pm$0.001) &- (196.17$\pm$0.06) &3.112$\pm$0.001& 786$\pm$109 &	1542$\pm$196 \\
7  & SDSS J1209+1138  &-    &-  &-	   & -  &-$\, \, \, \, \, \, \, \, \, \, \, \, \, \, \, \, \, \,\, \, \, \, \, \, \, \, \, \, \, \, \, \, \, \, \, \, $ & -$\, \, \, \, \, \, \, \, \, \, \, \, \, \, \, \, \, \, \, \, \, \, \, \, \, \, \, \, \, \, \, \, \, \, \, \,$ & - & -	& -\\
8  & UM683 	      &3.137$\pm$0.001 & 167.13$\pm$0.04&- &- & -$\, \, \, \, \, \, \, \, \, \, \, \, \, \, \, \, \, \,\, \, \, \, \, \, \, \, \, \, \, \, \, \, \, \, \, \, $   &-$\, \, \, \, \, \, \, \, \, \, \, \, \, \, \, \, \, \, \, \, \, \, \, \, \, \, \, \, \, \, \, \, \, \, \, \,$   &3.137$\pm$0.001& -423$\pm$76 & 915$\pm$46	 \\
13 & PKS-1017+109     &- & -&-  &-   &-$\, \, \, \, \, \, \, \, \, \, \, \, \, \, \, \, \, \,\, \, \, \, \, \, \, \, \, \, \, \, \, \, \, \, \, \, $  &-$\, \, \, \, \, \, \, \, \, \, \, \, \, \, \, \, \, \, \, \, \, \, \, \, \, \, \, \, \, \, \, \, \, \, \, \,$   &-& -	& -\\
18 & SDSS J1557+1540  &-  &-  &-  &-   &-$\, \, \, \, \, \, \, \, \, \, \, \, \, \, \, \, \, \,\, \, \, \, \, \, \, \, \, \, \, \, \, \, \, \, \, \, $  &-$\, \, \, \, \, \, \, \, \, \, \, \, \, \, \, \, \, \, \, \, \, \, \, \, \, \, \, \, \, \, \, \, \, \, \, \,$   & - & -	& -\\
29 & Q-0115-30        &-  &-  &3.223$\pm$0.001$^{\ddagger}$ & 191.65$\pm$0.06$^{\ddagger}$ & -$\, \, \, \, \, \, \, \, \, \, \, \, \, \, \, \, \, \,\, \, \, \, \, \, \, \, \, \, \, \, \, \, \, \, \, \, $  &-$\, \, \, \, \, \, \, \, \, \, \, \, \, \, \, \, \, \, \, \, \, \, \, \, \, \, \, \, \, \, \, \, \, \, \, \,$   &3.223$\pm$0.001& -11$\pm$120 & 508$\pm$172	 \\
39 & SDSS J0100+2105  &-   &  &3.100$\pm$0.001 &197.42$\pm$0.03 &-$\, \, \, \, \, \, \, \, \, \, \, \, \, \, \, \, \, \,\, \, \, \, \, \, \, \, \, \, \, \, \, \, \, \, \, \, $  &-$\, \, \, \, \, \, \, \, \, \, \, \, \, \, \, \, \, \, \, \, \, \, \, \, \, \, \, \, \, \, \, \, \, \, \, \,$ &3.100$\pm$0.001& -152$\pm$64 & 781$\pm$100	\\
43 & CTSH22.05        &3.110$\pm$0.002 &168.26$\pm$0.07 &-  &-   &-$\, \, \, \, \, \, \, \, \, \, \, \, \, \, \, \, \, \,\, \, \, \, \, \, \, \, \, \, \, \, \, \, \, \, \, \, $  &-$\, \, \, \, \, \, \, \, \, \, \, \, \, \, \, \, \, \, \, \, \, \, \, \, \, \, \, \, \, \, \, \, \, \, \, \,$  &3.110$\pm$0.002& 1236$\pm$133 &1068$\pm$142 \\ 
50 & SDSS J0819+0823  &3.210$\pm$0.001 & 164.26$\pm$0.02 & - &-   &-$\, \, \, \, \, \, \, \, \, \, \, \, \, \, \, \, \, \,\, \, \, \, \, \, \, \, \, \, \, \, \, \, \, \, \, \, $ &-$\, \, \, \, \, \, \, \, \, \, \, \, \, \, \, \, \, \, \, \, \, \, \, \, \, \, \, \, \, \, \, \, \, \, \, \,$  &3.210$\pm$0.001	 &-377$\pm$44 & 967$\pm$38 \\ \hline
\end{tabular}   
\begin{tablenotes}
\footnotesize
\item[\emph{a}]{Redshift obtained from the observed CO(6-5) emission line.}
\item[\emph{b}]{Redshift obtained from the observed [C{\sc i}](2-1) emission line.}
\item[\emph{c}]{Redshift obtained from the observed CO(7-6) emission line.}
\item[\emph{d}]{Quasar systemic redshift estimated using the observed molecular lines.}
\item[\emph{e}]{Velocity shift between the Ly$\alpha$ line peak of the nebulosities (extracted within the APEX beam, see Fig. \ref{fig:apexbeam}) and the systemic redshift $z_{\rm mol,sys}$.}
\item[\emph{$\ddagger$}]{These values have been calculated as the average between the two Gaussian components (see Fig. \ref{fig:co76spec}). For each component, the individual values of $\nu_{\rm C\textsc{i}(2-1)}$ are 191.91$\pm$0.11 GHz and 191.39$\pm$0.06 GHz.}
\end{tablenotes}                                                                                    
\end{threeparttable}
}
\end{table*}


\section{Molecular mass estimates}\label{section:allmasses}

In this section we estimate molecular gas masses $\rm M_{ H_{2}}$ by using different methods. Specifically, we compute i) carbon masses $\rm M_{\rm C{\textsc {i}}}$ and derive the respective $\rm M_{ H_{2}}$ 
by assuming a carbon abundance relative to H$_2$,  ii) $\rm M_{ H_{2}}$ for sources that have a measured constraint (upper limit) on the 
CO(7-6)/CO(6-5) ratio, 
iii) $\rm M_{ H_{2}}$ by combining the two previously obtained mass ranges from [C{\sc i}] and the CO ratio, and iv) $\rm M_{ H_{2}}$ for sources with no clear constraint on the CO ratio (i.e. sources with non-detections).

\subsection{Atomic carbon mass}\label{section:cimass}
The mass of the atomic carbon can be estimated from the [C{\sc i}](2-1) line luminosity through the formulation 
presented in \cite{2003weiss,2005weiss}, under the assumption that this [C{\sc i}] transition is optically thin:
\begin{equation}
\label{eqn:CI}
    \frac{M_{\rm C{\textsc {i}}}}{M_{\odot}} = 4.566\times 10^{-4}Q(T_{\rm ex})\frac{1}{5}e^{T_{\rm 2}/T_{\rm ex}}L'_{\rm [C{\sc I}]}
\end{equation}
where $Q(T_{\rm ex}$) = 1+3$e^{-T_{1}/T_{\rm ex}}$+5$e^{-T_{2}/T_{\rm ex}}$ corresponds to the C{\sc i} partition function, $T_{\rm ex}$ is the excitation temperature, and $T_{1}$ = 23.6 K and $T_{2}$ = 62.5 K correspond to the energies above the ground state. We used $T_{\rm ex}$=30 K as frequently found in high-redshift quasars 
(e.g., \citealt{2003weiss,2011walter}). 
The [C{\sc i}] line luminosity, $L'_{\rm [C{\sc I}]}$, can be estimated via \citep{1992solomon}

\begin{equation}\label{eq:luminosity}
    L'_{\rm line} = 3.25\times 10^{7}I_{\rm line} \nu_{\rm obs}^{-2} D_{\rm L}^{2}(1+z)^{-3}
\end{equation}
where $I_{\rm line}$ is the velocity-integrated line flux in units of Jy km s$^{-1}$, $\nu_{\rm obs}$ is the observed frequency in units of GHz, and $D_{\rm L}$ corresponds to the luminosity distance in Mpc. The final units of $L'_{\rm line}$ are K km s$^{-1}$ pc$^{2}$.

As noted in Section \ref{sec:detections}, the [C{\sc i}](2-1) emission line is detected for three sources of our sample: quasars with IDs 3, 29 and 39. Using equation~\ref{eqn:CI}, we found atomic carbon masses in the range of $\sim$4.0$\times$10$^{7}$M$_{\odot}$ - 3.1$\times$10$^{8}$M$_{\odot}$. For the five sources with upper limits on the [C{\sc i}](2-1) transition (quasars with IDs 7, 8, 13, 43 and 50), we found atomic carbon masses < 8.0$\times$10$^{7}$M$_{\odot}$.
The detected sources show higher values than usually reported for quasars in the literature. Indeed, it is common to find values of the order 10$^{6}$ - 10$^{7}$M$_{\odot}$, when the same excitation temperature of 30~K is assumed \citep[e.g.,][]{2011walter,2017venemans,2019yang}. The value for each source is listed in the third column of Table \ref{table:masses}. It is important to note that the assumption of a higher $T_{\rm ex}$ in equation \ref{eqn:CI}, for instance $T_{\rm ex}$=50 K, implies a  $M_{\rm C{\textsc {i}}}$ $\sim$38 $\%$ lower. 

The atomic carbon mass can be used 
to determine the molecular gas mass using the atomic carbon abundance relative to H$_{2}$:
\begin{equation}\label{eq:massfromci}
    \frac{X[\rm C{\textsc{i}}]}{X[\rm H_{ 2}]}= \frac{M_{\rm C{\textsc {i}}}}{6M_{\rm H_{2}}}.
\end{equation}
In this work 
we assumed $X[\rm C{\textsc{i}}]/X[\rm H_{2}]$= (8.4  $\pm$ 3.5) $\times$10$^{-5}$ as usually done in the literature for high-redshift quasars (e.g., \citealt{2011walter,2017venemans}). We note that this value is higher than what has been recently found for $z\sim 1-3$ main sequence galaxies ($\sim$1.7$\times$10$^{-5}$, \citealt{2018valentino,2020boogaard}).

Using this method, we constrained molecular gas masses for quasars with IDs 3, 29 and 39, which span from $\sim$8.0$\times$10$^{10}$M$_{\odot}$ to $\sim$6.1$\times$10$^{11}$M$_{\odot}$. For quasars with IDs 7, 8, 13, 43 and 50, we found molecular gas masses < 1.5$\times$10$^{11}$M$_{\odot}$. Most of these molecular gas masses are higher than the typical range ($\sim$10$^{9}$-10$^{11}$M$_{\odot}$) found in the literature for other high-$z$ quasars \citep[e.g.,][]{2003weiss,2011walter,2013anh,2019hill}. We list all these values in the fourth column of Table \ref{table:masses}. 

Importantly, we are assuming that all the [C{\sc i}](2-1) emission comes from the quasar hosts. If part of this emission comes from larger areas, the $T_{\rm ex}$ might break into different values (e.g., lower/higher on larger/smaller scales), affecting the molecular masses estimated in this work. To constraint $T_{\rm ex}$ around these quasars, resolved [C{\sc i}](2-1) and [C{\sc i}](1-0) observations are needed (e.g., ALMA, NOEMA).

\begin{table*}   
\centering
\caption{Molecular, carbon and dynamical mass estimations.} \label{table:masses}
\begin{threeparttable}
 \begin{tabular}{lccccccc}
\hline\hline

ID & Quasar          &  M$\rm_{C\textsc{i}}$ $^{a}$ 	&M$\rm_{H_{2}}$ $^{b}$	&M$\rm_{H_{2}}$ $^{c}$	&M$\rm_{H_{2}}$ $^{d}$ &	M$_{\rm dyn}{\rm sin}^2(i)$ $^{e}$\\
   &                 &  (10$^{8}$M$_{\odot}$)	      	&(10$^{11}$M$_{\odot}$)	&(10$^{11}$M$_{\odot}$)	&(10$^{11}$M$_{\odot}$) &(10$^{11}$M$_{\odot}$)	\\ \hline \\
3  & J 0525-233      &  2.3$\pm$0.6    			&4.6$\pm$2.3 		&<6.9		&2.3-6.9$^{ *}$ &	1.6$\pm$0.8	 \\
7  & SDSS J1209+1138 &  <0.6 	      			&<1.1 	 		&<3.2			&<1.1		 &-	\\
8  & UM683 	     &  <0.7    			&<1.5 	  	&0.7-64.1	        &0.7-1.5	& 1.5$\pm$0.8	 \\
13 & PKS-1017+109    &  <0.6	      			&<1.1         		&<3.2     	 	& <1.1 	&- \\
18 & SDSS J1557+1540 &   -		      		&-		 	&<0.6		 	&<0.6	&-	\\
29 & Q-0115-30       &  3.1$\pm$0.7    			&6.1$\pm$2.9  		&<4.4		 	&3.2-4.4	& 4.4$\pm$1.0 $^{\ddagger}$	 \\
 &     &   			& 		&		 	&	& 12.4$\pm$6.6 $^{\diamond }$	 \\
39 & SDSS J0100+2105 &  0.4$\pm$0.1   			&0.8$\pm$0.4 	 	&   <1.8		& 0.4-1.8$^{\dagger}$	& 0.4$\pm$0.3	\\
43 & CTSH22.05       &  <0.4	        		&<0.9 	 		&0.5-42.7 	        &0.5-0.9	& 1.6$\pm$1.4	 \\ 
50 & SDSS J0819+0823 &  <0.8	        		&<1.5 	 		&0.5-51.3	        &0.5-1.5	& 1.9$\pm$0.6	 \\ \hline
\end{tabular}                                          
\begin{tablenotes}
\footnotesize
\item Note: Upper limits of masses provided in each column are due to non-detections of [C{\sc i}] and/or CO (see Section \ref{section:allmasses}).
\item[\emph{a}]{Atomic carbon mass assuming an excitation temperature of 30 K (see Section \ref{section:cimass}).}
\item[\emph{b}]{Molecular gas mass derived from the atomic carbon mass (see Section \ref{section:cimass}, equation \ref{eq:massfromci}).}
\item[\emph{c}]{Molecular gas mass derived from CO (see Section \ref{section:massco}).}
\item[\emph{d}]{Molecular gas mass derived from applying both CO and [C{\sc i}] constraints (see Section \ref{sec:cociconstraints}).}
\item[\emph{e}]{ Inclination-dependent dynamical masses (see Section \ref{sec:dis:dynamicalmass}). The reported uncertainties include only the errors on the FWHM values.}
\item[\emph{$*$}]{ In the case the detected line for ID~3 is CO(7-6), we estimate $1.2 <M_{\rm H_{2}}/[10^{11} {\rm M_{\odot}}]< 4.5$ from the final CO SLEDs (see Fig. \ref{fig:models_literature}).}
\item[\emph{$\dagger$}]{ In the case the detected line for ID~39 is CO(7-6), we estimate $0.2 <M_{\rm H_{2}}/[10^{11} {\rm M_{\odot}}]< 0.6$ from the final CO SLEDs (see Fig. \ref{fig:models_literature}).}
\item[\emph{$\ddagger$}]{This value correspond to the dynamical mass for the disk scenario (see Section \ref{dis:q0115}).}
\item[\emph{$\diamond$}]{This value correspond to the dynamical mass for the merger scenario (see Section \ref{dis:q0115}).}
\end{tablenotes}                                                                                    
\end{threeparttable}
\end{table*}

\subsection{LVG Models and CO constraint}\label{section:massco}
The molecular gas mass can also be derived from CO following the equation
\begin{equation}\label{equation:massalpha}
    M_{\rm H_{2}} = \alpha L'_{\rm CO(1-0)}
\end{equation}
with $\alpha$ being the CO luminosity-to-gas mass conversion factor, and $L'_{\rm CO(1-0)}$ is the luminosity of the CO(1-0) that can be estimated from equation \ref{eq:luminosity}. In this work, we assume a value of $\alpha$ = 0.8 M$_{\odot}$ (K km s$^{-1}$ pc$^{2}$)$^{-1}$ \citep{1998downes}, which has been estimated for local ultra-luminous infrared galaxies and has been typically adopted to calculate molecular masses in high-redshift quasars \citep[e.g.,][]{2010wang,2017venemans}.

To estimate the molecular gas masses from the CO(6-5) and CO(7-6) transitions, we need to assume a CO spectral line energy distribution (hereafter SLED) to find the CO(1-0) line intensity. 
We did not find in the literature any quasar characterized by a SLED\footnote{There are only a handful of z$\sim$3 quasars with a well-characterized SLED (see Fig. \ref{fig:models_literature}).} that agrees with the observed constraints on the CO ratios.
For this reason, we modelled the CO SLED using the large velocity gradient (hereafter LVG) method, which has been widely applied to high-$z$ quasars by several authors \citep[e.g.,][]{2007weiss,2009riechers,2012schumacher}. These models include a velocity gradient (that indicates the change in the line of sight velocity in the turbulent medium) that is considerably larger than local thermal velocities of the gas, leading to photons being able to escape due to the different velocities along the cloud, following a photon escape probability. 
This method finds the populations of the molecular energy levels excited by collisions with H$_{2}$ (main collision partner for CO) for certain given parameters as CO abundance, kinetic temperature (T$_{\rm kin}$), H$_{2}$ density (n$_{\rm H_{2}}$) and velocity gradient (dv/dr). It is then possible to identify the 
physical parameters that best describe the conditions of the gas through the comparison of the model predictions to the observed line ratios \citep[e.g.,][]{2013carilli}.

\begin{figure}
    \centering
    \includegraphics[scale=0.55]{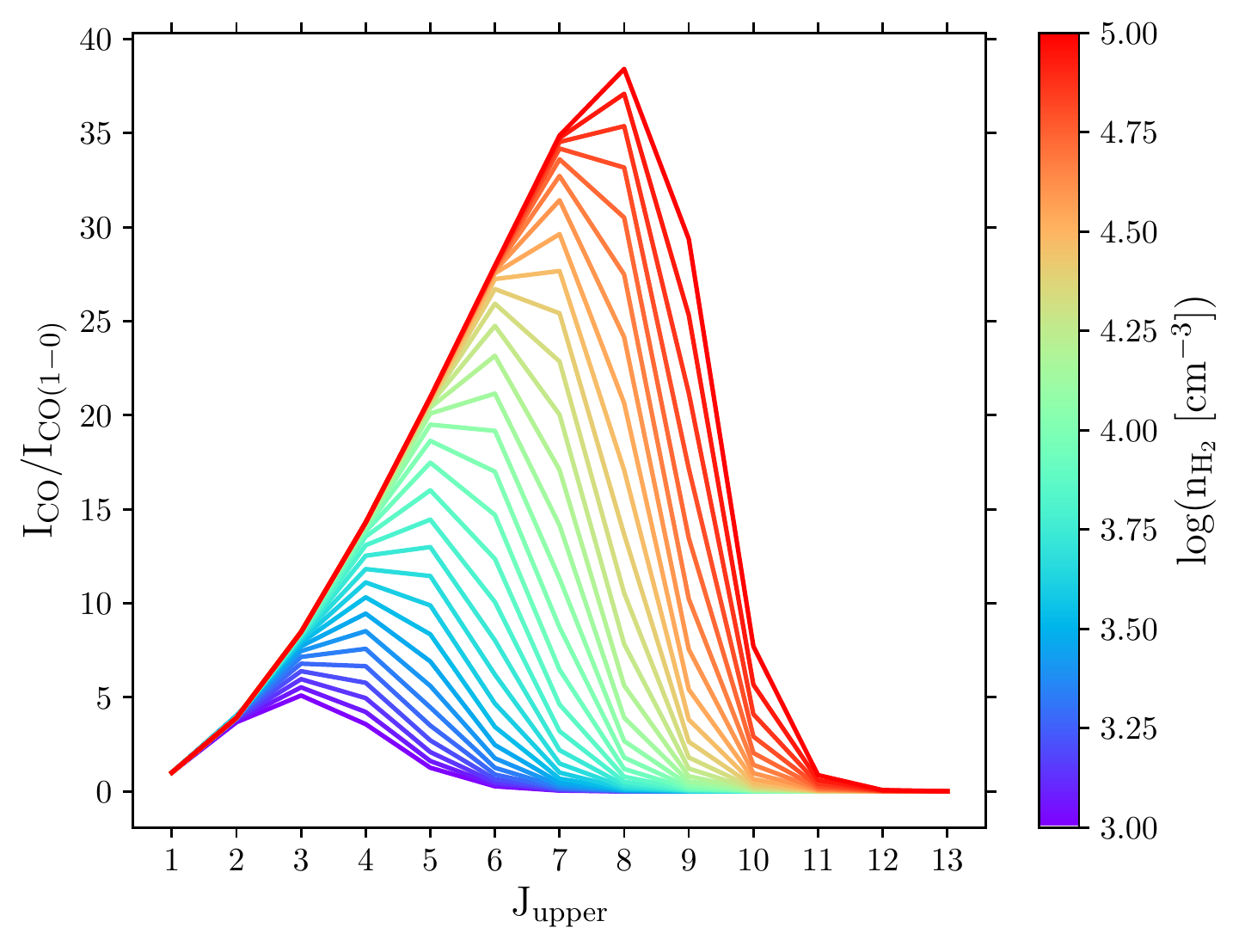}
    \caption{CO SLEDs obtained from LVG modelling, for T$_{\rm kin}$ = 30 K, using RADEX \citep{2007vandertak}.
    The CO line flux normalized to the CO(1-0) line  is plotted as a function of the upper rotational quantum number J.
    Different values in H$_{2}$ density are shown by the colour bar.}
    \label{fig:models_alldens}
\end{figure}

In this work, we have used the radiative transfer code \textsc{radex}\footnote{\url{https://home.strw.leidenuniv.nl/~moldata/radex.html}} \citep{2007vandertak}, considering a spherical and single component LVG model. For the calculations, we adopted an H2 ortho-to-para ratio of 3 and collision rates from \cite{2010yang}. We set the following input parameters and explore a grid of different models
in which we vary some of these parameters: T$_{\rm kin}$ equal to 30 K, n$_{\rm H_{2}}$ range of 10$^{3}$-10$^{5}$cm$^{-3}$ (typical for quasar host galaxies, \citealt{2013carilli}), background temperature of $\sim$11 K (cosmic microwave background at $z \sim$3), and a column density of the molecular gas given by:
\begin{equation}
     N_{\rm H_{2}} = 3.086\times 10^{18} n_{\rm H_{2}}\frac{\Delta V_{\rm turb}}{dv/dr}
\end{equation}
where $\Delta V_{\rm turb}$ corresponds to the turbulence line width fixed here at 100 km s$^{-1}$, and [CO]/$\rm dv/dr$ = 1$\times$10$^{5}$ pc (km s$^{-1}$)$^{-1}$, following procedures commonly adopted in the literature \citep[e.g.,][]{2007weiss,2012schumacher}.
LVG models are intrinsically degenerate in the parameters T$_{\rm kin}$ and n$_{\rm H_{2}}$, meaning that different combinations of these parameters can give the same SLED. We focused only on models with T$_{\rm kin}$= 30 K because the value of the kinetic temperature is expected to be comparable to the excitation temperature of neutral carbon \citep{2002israel}, which we assumed to be $\sim$30 K (see Section \ref{section:cimass}).  In total, 31 different CO ladders were modelled, which are shown in Fig. \ref{fig:models_alldens}. In this figure, the CO line flux normalized to the CO(1-0) line is shown as a function of rotational quantum number J, and the colour bar represents the different values of n$_{\rm H_{2}}$. 
The peak of the modelled ladders varies from J$\sim$3 to J$\sim$8, for n$_{\rm H_{2}}=$10$^{3}$ cm$^{-3}$ and 10$^{5}$cm$^{-3}$, respectively.
We note that  according to current observations, quasars are expected to show the SLED peak between J$\sim$6 and J$\sim$8  \citep[e.g.,][]{2009riechers,2011riechers,2013carilli,2018banerji,2019wang,2020bischetti}.

Importantly, the emitted CO flux is proportional to the source solid angle \citep[e.g.,][]{2007weiss}. As our observations are not able to constrain the size of the emitting source, we focus on line ratios in this paper. These LVG models will therefore represent only solutions for the component of highly excited gas, which likely coexist with a less excited component emitting the [C{\sc i}](2-1) emission that we observe in some of our targets. Indeed, it has been shown that emission from species with very different critical densities likely originate from gas at different densities (e.g., \citealt{2021harrington}), and also on different scales (e.g., \citealt{2014emonts,2018casey,2020spingola}).

We first find models that reproduce the observed upper limits on the CO(7-6)/CO(6-5) line ratios.
In our dataset  there are only 
three sources for which it was possible to estimate an observed $I_{\rm CO(7-6)}/I_{\rm CO(6-5)}$ 
upper limit: the quasars with IDs 8, 43 and 50 (see Table \ref{table:fluxes}). For these sources, we selected LVG models constrained by their observed $I_{\rm CO(7-6)}/I_{\rm CO(6-5)}$ ratio and found densities in the range n$_{\rm H_{2}} \sim$10$^{3}$-10$^{4.4}$cm$^{-3}$. These models constrained the peak of the SLED to be at J$\sim$3-7, and the molecular gas masses to be (0.5 - 64.1)$\times$10$^{11}$M$_{\odot}$ when considering the predicted $I_{\rm CO(1-0)}$ from the SLEDs.  
These results are shown in the fifth column of Table \ref{table:masses}. 

It is noteworthy that using a higher T$_{\rm kin}$ in our LVG models, for instance T$_{\rm kin}$ = 50 K (found in some high-$z$ quasars, e.g., \citealt{2007bweiss,2011riechers}), implies an upper limit in the molecular masses $\sim$32$\%$ lower, i.e. this difference does not alter our results significantly. In contrast, using a lower T$_{\rm kin}$, for example T$_{\rm kin}$ = 20 K, implies non-physical values for the upper limit in the molecular masses of up to $\sim$10$^{13}$M$_{\odot}$.

\subsection{Molecular mass constraints from a joint CO and [C{\sc i}] analysis} \label{sec:cociconstraints} 

After applying the CO constraint explained above, we set another condition based on the molecular gas masses already estimated from the [C{\sc i}](2-1) transition (see Section \ref{section:cimass}). As $\alpha$ is fixed, this condition is identical to impose a constraint on the CO(1-0) luminosity. We caution that this approach may introduce a bias in our calculation, which depends on the goodness of the assumed parameters to model the molecular mass from [C{\sc i}], and on $\alpha$ itself with respect to the physical conditions in each individual source. Observations of the CO(1-0) transition for these sources are definitely needed to confirm this methodology. 

\begin{figure} 
    \centering
    \includegraphics[scale=0.55]{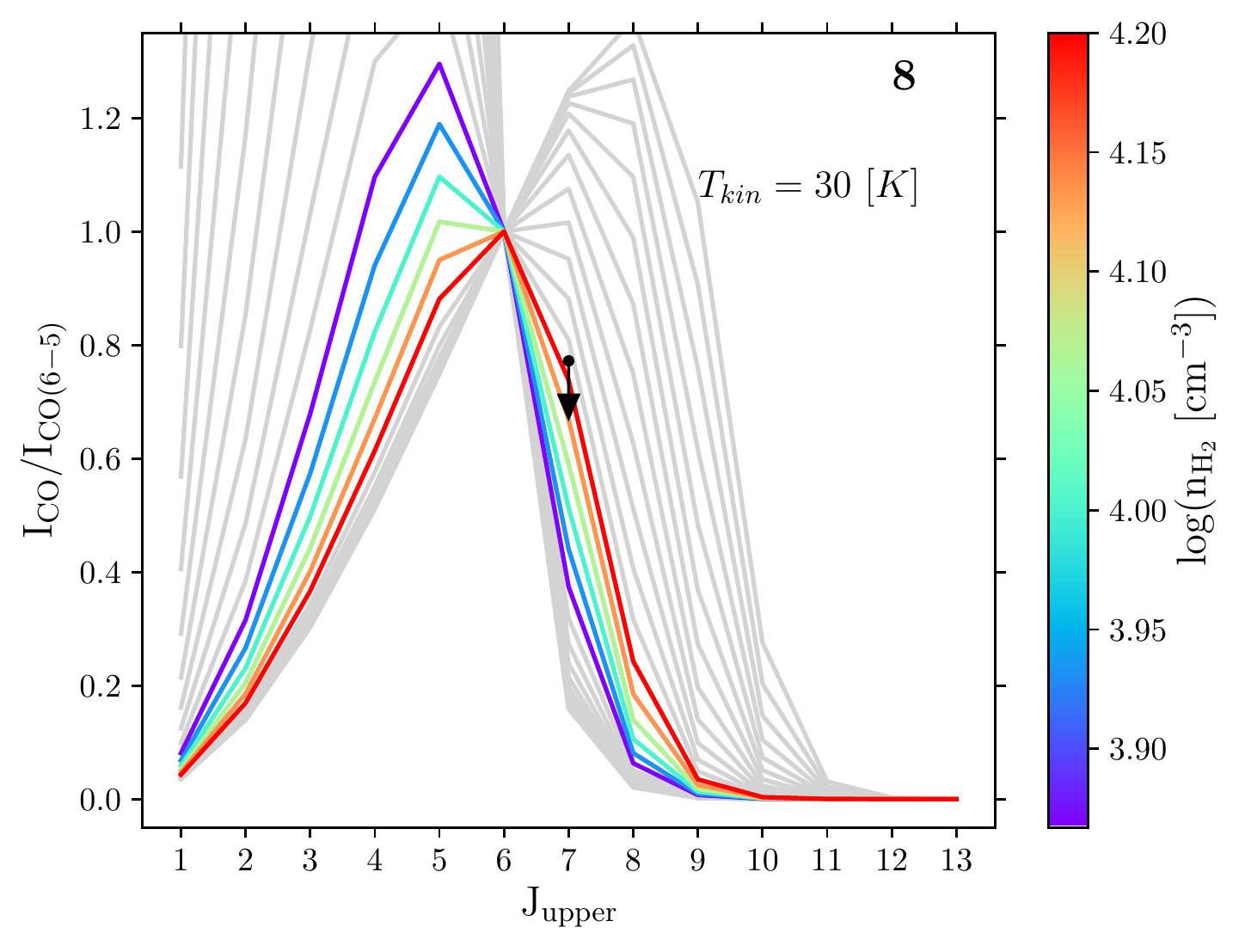}
    \includegraphics[scale=0.55]{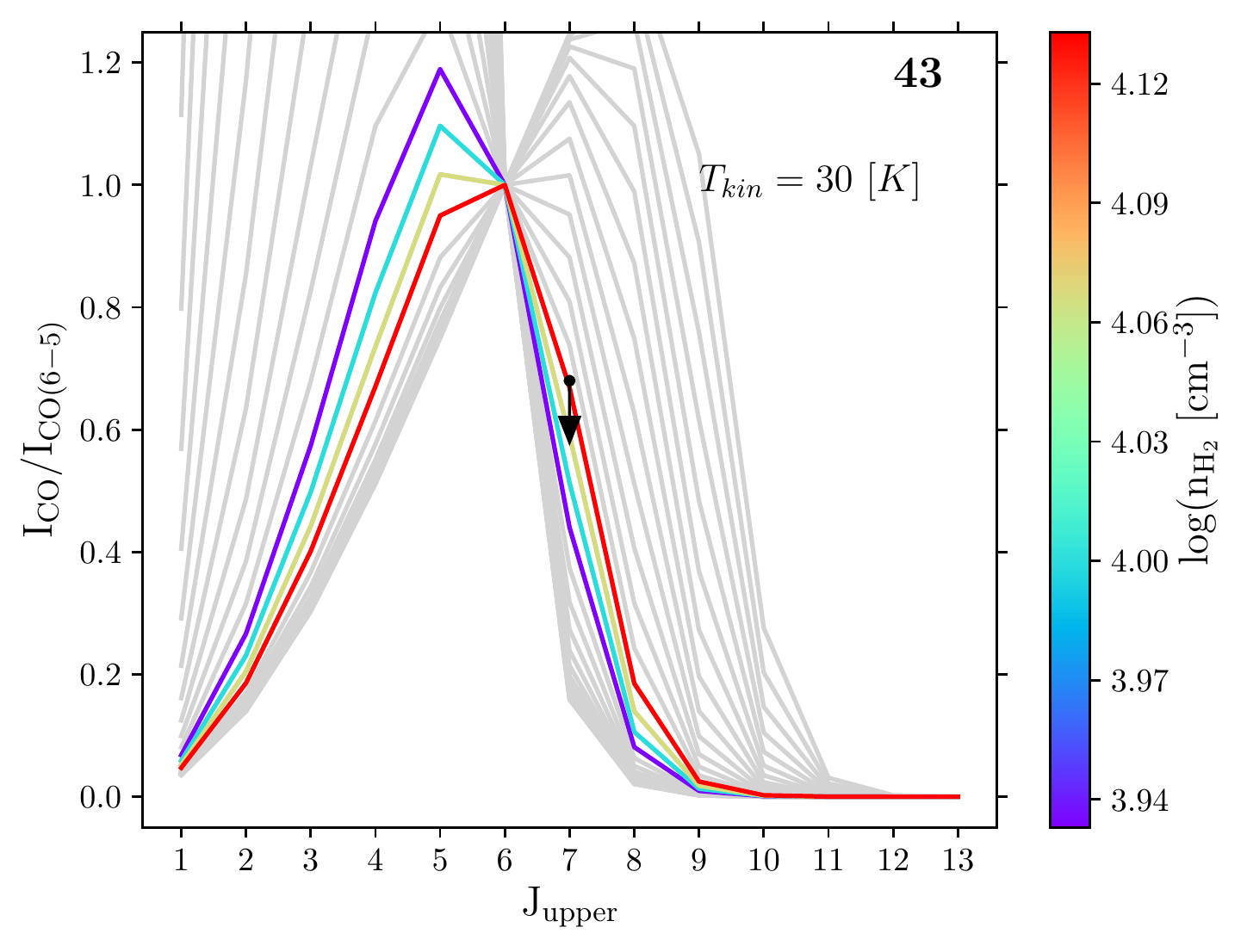}
    \includegraphics[scale=0.55]{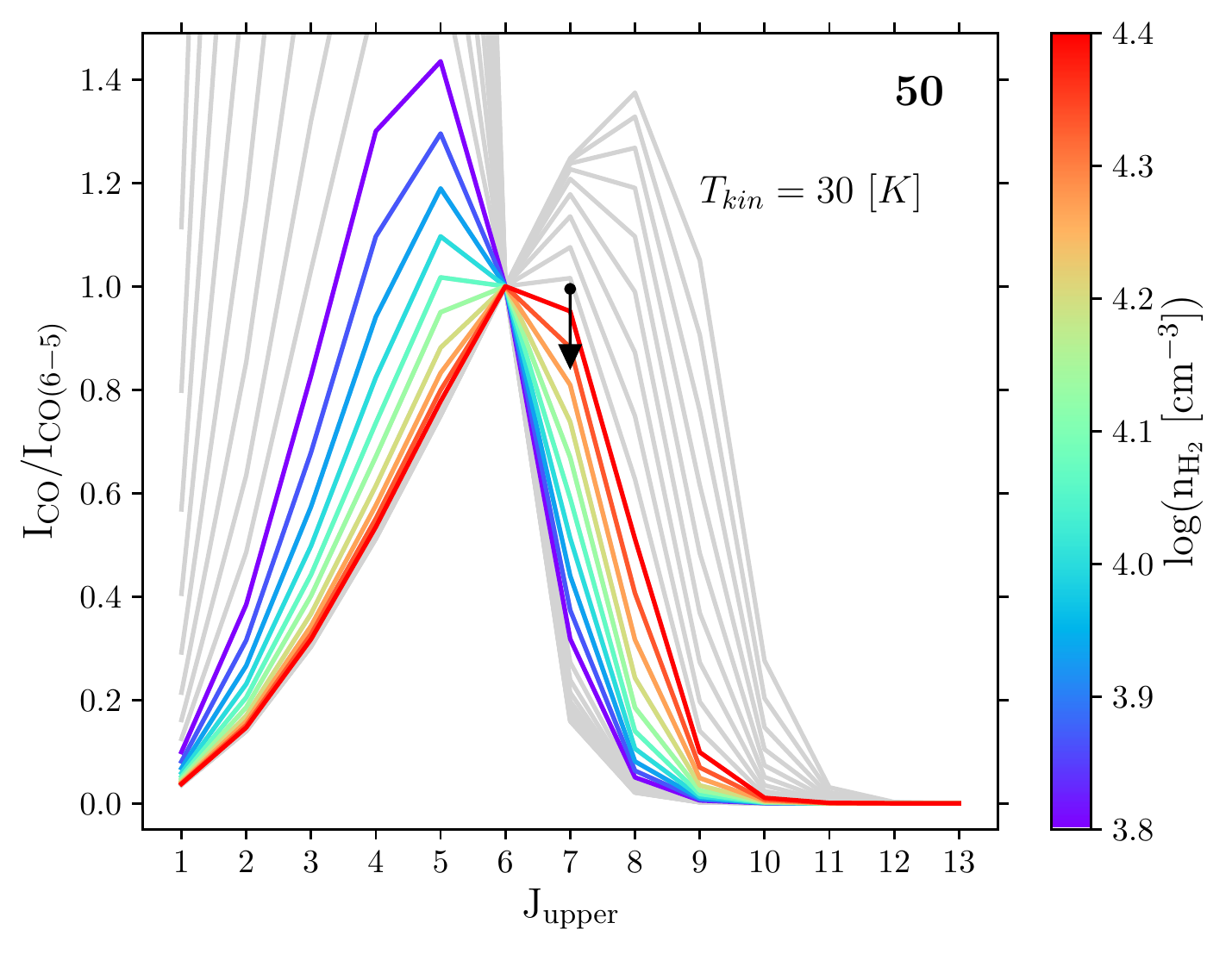}
    \caption{CO SLEDs (at T$_{\rm kin}$ = 30 K) obtained with the joint CO and [C{\sc i}] constraints, for the sources with IDs 8, 43 and 50 (top, middle and bottom panels, respectively), the SLEDS that can fit the joint constraints are represented by the coloured curves. The
    ladders show the CO line flux normalized to the CO(6-5) line as a function of rotational quantum number J, and the vertical black arrow represents the observed upper limit on the CO(7-6)/CO(6-5) ratio for each source. The different values in H$_{2}$ density are shown by the colour bar.  
    The grey curves in the background show the discarded CO SLEDs (from Fig. \ref{fig:models_alldens}).}
    \label{fig:models_3sources}
\end{figure}

For each source, this step excluded some of the LVG models at the lowest densities selected in Section \ref{section:massco}. In this way, we obtained the final CO SLEDs from the union of the  
CO and [C{\sc i}] constraints. 
We show these final constrained CO SLEDs in Fig. \ref{fig:models_3sources} for quasars with IDs 8, 43 and 50. 
Their ranges of molecular masses using these joint constraints are (0.7-1.5)$\times 10^{11} \rm M_{\odot}$, (0.5-0.9)$\times 10^{11} \rm M_{\odot}$ and (0.5-1.5)$\times 10^{11} \rm M_{\odot}$, respectively.  
These values are also listed in the second to last column of Table \ref{table:masses}.

\begin{figure}
    \centering
    \includegraphics[scale=0.55]{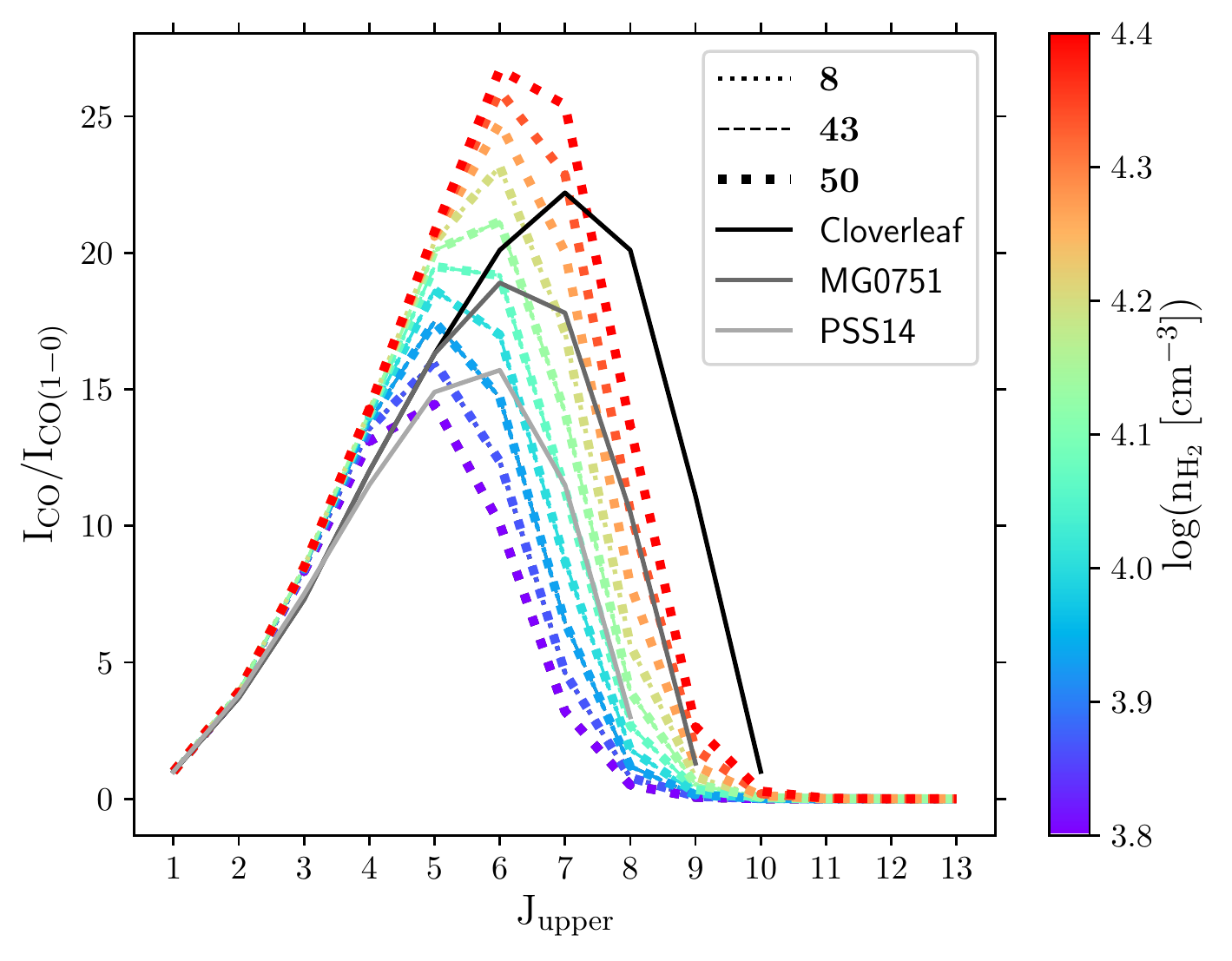}
    \caption{CO SLEDs for the sources with IDs 8, 43 and 50 (discontinuous lines; see legend) compared to
    the SLEDs for the quasars Cloverleaf, MG0751 and PSS14, obtained by \citet{2007bweiss} (continuous grey lines). The  ladders correspond to the CO line flux normalized to the CO(1-0) line as a function of rotational quantum number J.}
    \label{fig:models_literature}
\end{figure}

Fig. \ref{fig:models_literature} shows the CO SLEDs obtained in this work (discontinuous lines) in comparison to 
three $z\sim$3 quasars with well sampled  
SLEDs (grey continuous lines), Cloverleaf ($z$ = 2.6, \citealt{2007bweiss,2009bradford}), MG 0751+2716 ($z$ = 3.2, \citealt{2007bweiss}) and PSS1409 ($z$ = 2.6, \citealt{2007bweiss}). From this figure, we see that none of the SLEDs obtained in this work are similar in shape to those three obtained previously,
indicating that our quasars have different physical properties.  Our results yield SLEDs with J peak somewhat lower than the expected range for quasars (J$\sim$6-8), varying from J$\sim$5 to J$\sim$7. 

The six sources with IDs 3, 7, 13, 18, 29 and 39, have only upper limits for the targeted CO transitions. 
Our observations are therefore not able to put any constraint on the SLED of these sources. To compute the upper limits for the CO(1-0) luminosity, we assumed that the models selected for the previous three sources with IDs 8, 43 and 50, apply to the full observed sample.

We first selected the CO SLEDs found for the quasars with IDs 8, 43 and 50 (shown in Fig. \ref{fig:models_literature}).  
Depending on which emission line upper limit flux was measured for each of the six sources mentioned above, we used the CO(1-0)/CO(6-5) and/or CO(1-0)/CO(7-6) ratios.  
Then, we multiplied these CO ratios by the observed upper limit flux for each of them. After this,  we estimated upper limits for the  molecular gas masses of 
the six quasars with IDs 3, 7, 13, 18, 29 and 39.
For the sources that have flux upper limits for both CO lines (quasars with IDs 13, 29 and 39), two upper limits in 
molecular gas mass were derived and the higher value was selected.

To summarize: the resulting molecular masses of our sample, obtained from applying only the CO constraints, are in the range of (0.5 - 64.1)$\times$10$^{11}$M$_{\odot}$ for the sources with IDs 8, 43 and 50, and are <6.9$\times$10$^{11}$M$_{\odot}$ for the objects with IDs 3, 7, 13, 18, 29 and 39\footnote{ We note that the latter upper limit is based on the final CO SLEDs obtained for the sources with IDs 8, 43, and 50 (Figure~\ref{fig:models_literature}).}. We note that the range mentioned includes masses of the order of 10$^{12}$M$_{\odot}$, which are implausible given the expected dark-matter halo mass for these quasars ($M_{\rm DM\sim}10^{12.5}$~M$_{\odot}$, e.g., \citealt{2012white}).
These values are also much higher than the values found in the literature for other high-$z$ quasars ($\sim$10$^{9}$-10$^{11} \rm M_{\odot}$, e.g., \citealt{2003weiss,2011walter,2013anh,2019hill}). These high masses are caused by models with 
densities of n$_{\rm H_{2}} \sim$10$^{3}$-10$^{3.8}$cm$^{-3}$. This disagreement with the literature values suggests that
such low densities are not plausible in explaining our current constraints on the emission from the high-J CO transitions. We note that the molecular gas masses obtained from applying the CO and [C{\sc i}] constraints jointly, remove such models.\\

We emphasize that the final selected densities (n$_{\rm H_{2}} \sim$10$^{3.8}$-10$^{4.4}$cm$^{-3}$, see Fig. \ref{fig:models_literature}) represent those values needed by the high-J CO lines to match the molecular masses estimated from [C{\sc i}](2-1). These densities should not be regarded as representative of those emitting the [C{\sc i}](2-1) line, unless this emission is more extended than the high-J CO transitions.\footnote{ For completeness, we report here the predictions of the [C{\sc i}](2-1)/CO(6-5) ratio using our 10 finally selected LVG models. These models with ${\rm T_{\rm kin}=30}$~K are characterized by [C{\sc i}](2-1)/CO(6-5)<1.33. In the case of ID~29, this value is lower than the observed [C{\sc i}](2-1)/CO(6-5)>4.36. This apparent discrepancy could be solved by invoking a [C{\sc i}](2-1) solid angle $>~ 3\times$ the high-J CO solid angle (which corresponds to $>~ 1.7\times$ the high-J CO source radius) and/or with multi-component models (e.g., \citealt{2003weiss,2021harrington}).} \\

The molecular masses that are finally derived after applying a more restricted set of
models are in the range of 0.4 - 6.9$\times$10$^{11}$M$_{\odot}$ for quasars with IDs 3, 8, 29, 39, 43 and 50, and  <1.1$\times$10$^{11}$M$_{\odot}$ for quasars with IDs 7, 13 and 18. The brightest of our detected sources  still has an estimated molecular gas mass
that is higher than typically found in the literature. We will discuss this further in the next sections.

We finally check the molecular gas mass estimate for ID~3 and ID~39, for which we identify [C{\sc i}](2-1) based on the uncertain \ion{C}{iv} systemic redshift (section~\ref{sec:detections}). In the case the detected lines are CO(7-6) instead of [C{\sc i}](2-1), the final models shown in Figure~\ref{fig:models_literature} would imply molecular gas masses of (1.2 - 4.5)$\times$10$^{11}$M$_{\odot}$ and (0.2 - 0.6)$\times$10$^{11}$M$_{\odot}$, respectively. These values would still confirm our findings, with ID~3 having a large molecular mass. 

\section{Comparison with the large-scale Ly$\alpha$ emission} \label{section:comparisonlya} 

To investigate any relation between the molecular gas phase and the cool halo gas, in this section we compare the APEX observations 
with the Ly$\alpha$ line properties of the associated large-scale nebulae discovered with MUSE \citep{2019arrigoni}. First, we focus on finding any trend in the properties of each phase when compared with the other. Secondly, we study the line emission locations and profiles.

\subsection{Molecular lines versus Ly${\rm \alpha}$ line properties} 
\label{sec:lyaphysicalprop}

The FWHM of molecular lines from high-redshift galaxies are frequently used as a dynamical mass estimator \citep[assuming a rotating disk geometry, e.g.,][]{2003walter,2009narayanan,2013wang}, and hence their stellar mass and therefore their halo mass can be determined (see Section \ref{sec:dis:dynamicalmass} for further discussion). As halos of different masses are expected to be characterised by different fractions of cool and hot gas \citep[e.g.,][]{2006dekel}, the FWHM of molecular lines could show important trends with the Ly$\alpha$ properties. For instance, more massive halos should in principle show larger FWHM of the molecular lines and smaller fractions of cool gas, with consequently smaller Ly$\alpha$ luminosities (L$_{\rm Ly\alpha}$) and Ly$\alpha$ areas compared to less massive halos.
We start by comparing the FWHM of the observed molecular lines with the total L$_{\rm Ly\alpha}$, the average Ly$\alpha$ surface brightness SB$_{\rm Ly\alpha}$, and the area encompassed by the Ly$\alpha$ emission by the nebulae surrounding the quasars in our sample.

\begin{figure*}
    \centering
    \includegraphics[width=\linewidth]{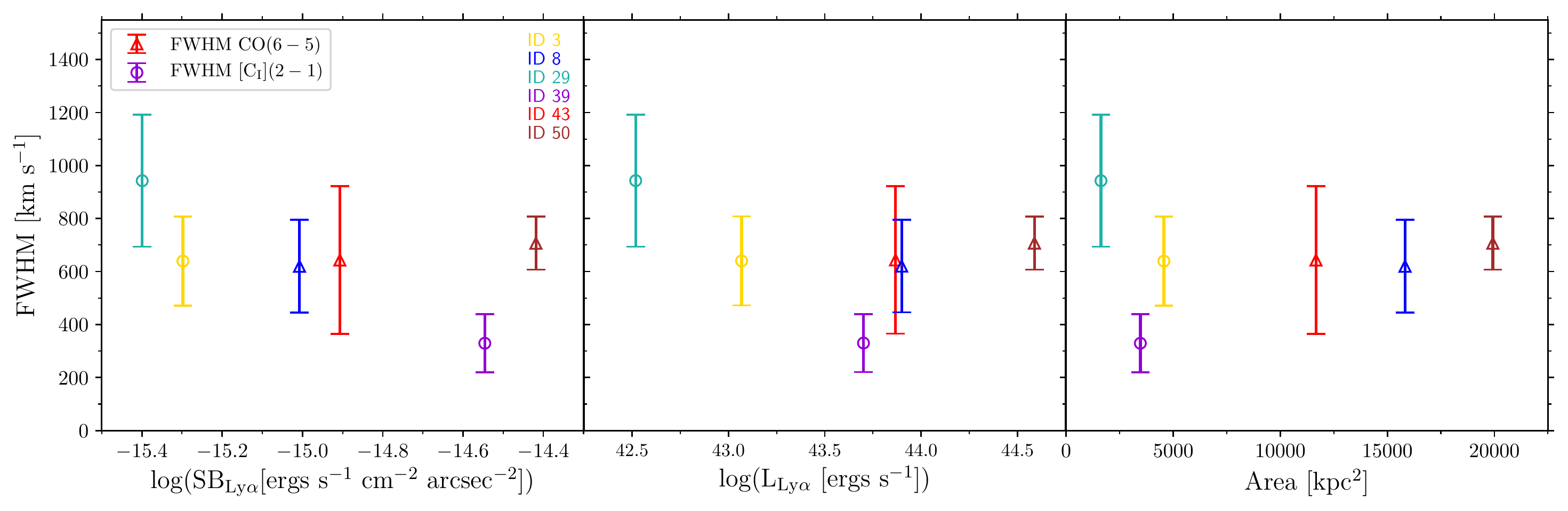}
    \caption{FWHM of the detected CO(6-5) (triangles) and [C{\sc i}](2-1) (circles) emission lines versus Ly$\alpha$ surface brightness corrected for cosmological dimming (first panel), the total Ly$\alpha$ luminosity (second panel),  and physical area (third panel) for the large-scale Ly$\alpha$ nebulae discovered around the quasars in our sample 
    \citep{2019arrigoni}. 
    }
    \label{fig:fwhmlya}
\end{figure*}

Figure \ref{fig:fwhmlya} shows this comparison for the different sources in our sample (represented by different colours). Specifically, 
the left panel shows the FWHM of the molecular lines as a function of SB$_{\rm Ly\alpha}$ (corrected for the cosmological dimming), the central panel shows the FWHM versus L$_{\rm Ly\alpha}$, and the right panel shows the FWHM versus the area of the Ly$\alpha$ nebulae. The legend in each panel indicates the different markers used for the different molecular lines (CO(6-5) and [C{\sc i}](2-1)).

From this figure, we find that there is no clear correlation between FWHM of the molecular lines and Ly$\alpha$ properties. Also, we note that all the quasars 
show similar values of FWHM (in the range of 329 - 943 km s$^{-1}$, average FWHM 647$\pm$129 km s$^{-1}$), considering the uncertainties. 
In the literature, the molecular linewidths (CO and [C{\sc i}] lines) found for high-$z$ quasars have values between $\sim$150 - 450 km s$^{-1}$ \citep[e.g.,][]{2003weiss,2007weiss,2011walter,2012schumacher,2017venemans}. The values found for our quasars are larger
on average. For example, quasars with IDs 29 and ID 50 have a FWHM of  943$\pm$249 km s$^{-1}$ ([C{\sc i}](2-1) emission) and 707$\pm$100 km s$^{-1}$ (CO(6-5) emission) respectively, suggesting that they have a different kinematics and/or physical properties 
compared to the quasars studied in the literature.

As molecular gas is expected to form from the cooling of 10$^{4}$K gas \citep[e.g.,][]{1972dalgarno}, one may naively expect to observe the largest molecular reservoirs in sources with the largest and brightest Ly$\alpha$ nebulae.
As a next step, we  focus on the molecular gas masses obtained from the joint constraints from CO and [C{\sc i}] (see Section \ref{sec:cociconstraints}), and compare them to L$_{\rm Ly\alpha}$, SB$_{\rm Ly\alpha}$ and the area of the large-scale nebulae.

\begin{figure*}
    \centering
    \includegraphics[width=\linewidth]{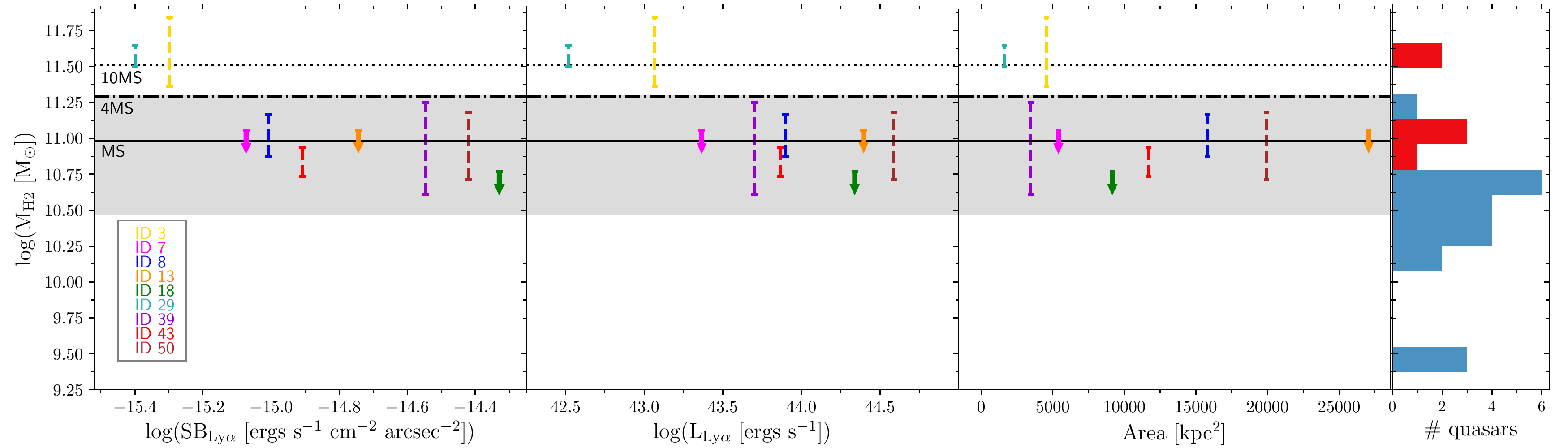}
    \caption{Molecular gas mass $\rm M_{H2}$ derived from the joint CO and [C{\sc i}] constraints
    versus Ly$\alpha$ surface brightness corrected for cosmological dimming (first panel), the total Ly$\alpha$ luminosity (second panel),  and physical area (third panel) for the large-scale Ly$\alpha$ nebulae discovered around the quasars in our sample (\citealt{2019arrigoni}). 
    The ID numbers of the sources are shown in the first panel. The vertical dashed lines represent ranges of $\rm M_{H2}$, and the arrows indicate upper limits (see Table~\ref{table:masses}). The horizontal lines show the expected location of massive star-forming main sequence MS, 4$
    \times$MS and 10$
    \times$MS galaxies using the gas fractions in \citet{2018tacconi} and the halo mass-stellar mass relation in \citet{2018moster}. The grey area is the uncertainty on this expectation, for 1$\times$MS. The histogram show the $\rm M_{H2}$ distribution for our sample considering the average values of the mass ranges (red), and for a sample of $z\sim$2.5 - 3 quasars (blue) extracted from the literature (\citealt{2002barvainis,2003weiss,2004beelen,2011walter,2012schumacher,2019hill,2020bischetti}), whose $\rm M_{H2}$ have been estimated with different molecular gas tracers and have a typical uncertainty of $\sim\ 0.1$ dex. 
    }
    \label{fig:masseslya}
\end{figure*}

The results of this analysis are presented  in Figure \ref{fig:masseslya}, where 
the colours represent different sources, the vertical dashed lines are ranges and the arrows upper limits of the estimated molecular gas masses.
The horizontal black lines represent the expected molecular mass for the quasar host galaxies on the main sequence (MS) of star-forming galaxies (log($M_{\rm H_{2}}[M_{\odot}]$) = 10.98$^{+0.32}_{-0.51}$), on 4$\times$MS (log($M_{\rm H_{2}}[M_{\odot}]$) =11.29) and 10$\times$MS (log($M_{\rm H_{2}}[M_{\odot}]$) =11.51). These values were computed under the following assumptions:
i) a quasar halo mass of log($M_{\rm halo}[M_{\odot}]$) = 12.68$^{0.81}_{-0.67}$ (for z$\sim$3 quasars), estimated in the study of  \cite{2008kim} from Ly$\alpha$ forest statistics (this value encompasses estimates from quasar clustering, e.g., \citealt{2007shen,2012white,2018timlin}), 
ii) the $M_{\rm halo}-M_{*}$ relation from \cite{2018moster} to estimate the stellar mass
of the objects in our sample, and iii) a molecular gas fraction expected for objects
on MS, 4$\times$MS and 10$\times$MS of star-forming galaxies, 
as defined in the empirical relation of \cite{2018tacconi}. The grey shaded region represents the large uncertainties 
in the calculation for 1$\times$MS. We note that quasars hosts are estimated to have star formation rates higher or similar to 
MS galaxies at the same redshift (e.g., \citealt{2016zhang,2020shangguan,2020jarvis,2021circosta}).
The histogram shows the $M_{\rm H_{2}}$ distribution for our sample (red) and for a
sample of $z\sim$2.5 - 3 quasars (blue) extracted from the literature as explained in the figure caption. The molecular masses reported for these quasars are overall below the computed predictions (see histogram in Fig.~\ref{fig:masseslya}), possibly indicating some gas depletion with respect to similarly massive star-forming galaxies.

It is clear that the molecular gas masses of the quasars with APEX detections are in agreement with the expected values for MS galaxies, 
with  the exception of  quasars with IDs 3 and 29, which have masses well above the MS. 
Intriguingly, these two objects characterised by the highest molecular gas masses
are associated with the Ly$\alpha$ nebulae with the lowest values of L$_{\rm Ly\alpha}$ and SB$_{\rm Ly\alpha}$.  
The third panel of Fig. \ref{fig:masseslya} 
shows that the most massive molecular reservoirs are associated with some of the smallest nebulae.
They also have the highest FWHM  [C{\sc i}](2-1) emission lines (see Fig. \ref{fig:fwhmlya}). We will discuss this 
result in more detail in Section \ref{sec:dis:molmassneb}.

\subsection{Molecular tracers vs Ly$\alpha$ line profiles} 
\label{sec:lyacoprofiles}

Now that we have a more robust estimate of the systemic redshift $z_{\rm mol,sys}$ for some of our sources, we can compare its location with the redshift of the discovered Ly$\alpha$ nebulae, assessing if the Ly$\alpha$ emission spans similar velocity ranges with respect to the molecular phase. For this purpose, we compare the molecular and Ly$\alpha$ line profiles extracted within the same aperture (the APEX beam; see Fig.~\ref{fig:apexbeam}), and also compare those with the quasar spectra.  

\begin{figure*}
    \centering
    \includegraphics[scale=0.75]{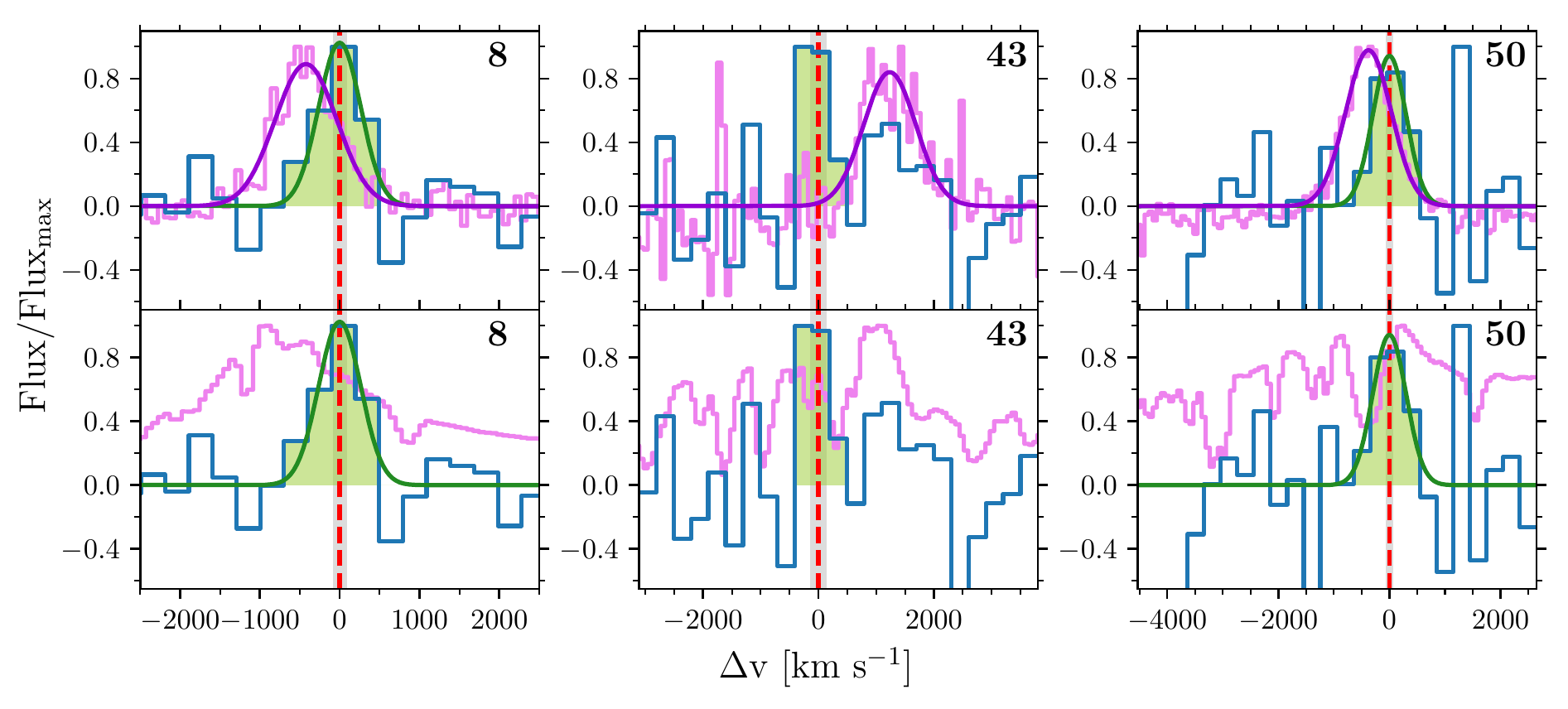} 
    \caption{\emph{Top panels:} Normalized APEX CO(6-5) emission (blue) of the three detected sources, ID~8, ID~43, and ID~50 compared with the nebulae Ly$\alpha$ emission from MUSE (magenta) extracted within the APEX beam (see Figure~\ref{fig:apexbeam}). The APEX spectra are shown with a bin size of 300 km s$^{-1}$.  Each quasar ID number is shown in the top right corner of each panel. 
    The red vertical lines with grey shaded area correspond to the systemic redshift estimated in this work with its uncertainty. 
    The dark green curves and green shaded areas follow the same notation as in Figure \ref{fig:co65spec}. 
    To guide the eye, the purple curves represent a Gaussian fit applied to the Ly$\alpha$ emission lines. \emph{Bottom panels:} same as above, but comparing with the quasar spectra extracted 
    within a circular aperture of 1.5 arcsec radius.}
    \label{fig:lyaco65spec}
\end{figure*}

\begin{figure*}
    \centering
    \includegraphics[scale=0.75]{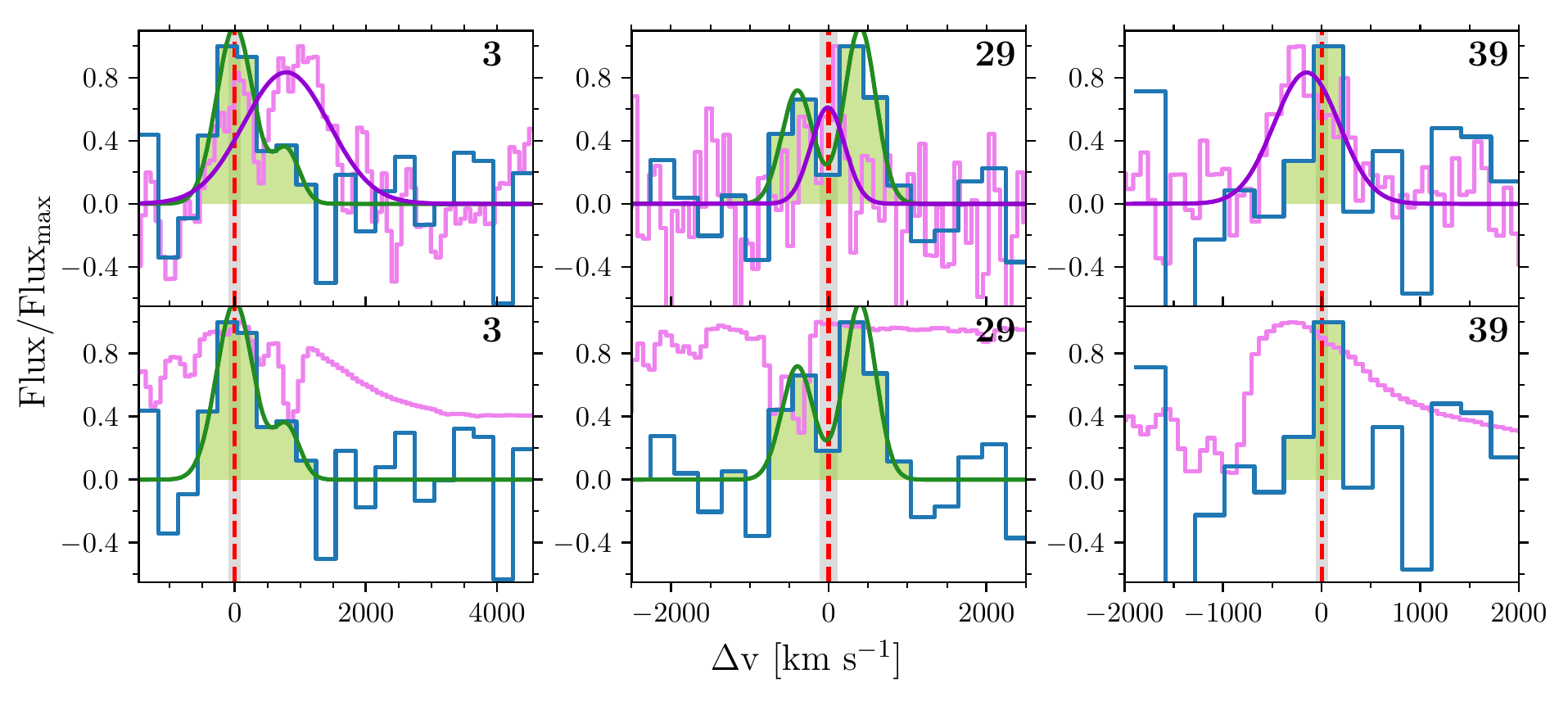}
    \caption{Same as Figure \ref{fig:lyaco65spec}, but for the APEX [C{\sc i}](2-1) detections (blue).}
    \label{fig:lyaco76spec}
\end{figure*}

Figures \ref{fig:lyaco65spec} and \ref{fig:lyaco76spec} show the normalized APEX CO(6-5) and [C{\sc i}](2-1) 
detections (blue) in comparison to their normalized Ly$\alpha$ nebula spectrum (upper row) and quasar spectrum (bottom row).
The quasar spectra  have been  extracted within circular apertures of 1.5 arcsec radius. The vertical red dashed line represents the current systemic redshift estimated from the molecular lines and the grey shaded area corresponds to its uncertainty (see Section \ref{section:redshifts}).

From both figures, we note that some of the peaks of the Ly$\alpha$ nebulae present a significative shift from the current systemic redshift.  For each object, we listed these velocity shifts $\Delta v_{\rm Ly\alpha Neb-mol}$ between the quasar systemics $z_{\rm mol, sys}$ and the Ly$\alpha$ of the nebulosities in the last column of Table \ref{table:redshifts}. The values of $\Delta v_{\rm Ly\alpha Neb-mol}$ are in the range of -423 to 1236 km s$^{-1}$, with the quasars with ID~8 (or UM683) and ID 43 (or CTSH22.05) having the bluest and the reddest shift, respectively. The velocity shifts do not show any trend with respect to the Ly$\alpha$ physical properties explored in this paper.
Considering that the Ly$\alpha$ photons experiment changes in frequency due to the scattering processes, these shifts can be an indication of bulk inflows (blueshift) or outflows (redshift) \citep[e.g.,][]{2015prescott,2017dijkstra}. 

\begin{figure}
    \centering
    \includegraphics[width=0.95\columnwidth]{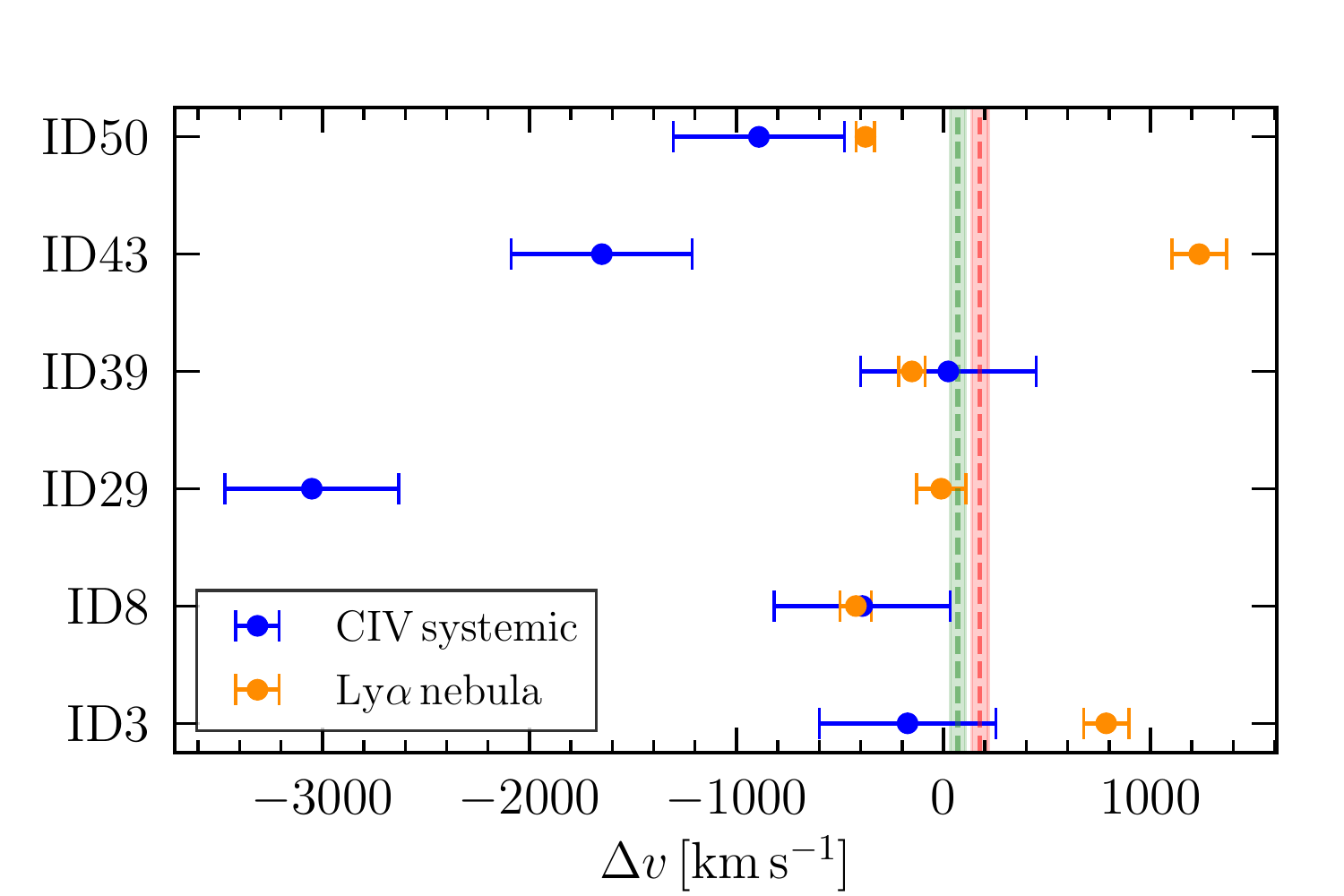}
    \caption{Velocity shifts between $z_{\rm sys}$ (from C{\sc iv}) and $z_{\rm mol,sys}$ (blue), and redshift of Ly$\alpha$ nebulae and $z_{\rm mol,sys}$ (orange). The green and red vertical dashed lines with errors are the average shift for Ly$\alpha$ nebulae for the sample in \citet{2019farina}
    (69$\pm$36 km~s$^{-1}$; for 9 z$\sim$6 quasars with [C{\sc ii}] systemic redshifts), and for our work (176$\pm$39 km~s$^{-1}$), 
    respectively.}
    \label{fig:all_shifts}
\end{figure}

Figure \ref{fig:all_shifts} shows the velocity shifts  between $z_{\rm sys}$ and $z_{\rm mol,sys}$ (blue), and redshift of the Ly$\alpha$ nebulae and $z_{\rm mol,sys}$ (orange). The now obtained systemic redshifts from molecular tracers are found to be, in most of the cases, more consistent with the Ly$\alpha$ nebular redshifts, showing an average shift of 176$\pm$39~km s$^{-1}$ (red vertical line). We note that this shift is larger than the average shift found at $z\sim6$, 69$\pm$36~km~s$^{-1}$ \citep{2019farina}, obtained by comparing [C{\sc ii}] redshifts and the nebular redshifts for nine sources (green vertical line). 
The small number statistics hampers any conclusion from this comparison. For example, if we remove from our sample the radio-loud object, ID~3, we would get a smaller average shift of 46$\pm$35 km~s$^{-1}$, which is consistent with the work by \cite{2019farina}.

This analysis shows that the systemic redshifts obtained from C{\sc iv} are not reliable in 4 out of 6 cases ($\sim70$\%). This poor reliability was already indicated by the large velocity shifts of the extended nebular Ly$\alpha$ emission and the C{\sc iv} redshifts in surveys targeting $z\sim2-3$ quasars, reporting values as high as $\sim6000$ km~s$^{-1}$ \citep{2019arrigoni,2019cai}. The same works found that the peak of the nebular emission has smaller velocity shifts with respect to the observed peak of the Ly$\alpha$ emission of the quasar. Figures \ref{fig:lyaco65spec} and \ref{fig:lyaco76spec} indicate that for 3 out of 6 quasars (ID 3, 29, and 50), this was the case because of that peak being at the real systemic.
We discuss the most significant shifts between  $z_{\rm mol,sys}$ and the redshift of the Ly$\alpha$ nebulae in Section~\ref{sec:dis:radiative}.

\section{Discussion}\label{section:discussion}

In this section, we discuss different scenarios regarding the quasar host properties found in this work, and we 
try to interpret the link between the molecular gas content and the large scale Ly$\alpha$ emission. Future studies are essential to confirm or reject what we propose in the following sections.

\subsection{Quasar host properties}
\subsubsection{Dynamical masses}\label{sec:dis:dynamicalmass}

For high-$z$ galaxies, it is possible to derive the dynamical mass of the system using the formula commonly used in the literature \citep[e.g.,][]{2003walter,2013wang,2016venemans}, assuming rotational support: $ M_{\rm dyn} = 2.33\times 10^{5} v^{2}_{\rm circ} R [M_{\odot}]$, where $R$ is the disk radius from the observed line and $v_{\rm circ}$ is the maximum circular velocity of the gas disk, which is calculated as $v_{\rm circ} = 0.75\times {\rm FWHM}/{\rm sin}(i)$, i.e. it depends on the FWHM of the observed molecular line 
and inclination $i$ of the galaxy (with respect to the plane of the sky). From our molecular data it is not possible to estimate the parameters $R$ and $i$, for this reason we proceed with a set of assumptions.  Arrigoni Battaia et al. in preparation studied the quasar with ID 13 (or PK-1017+109) through ALMA observations, and found  a radius of $\sim$3~kpc for this object. This is on the high side of 
the typical disk radii range found in literature (from molecular gas) for high-z quasar hosts ($\sim$0.5 - 2.5 kpc, e.g.,\citealt{2003carilli,2004walter,2008aravena,2009riechers,2011polletta,2012schumacher,2021molina,2021stacey}). 
Here we adopt $R$ = 3 kpc, which is
in agreement with our ALMA observations of one of these systems. 

Applying the above to five sources in our sample with FWHM estimations, excluding quasar ID 29 that is discussed in Section \ref{dis:q0115}), we find inclination-dependent dynamical masses in the range of $ M_{\rm dyn}{\rm sin}^2(i) \sim (0.4 - 1.9)\times 10^{11} \rm M_{\odot}$ for quasars with IDs 3, 8, 39, 43 and 50.
The quasar with ID 39 (or SDSS J0100+2105)  is the least massive 
and the quasar with ID 50 (or SDSS J0819+0823) the most massive. The specific values for these dynamical masses are tabulated in the last colum of Table \ref{section:allmasses}, with uncertainties considering only the errors from their FWHMs. Even though these dynamical masses are highly uncertain, we can get a rough idea on the inclination. Indeed, assuming that the amount of dark matter is negligible on host scales, $M_{\rm dyn} \approx M_{*} + M_{\rm H_{2}}$, where $ M_{*}$ is the stellar mass. Assuming that the stellar mass is at least as high as the highest allowed molecular mass from our calculations, we find that two sources, IDs~3 and 39, require relatively small angles ($\sim20^\circ$), while the other four sources, IDs~8, 29, 43 and 50, require an inclination ($\sim50^\circ-60^\circ$) close to the mean expected value of ${\rm sin}(i)=0.79$ ($i$ = 57.3$^{\circ}$, \citealt{2009law}).

One  caveat is that the estimates are  subject to the different molecular lines used.
We computed that the CO lines reported in
the literature for $z\sim 3$ quasars are on average larger than [C{\sc i}] lines by 112$\pm$43 km~s$^{-1}$ (e.g., \citealt{2003weiss,2012schumacher,2018banerji,2019yang}).
As these transitions have different critical densities, this could be an indication that the lines trace different gas components, which can 
be the case if [C{\sc i}] and the  high-$J$ CO transitions (as CO(6-5)) 
traces gas also on different scales, with [C{\sc i}] extending to outer scales  \citep{2012schumacher}.

\subsubsection{The case of Q-0115-30 (ID 29): disk or merger?}\label{dis:q0115}

\begin{figure}
    \centering
    \includegraphics[width=0.95\columnwidth]{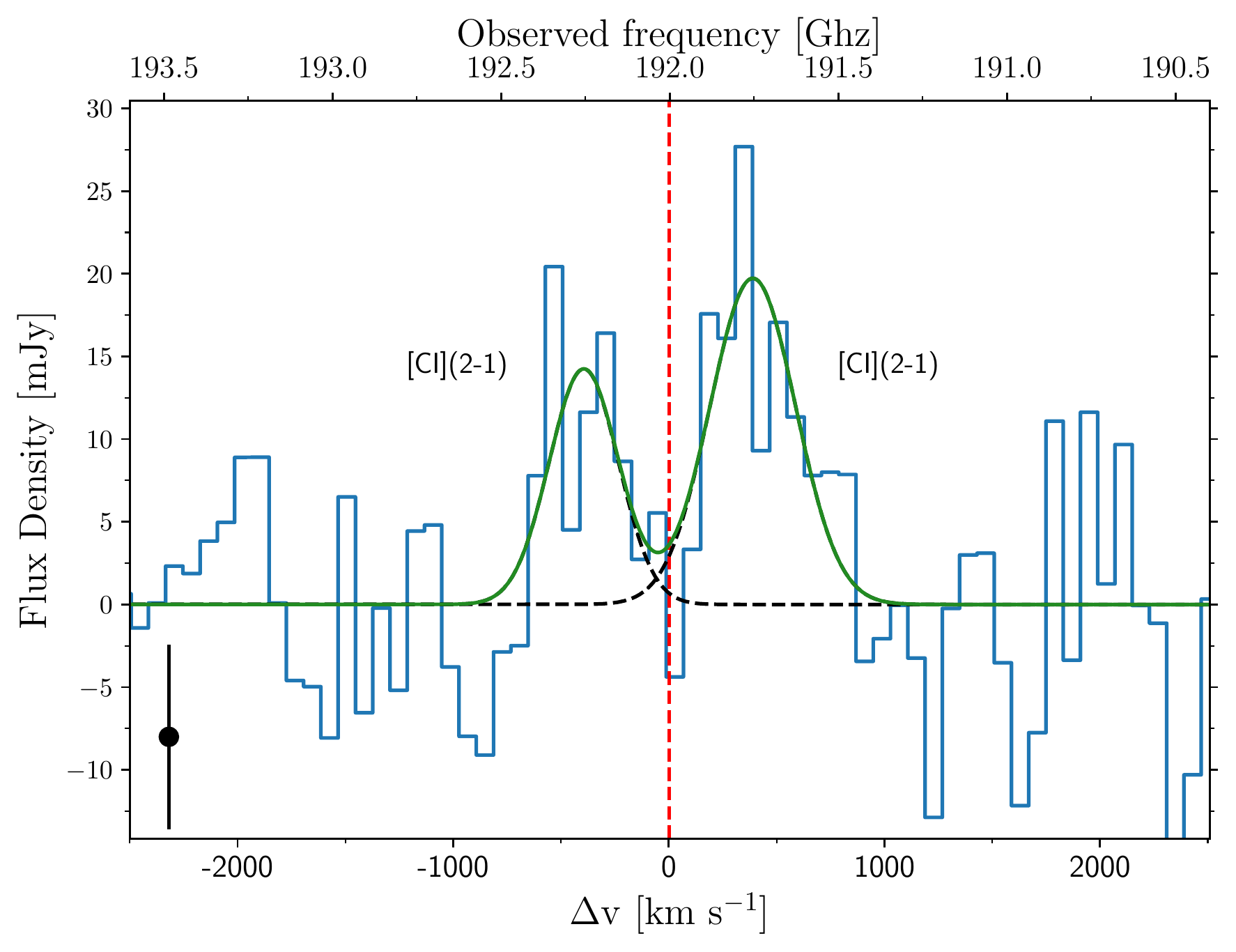} 
    \caption{APEX [C{\sc i}](2-1) spectral line observation for Q-0115-30 (quasar with ID~29). The spectrum is shown with a bin size of 80 $\rm km\ s^{-1}$, the highest resolution possible. The red vertical line represent the systemic redshift estimated from the molecular lines. The curves and symbol follow the same notation as Figure \ref{fig:co76spec}.}
    \label{fig:spectrumID29}
\end{figure}

In Section \ref{sec:detections}, we showed that the quasar with ID 29 (or Q-0115-30) has a [C{\sc i}](2-1) line emission consisting
of two Gaussian components separated by $\sim$800 km~s$^{-1}$. While this line identification needs further confirmation, in this section we discuss their possible origins. 
This double-peaked feature is unique within our sample and could be 
produced by a rotating disk or a merger between two molecular gas reservoirs \citep[e.g.,][]{2003neri,2006narayanan,2008greve,2011polletta}.
In Figure \ref{fig:spectrumID29} we show this double-peaked line emission at the highest resolution possible (bin size of 80 $\rm km\ s^{-1}$) 
for which there is still a  decent peak S/N of $\sim$2.5 and $\sim$3.5.

We now calculate the  dynamical masses, under the assumption that the system is virialized.
For the case of the gas distributed in a rotating disk, we estimate the dynamical mass $M_{\rm dyn}^{\rm disk}$ as explained in Section \ref{sec:dis:dynamicalmass}, but considering the velocity difference between the two [C{\sc i}](2-1) peaks $\Delta v$, following the formalism of \cite{2003neri}.
For the disk radius, we assume a value of $R$ = 3 kpc. We obtain $M_{\rm dyn}^{\rm disk}\ sin^{\rm 2}(i) \sim 4.4 \times 10^{11} {\rm M_{\odot}}$.
This quasar has the highest dynamical mass of the sample, which is somewhat expected given its larger molecular mass. A larger molecular mass and dynamical mass with respect to the other quasars could also imply a larger dark-matter halo mass.

The combination of an almost edge-on disk and a larger halo mass, could explain
a low level of extended Ly$\alpha$ emission as observed for this system. In a unification scenario for AGN (e.g., \citealt{1995urry})
radiation from the quasar escapes through two ionization cones determined by 
obscuration from the dust distribution on small scales. Since we observe this object as a quasar, 
our line-of-sight is  inside its ionizing cones. A large fraction of the quasar emission 
will pass through its massive host molecular disk, 
and will be absorbed before reaching the CGM. In other words, the obscuration from such a misaligned host galaxy decreases the budget of ionizing photons able to boost the Ly$\alpha$ emission on CGM scales, resulting in a dimmer and less extended Ly$\alpha$ nebula. Any Ly$\alpha$ emission boosted on the scales of the host galaxy likely experience a large number of scatterings and final absorption due to dust, expected to be present in large quantities in such a large molecular reservoir (e.g., \citealt{2017venemans}; see Section~\ref{sec:dis:molmassneb} for a  rough indication of the dust mass). 

On the other hand, a large halo mass implies a smaller fraction of cool gas able to survive the shock-heating process and penetrate the halo down to the host galaxy (e.g., \citealt{2006dekel}). In turn, warmer halo and a smaller cool CGM reservoir would result in a lower level of Ly$\alpha$ emission. We therefore conclude that in this scenario, the mass of the halo together with the geometry of the system and Ly$\alpha$ radiative transfer effects could determine the small extent and low SB level of the Ly$\alpha$ nebula surrounding this quasar. Radiative transfer calculations possibly coupled to cosmological hydrodinamical simulations with dust implementations are needed to confirm this interpretation.

For the merger scenario, we can estimate the dynamical mass $M_{\rm dyn}^{\rm merger}$ following \cite{2008greve}: $ M_{\rm dyn}^{\rm merger} = 2\times (2.33\times 10^{5}) {\rm FWHM}^{2} R [{\rm M_{\odot}}] $, where ${\rm FWHM}$ is the full width half maximum of the observed [C{\sc i}](2-1) emission lines, and $R$ is the half projected distance between the two merging objects. We used the same $R$ value assumed in the disk scenario (to obtain a lower limit in $M_{\rm dyn}^{\rm merger}$)\footnote{In this scenario the companion object could be located anywhere inside the APEX beam, with the 800~km~s$^{-1}$ velocity shift representing a large peculiar velocity or a distance of at maximum $\sim1.6$~Mpc within the Hubble flow. Given the expected virial radius of such massive objects, the latter scenario will represent a merger in its very early phases or two objects still separated in the Hubble flow.}, and  $\rm FWHM = 943 \pm 249\ km s^{-1}$ (see Table \ref{table:fluxes}), which was calculated as the sum of each single Gaussian fit corresponding to each [C{\sc i}](2-1) observed line. 
We obtain a mass of $M_{\rm dyn}^{\rm merger} = 1.2 \times 10^{12} {\rm M_{\odot}}$. Assuming that the amount of dark matter is negligible on host scales, the stellar mass is given by $M_{*} \approx M_{\rm dyn}^{\rm merger} - M_{\rm H_{2}} $. Then, using the molecular gas mass range derived from applying the CO and [C{\sc i}] constraints, we estimate a total stellar mass in the range 
$M_{*} = (7.6-8.8) \times 10^{11} {\rm M_{\odot}}$.   
Following \cite{2018moster}, this stellar mass implies an extremely large dark matter halo mass of $\rm M_{halo} \sim 10^{15.0}- 10^{15.1}\ M_{\odot}$.

\begin{figure}
    \centering
    \includegraphics[width=0.95\columnwidth]{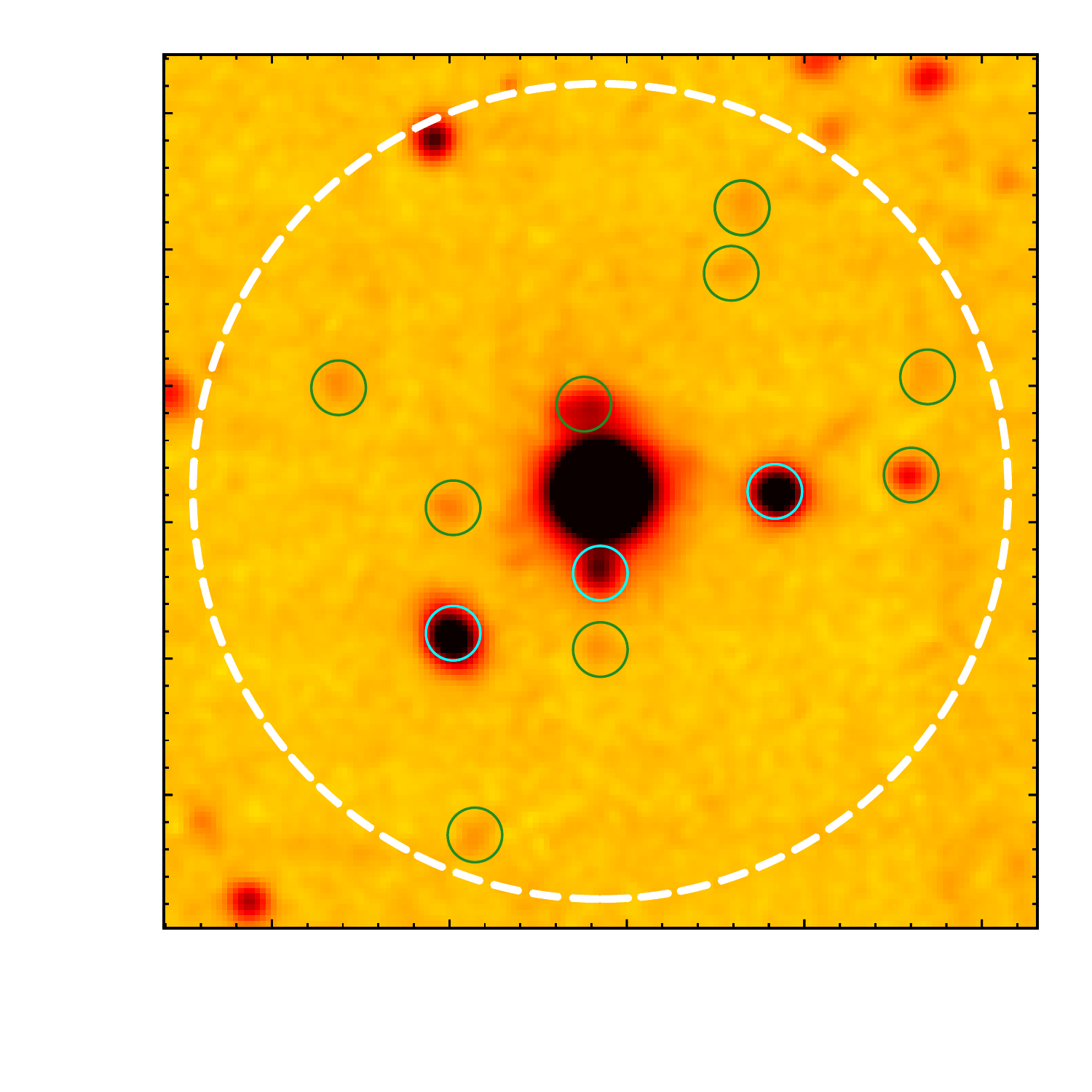} 
    \caption{MUSE white-light image for the field around ID~29 (or Q-0115-30, at the center) highlighting the location of continuum sources within the APEX beam for the [C{\sc i}](2-1) observations (white dashed circle). Low-$z$ interlopers (cyan circles) and sources with unknown redshift (green circles) are indicated (see Section~\ref{dis:q0115} for details).}
    \label{fig:contID29}
\end{figure}

We also estimated the individual stellar masses of each component of the merging system.
The molecular gas masses for the two different objects computed from the atomic carbon masses are $M_{\rm H_{2}}^{\rm obj1} \sim 2.4 \times 10^{11} {\rm M_{\odot}}$ and $ M_{\rm H_{2}}^{\rm obj2} \sim 3.7 \times 10^{11} {\rm M_{\odot}}$, respectively.
Then, assuming $M_{\rm H_{2}}^{\rm obj1}/M_{\rm H_{2}}^{\rm obj2}=M_{*}^{\rm obj1}/M_{*}^{\rm obj2}$ 
, we obtain $M_{*}^{\rm obj1} \sim 2.3 \times 10^{11} {\rm M_{\odot}}$ and $ M_{*}^{\rm obj2} \sim 3.6 \times 10^{11} {\rm M_{\odot}}$. 
The current MUSE data are not able to  verify whether such a massive companion object exists, 
but this field is characterized by nine faint continuum objects with unknown redshifts within the beam of our APEX observations (Figure~\ref{fig:contID29}). These objects do not have any emission lines or absorption features that  allow us to verify their redshifts. 
In particular, we note the presence of one continuum source at $\sim 3\arcsec$ (or $\sim 23$~projected kpc) from the quasar.
In summary, given the brightness of the quasar and the seeing-limited MUSE observations, we are not able to assess whether there is an ongoing merger on very small scales.  A strongly dust-obscured and massive companion could be missed by our optical observations (e.g., \citealt{1996omont}).
Finally. we note that within the APEX beam we also find three low-$z$ interlopers
(cyan circles in Figure~\ref{fig:contID29}), whose redshifts can be clearly determined ($z\sim0.52,1.05,0.07$). 
These redshifts ensure that the molecular emission from our targeted object is not contaminated by these
sources.   

Our consideration of ID~29 shows that its dynamical mass is very large, independent of the origin of the double-peaked [C{\sc i}](2-1) line emission. Observing this system with an interferometer (e.g., ALMA, NOEMA) would allow us to map the [C{\sc i}](2-1) emission and ascertain whether ID~29 is associated with an exceptionally massive molecular disk or is merging with a similarly massive companion. Such observations, by probing the $\sim 1-2$~mm continuum, would be in turn able to verify the reason why the Ly$\alpha$ nebula around ID~29 is dimmer and less extended than similarly bright quasars.

\subsection{Ly$\alpha$ nebulae kinematics with respect to molecular gas} 
\label{sec:dis:radiative}

The use of Ly$\alpha$ emission as a tracer of gas kinematics is a complex task due to its resonant nature (e.g., \citealt{1990neufeld}). Ly$\alpha$ photons are expected to interact several times with Hydrogen gas before escaping most  astrophysical systems (e.g., \citealt{2017dijkstra}). During this process, scattered photons can experience both large changes in their frequency and a large displacement in space,
possibly washing out any information on gas kinematics. Also, the larger the number of scatterings in a medium, 
the higher the probability for a Ly$\alpha$ photon to be absorbed by dust. If the dust distribution is not homogeneous, 
the Ly$\alpha$ line shape could be affected, possibly hiding information on the kinematics of the system.

In Section \ref{sec:lyacoprofiles} we compared the molecular emission lines  observed with APEX, with the Ly$\alpha$ emission obtained with MUSE and integrated within the APEX beam.
The main result is that we find velocity shifts between the Ly$\alpha$ emission and 
the systemic redshift $z_{\rm sys,mol}$ in the range -423 to 1236~km~s$^{-1}$. 
The presence of a velocity shift between this Ly$\alpha$ emission line 
integrated on halos scales and the systemic redshift of the quasar host galaxy,
could  be associated with a variety of physical processes occurring in the halo, e.g.
substructures or gas infalling onto the central quasar, large-scale outflows,
rotating structures or projection effects along the line-of-sight \citep[see][and references therein]{2018arrigoni}. 
In the following discussion we focus on the possibility that these shifts represent bulk inflows
or outflows.

In particular, as indicated by Ly$\alpha$ radiative transfer modelling \citep[e.g,][]{2006verhamme,2009laursen}, 
photons scattering through outflowing or inflowing gas should appear redshifted
or blueshifted, respectively, from the systemic redshift. 
The quasar with ID~29 has a shift consistent with $z_{\rm sys,mol}$. We cannot therefore draw any firm conclusion for this source with the current dataset. It could be that Ly$\alpha$ radiative transfer effects and a balanced interplay of outflows and inflows result in the absence of a strong line shift or that very fast outflows or inflows bring the gas out of resonance, allowing the observation of Ly$\alpha$ at $z_{\rm mol,sys}$.
For the remaining four sources, IDs 3, 8, 39, 43 and 50, we find significant offsets. In particular, IDs 3 and 43 present the largest shifts and  also show the largest values of FWHM of the Ly$\alpha$ emission lines in the sample (with ${\rm FWHM_{\rm Ly\alpha}}>1000$~km~s$^{-1}$), which is 
typically characterized by relatively quiescent kinematics (median $\rm FWHM_{Ly\alpha} = 915\ km\ s^{-1}$)\footnote{The FWHM$_{\rm Ly\alpha}$ values reported in this work are from Gaussian fits of the Ly$\alpha$ emission integrated within the APEX beam. They therefore differ from the 2D first moment analysis on resolved maps presented in \citet{2019arrigoni}. The FWHM$_{\rm Ly\alpha}$ values are not corrected for the instrument resolution.}. In the following, we review these four sources.

The quasars with IDs 8, 39 and 50 show negative shifts of -423 $\pm$ 76, -152 $\pm$ 64 and -377 $\pm$ 44 km s$^{-1}$, respectively, which could indicate an overall inflow signature from CGM scales onto the central quasar.
The quasar with ID~39 (or SDSS~J0100+2105) is surrounded by one of the less extended Ly$\alpha$ nebulae in our sample (see Fig. \ref{fig:masseslya}; Ly$\alpha$  extends
out to $\sim50$ projected kpc). The observed blueshift for this source  is lower than the inflow velocities ($v_{\rm in}\approx0.9 v_{\rm vir}\sim320-460$~km~s$^{-1}$; \citealt{2015goerdt}) expected for the cool gas in quasar host halos ($10^{12}-10^{13}$~M$_{\odot}$, e.g., \citealt{2012white,2018timlin} and references therein). This could be due to projection effects or to the presence of other violent motions (e.g.  winds, turbulences) that could wash out the inflow signature.

The observed blueshift for the quasars with ID~8 (or UM683) and ID~50 (or SDSS J0819+0823) is comparable with the inflow velocities 
expected for the cool gas in quasar host halos. 
The quasar with ID~8 has an average molecular reservoir and a Ly$\alpha$ nebula with 
intermediate surface brightness, extending out to $\sim 145$ kpc (\citealt{2019arrigoni}.)
Interestingly, ID~50 is surrounded by one of the brightest and more extended (in area) Ly$\alpha$ nebulae 
in the QSO museum sample (see Fig. \ref{fig:masseslya}; Ly$\alpha$  extends
out to $\sim130$ projected kpc), with a clearly asymmetric morphology. 
The widths of their integrated Ly$\alpha$ lines, FWHM${_{\rm Ly\alpha} = 915\pm46}$~km~s$^{-1}$ and FWHM${_{\rm Ly\alpha} = 967\pm38}$~km~s$^{-1}$ (for IDs~8 and 50, respectively), may therefore be explained by gravitational motions in such massive halos (${\rm FWHM}_{\rm 1D}^{\rm DM}\sim580-870$~km~s$^{-1}$; \citealt{2019arrigoni}) and by the presence of large turbulence within the cool gas reservoir. 

In contrast, the quasars with IDs 3 and 43 show positive shifts of 786$\pm$109 km s$^{-1}$ 
and 1236$\pm$133 km s$^{-1}$, respectively, which are suggestive of bulk large-scale outflows.
These velocity shifts are overall higher than what has been found in Ly$\alpha$-emitting galaxies 
(average of $\Delta v\sim$ 200 km s$^{-1}$, e.g., \citealt{2015trainor,2015prescott,2018verhamme}),
Lyman-break galaxies (average of $\Delta v\sim$ 460 km s$^{-1}$, see \citealt{2017dijkstra} and references therein),
and are also higher than the values found for other Ly$\alpha$ nebulae, the so called Ly$\alpha$ blobs ($\sim$0 - 230 km s$^{-1}$, e.g., \citealt{2011yang,2013mclinden,2015prescott}). 

The quasar with ID 3 is radio-loud, characterized by a large reservoir of molecular gas and has one of the Ly$\alpha$ nebulae with the lowest surface brightness, extending out to a maximum of $\sim70$~projected kpc (\citealt{2019arrigoni}). Its integrated Ly$\alpha$ emission shows the most active kinematics of the sample ($\rm FWHM_{Ly\alpha} = 1542\pm196\ \rm km\ s^{-1}$), which together with the observed large Ly$\alpha$ shift likely indicates that small-scale radio feedback is affecting the surrounding gas distribution in this system.
The quasar with ID 43 has an average molecular reservoir and a Ly$\alpha$ nebula with 
intermediate surface brightness, extending out to $\sim118$ kpc (\citealt{2019arrigoni}). 
Its radio nature is not clear because there are no radio observations of this target in the literature.
Its  relatively active Ly$\alpha$ kinematics ($\rm FWHM_{Ly\alpha} = 1068\pm142 km\ s^{-1}$),
together with the large observed positive shift of the Ly$\alpha$ line with respect to CO(6-5) 
and the asymmetric morphology of the nebula with a northern bright region,
might all be hints of  outflowing gas in this system.

\subsection{Relation between large molecular reservoirs and Ly$\alpha$ nebulae} \label{sec:dis:molmassneb}

In section \ref{sec:lyaphysicalprop}, we found a tentative trend indicating that the two quasars with the lowest surface brightness for their Ly$\alpha$ nebulae are associated with the highest molecular gas masses. As the molecular gas content is also closely linked to the amount of dust\footnote{To have an idea of the dust masses of our sample, one could assume a gas-to-dust ratio of 70 \citep[e.g.,][]{2013riechers,2017venemans} and considered that the $\sim$75$\%$ of the total gas mass corresponds to molecular gas \citep[e.g.,][]{2016wang}. Under these very simple assumptions, our quasars should have a dust mass in the range of $M_{\rm dust}$ = 0.8 - 12.9$\times$10$^{9} \rm M_{\odot}$ for the sources with IDs 3, 8, 29, 39, 43 and 50, and are <2.1$\times$10$^{9}$M$_{\odot}$ for the objects with IDs 7 and 18. Under these assumptions objects with ID 3 and 29 with the most massive molecular reservoirs also present the highest dust masses.}, this could imply that the radiation of these quasars is more obscured by their host galaxies, therefore
reducing the reprocessed Ly$\alpha$ emission from the surrounding gas distribution. Two effects could then be in play in the radiative transfer of these systems: first, scattered Ly$\alpha$ photons on small scales (tens of kpc) encounter a very high column density and dusty medium resulting in  absorption of the emission, and second,  most ionizing photons are not able to escape the massive host galaxy towards the CGM, limiting the detectability of cool halo gas in emission.

On top of this, the halo mass could play an important role in determining the reservoir of cool gas on CGM scales, with more massive halos having a smaller fraction of their gas in this 
phase. Taking also this aspect into consideration, there are at least two possible interpretations for this tentative trend: 1) IDs 3 and 29 (which present the most massive molecular reservoirs together with dimmer nebulae) occupy similar massive dark matter halos as the rest of quasars in the sample, but their molecular gas fraction (and consequently dust) are higher, resulting in smaller escape of ionizing  and Ly$\alpha$ photons, 
2) IDs 3 and 29 live in more massive dark matter halos, which are warmer and characterized by a smaller cool gas fraction, hence lower levels of Ly$\alpha$.

Unfortunately, we cannot unambiguously infer the halo masses with the available observations, nor we can firmly quantify obscuration in these systems.  
Future analysis based on high-resolution interferometric observations and spectra from other CO transitions (e.g., with ALMA, NOEMA) are needed to better constrain the nature of these sources. Such follow-up studies together with constraints on the far-infrared continuum will be able to 
accurately determine the dust and molecular content of these systems, allowing a better description of the radiative transfer around these quasars.

\section{Summary} \label{section:summary}

With the purpose of gaining a new leverage on the gas cycle around $z\sim3$ quasars, we exploit the frequency window opened-up by SEPIA180 at APEX, targeting 
the CO(6-5), CO(7-6) and [C{\sc i}](2-1) transitions for a sample of 9 $z\sim3$ quasars already observed with MUSE/VLT. Specifically, these 9 targets (quasars with IDs 3, 8, 13, 18, 29, 39, 43 and 50) cover the most of the parameter space of Ly$\alpha$ nebulae properties in the QSO MUSEUM survey \citep{2019arrigoni}. 
These data (average rms of 2.6~mJy)
allowed us to investigate any relation between the molecular gas content and the large-scale atomic phase ($\sim$10$^{4}$K) as traced by the Ly$\alpha$ emission. The main results of this work are summarized as follows.\\

\begin{itemize}
\item CO(6-5) emission is detected in three sources (quasars with IDs 8, 43 and 50) with fluxes of $3.4\leq \rm I_{CO(6-5)} \leq 5.1$~Jy~km~s$^{-1}$ and $620\leq$~FWHM~$\leq707$~km~s$^{-1}$, while the [C{\sc i}](2-1) emission is detected in other three sources 
(quasars with IDs 3, 29, and 39) 
with fluxes of $2.3 \leq \rm I_{[C\textsc{i}](2-1)} \leq 15.7$~Jy~km~s$^{-1}$ and $329 \leq$~FWHM~$\leq943$~km~s$^{-1}$. 

\item The brightest and widest detection in the sample is the [C{\sc i}](2-1) line ($\rm I_{[C\textsc{i}](2-1)}\sim$15.7~Jy~km~s$^{-1}$, FWHM$\sim$943~km~s$^{-1}$) associated to ID~29, which also presents the dimmest and smallest Ly$\alpha$ nebula in the sample.

\item We obtained 
molecular gas masses $M_{\rm H_{2}}$ from 
the CO 
and [C{\sc i}](2-1)  
constraints (Sections~\ref{section:cimass} and \ref{section:massco}). 
Applying the two estimates together, the molecular masses for sources with detections are in the range   $\rm M_{H_{2}} = (0.4-6.9) \times 10^{11} M_{\odot}$, 
while for quasars with non-detections, the upper limits are $\rm M_{H_{2}} < 1.1\times 10^{11} M_{\odot}$. 
These masses are on the high side of the typical ranges found in the literature ($\sim$10$^{9}$-10$^{11}$M$_{\odot}$) for other $z\sim 3$ quasars, 
with ID~29 and ID~3 being outliers.

\item The systemic redshift of 6 quasars is refined using the CO(6-5) and [C{\sc i}](2-1) 
detections. These new systemic redshifts have an average uncertainty of 74.8 km s$^{-1}$.

\item We found significant offsets between the extended Ly$\alpha$ nebulae (Ly$\alpha$ emission extracted within the APEX beam) and the newly estimated systemic redshifts in five sources (quasars with IDs 3, 8, 39, 43 and 50). Two of these sources also show the largest values of FWHM$_{\rm Ly\alpha}$ in our sample ($>$ 1000 km s$^{-1}$), which is otherwise characterized by relatively quiescent kinematics (median FWHM$_{\rm Ly\alpha}$ = 915 km s$^{-1}$). In Section~\ref{sec:dis:radiative} we discuss how these velocity shifts could be signatures of bulk inflows/outflows.
For the quasars with IDs 8, 39 and 50, the nebular Ly$\alpha$ line shows a blueshift of -423 $\pm$ 76, -152 $\pm$ 64 and -377 $\pm$ 44 km s$^{-1}$, respectively, which could indicate a large-scale inflow. 
In contrast, quasars with IDs 3 and 43 have the bulk of the Ly$\alpha$ line redshifted by  
786 $\pm$ 109 km s$^{-1}$ and 1236 $\pm$ 133 km s$^{-1}$, respectively, which can indicate that the bulk of the large-scale gas is outflowing in both cases. For ID~3, this scenario is further strengthened by the fact that this quasar is radio-loud. 

\item We found that the two most massive molecular reservoirs in our sample (quasars with IDs 3 and 29, $\rm M_{H_{2}} = (2.3-6.9) \times 10^{11} M_{\odot}$ and $\rm M_{H_{2}} = (3.2-4.4) \times 10^{11} M_{\odot}$, respectively) are associated with the dimmest and smallest Ly$\alpha$ nebulae. 
This suggests that the quasar host galaxy properties are key in understanding the powering and physics of surrounding Ly$\alpha$ nebulae. Obscuration from the host galaxy, due to physical properties (e.g., higher molecular gas fraction, higher dust fraction) or geometry, could reduce the escape of ionizing and Ly$\alpha$ photons emitted by the quasar, ultimately reducing the emission from the cool CGM. This interpretation is further strengthened
by the detection of a double-peaked [C{\textsc{i}}](2-1) line emission for ID~29, which could be due to an inclined massive molecular disk, 
likely misaligned with the ionization cones of ID~29 (see Section \ref{dis:q0115}). Another possibility 
is that these quasars with more massive molecular reservoirs could be hosted by more massive dark-matter halos.
Such halos are expected to be warmer and to show a smaller fraction of cool CGM gas, which in turn would result in lower values of Ly$\alpha$ emission.  

\end{itemize}

These APEX observations therefore suggest  that the quasar host-galaxy properties could affect the Ly$\alpha$ radiative transfer, thus regulating the amount of detectable emission on large scales around quasars.
Overall, this work stresses the importance of simultaneously studying the physical properties 
of the multiphase  gas reservoirs surrounding quasars. By characterizing the quasar host galaxy
properties and system geometry (e.g., systemic redshift, inflow/outflow), future studies will be able to assess 
the radiative transfer effects affecting the quasar radiation and ultimately better constrain the powering mechanisms 
for extended gas reservoirs. Deep high-resolution interferometric observations (e.g., with ALMA, NOEMA) are 
indispensable to map the extent and geometry of the  molecular reservoirs, ultimately linking its physics to the other CGM phases.

\section*{Acknowledgements}
Based on observations taken with the Atacama Pathfinder Experiment, which is a collaboration between the Max- Planck-Institut fur Radioastronomie, the European Southern Observatory (ESO), and the Onsala Space Observatory, under ESO programmes 0102.A-0394A and 0103.A-0306A. We thank the APEX observing staff and ESO astronomers for their support and hospitality during the collection of this large dataset. We further thank Aura Obreja for participating in the acquisition of these observations.

\section*{Data Availability}
The data used in this work are available from the ESO Science Archive Facility (\url{https://archive.eso.org/}). The fully reduced dataset will be shared upon reasonable request to the corresponding author.




\bibliographystyle{mnras}
\bibliography{paper} 

\begin{thebibliography}{}
\makeatletter
\relax
\def\mn@urlcharsother{\let\do\@makeother \do\$\do\&\do\#\do\^\do\_\do\%\do\~}
\def\mn@doi{\begingroup\mn@urlcharsother \@ifnextchar [ {\mn@doi@}
  {\mn@doi@[]}}
\def\mn@doi@[#1]#2{\def\@tempa{#1}\ifx\@tempa\@empty \href
  {http://dx.doi.org/#2} {doi:#2}\else \href {http://dx.doi.org/#2} {#1}\fi
  \endgroup}
\def\mn@eprint#1#2{\mn@eprint@#1:#2::\@nil}
\def\mn@eprint@arXiv#1{\href {http://arxiv.org/abs/#1} {{\tt arXiv:#1}}}
\def\mn@eprint@dblp#1{\href {http://dblp.uni-trier.de/rec/bibtex/#1.xml}
  {dblp:#1}}
\def\mn@eprint@#1:#2:#3:#4\@nil{\def\@tempa {#1}\def\@tempb {#2}\def\@tempc
  {#3}\ifx \@tempc \@empty \let \@tempc \@tempb \let \@tempb \@tempa \fi \ifx
  \@tempb \@empty \def\@tempb {arXiv}\fi \@ifundefined
  {mn@eprint@\@tempb}{\@tempb:\@tempc}{\expandafter \expandafter \csname
  mn@eprint@\@tempb\endcsname \expandafter{\@tempc}}}

\bibitem[\protect\citeauthoryear{{Alaghband-Zadeh} et~al.,}{{Alaghband-Zadeh}
  et~al.}{2013}]{2013alaghbandzadeh}
{Alaghband-Zadeh} S.,  et~al., 2013, \mn@doi [\mnras] {10.1093/mnras/stt1390},
  \href {https://ui.adsabs.harvard.edu/abs/2013MNRAS.435.1493A} {435, 1493}

\bibitem[\protect\citeauthoryear{{Anh}, {Boone}, {Hoai}, {Nhung}, {Wei{\ss}},
  {Kneib}, {Beelen}  \& {Salom{\'e}}}{{Anh} et~al.}{2013}]{2013anh}
{Anh} P.~T.,  {Boone} F.,  {Hoai} D.~T.,  {Nhung} P.~T.,  {Wei{\ss}} A.,
  {Kneib} J.~P.,  {Beelen} A.,   {Salom{\'e}} P.,  2013, \mn@doi [\aap]
  {10.1051/0004-6361/201321363}, \href
  {https://ui.adsabs.harvard.edu/abs/2013A&A...552L..12A} {552, L12}

\bibitem[\protect\citeauthoryear{{Antonucci}}{{Antonucci}}{1993}]{1993antonucci}
{Antonucci} R.,  1993, \mn@doi [\araa] {10.1146/annurev.aa.31.090193.002353},
  \href {https://ui.adsabs.harvard.edu/abs/1993ARA&A..31..473A} {31, 473}

\bibitem[\protect\citeauthoryear{{Aravena} et~al.,}{{Aravena}
  et~al.}{2008}]{2008aravena}
{Aravena} M.,  et~al., 2008, \mn@doi [\aap] {10.1051/0004-6361:200810628},
  \href {https://ui.adsabs.harvard.edu/abs/2008A&A...491..173A} {491, 173}

\bibitem[\protect\citeauthoryear{{Arrigoni Battaia}, {Hennawi}, {Prochaska}  \&
  {Cantalupo}}{{Arrigoni Battaia} et~al.}{2015}]{2015arrigoni}
{Arrigoni Battaia} F.,  {Hennawi} J.~F.,  {Prochaska} J.~X.,   {Cantalupo} S.,
  2015, \mn@doi [\apj] {10.1088/0004-637X/809/2/163}, \href
  {https://ui.adsabs.harvard.edu/abs/2015ApJ...809..163A} {809, 163}

\bibitem[\protect\citeauthoryear{{Arrigoni Battaia}, {Prochaska}, {Hennawi},
  {Obreja}, {Buck}, {Cantalupo}, {Dutton}  \& {Macci{\`o}}}{{Arrigoni Battaia}
  et~al.}{2018a}]{2018arrigoni}
{Arrigoni Battaia} F.,  {Prochaska} J.~X.,  {Hennawi} J.~F.,  {Obreja} A.,
  {Buck} T.,  {Cantalupo} S.,  {Dutton} A.~A.,   {Macci{\`o}} A.~V.,  2018a,
  \mn@doi [\mnras] {10.1093/mnras/stx2465}, \href
  {https://ui.adsabs.harvard.edu/abs/2018MNRAS.473.3907A} {473, 3907}

\bibitem[\protect\citeauthoryear{{Arrigoni Battaia} et~al.,}{{Arrigoni Battaia}
  et~al.}{2018b}]{2018arrigoni2}
{Arrigoni Battaia} F.,  et~al., 2018b, \mn@doi [\aap]
  {10.1051/0004-6361/201834195}, \href
  {https://ui.adsabs.harvard.edu/abs/2018A&A...620A.202A} {620, A202}

\bibitem[\protect\citeauthoryear{{Arrigoni Battaia}, {Hennawi}, {Prochaska},
  {O{\~n}orbe}, {Farina}, {Cantalupo}  \& {Lusso}}{{Arrigoni Battaia}
  et~al.}{2019a}]{2019arrigoni}
{Arrigoni Battaia} F.,  {Hennawi} J.~F.,  {Prochaska} J.~X.,  {O{\~n}orbe} J.,
  {Farina} E.~P.,  {Cantalupo} S.,   {Lusso} E.,  2019a, \mn@doi [\mnras]
  {10.1093/mnras/sty2827}, \href
  {https://ui.adsabs.harvard.edu/abs/2019MNRAS.482.3162A} {482, 3162}

\bibitem[\protect\citeauthoryear{{Arrigoni Battaia} et~al.,}{{Arrigoni Battaia}
  et~al.}{2019b}]{2019arrigoni2}
{Arrigoni Battaia} F.,  et~al., 2019b, \mn@doi [\aap]
  {10.1051/0004-6361/201936211}, \href
  {https://ui.adsabs.harvard.edu/abs/2019A&A...631A..18A} {631, A18}

\bibitem[\protect\citeauthoryear{{Ba{\~n}ados} et~al.,}{{Ba{\~n}ados}
  et~al.}{2018}]{2018banados}
{Ba{\~n}ados} E.,  et~al., 2018, \mn@doi [\nat] {10.1038/nature25180}, \href
  {https://ui.adsabs.harvard.edu/abs/2018Natur.553..473B} {553, 473}

\bibitem[\protect\citeauthoryear{{Bacon} et~al.,}{{Bacon}
  et~al.}{2010}]{2010bacon}
{Bacon} R.,  et~al., 2010, in {McLean} I.~S.,  {Ramsay} S.~K.,   {Takami} H.,
  eds,  Society of Photo-Optical Instrumentation Engineers (SPIE) Conference
  Series Vol. 7735, Ground-based and Airborne Instrumentation for Astronomy
  III. p. 773508, \mn@doi{10.1117/12.856027}

\bibitem[\protect\citeauthoryear{{Banerji}, {Carilli}, {Jones}, {Wagg},
  {McMahon}, {Hewett}, {Alaghband-Zadeh}  \& {Feruglio}}{{Banerji}
  et~al.}{2017}]{2017banerji}
{Banerji} M.,  {Carilli} C.~L.,  {Jones} G.,  {Wagg} J.,  {McMahon} R.~G.,
  {Hewett} P.~C.,  {Alaghband-Zadeh} S.,   {Feruglio} C.,  2017, \mn@doi
  [\mnras] {10.1093/mnras/stw3019}, \href
  {https://ui.adsabs.harvard.edu/abs/2017MNRAS.465.4390B} {465, 4390}

\bibitem[\protect\citeauthoryear{{Banerji}, {Jones}, {Wagg}, {Carilli},
  {Bisbas}  \& {Hewett}}{{Banerji} et~al.}{2018}]{2018banerji}
{Banerji} M.,  {Jones} G.~C.,  {Wagg} J.,  {Carilli} C.~L.,  {Bisbas} T.~G.,
  {Hewett} P.~C.,  2018, \mn@doi [\mnras] {10.1093/mnras/sty1443}, \href
  {https://ui.adsabs.harvard.edu/abs/2018MNRAS.479.1154B} {479, 1154}

\bibitem[\protect\citeauthoryear{{Barvainis}, {Alloin}  \&
  {Bremer}}{{Barvainis} et~al.}{2002}]{2002barvainis}
{Barvainis} R.,  {Alloin} D.,   {Bremer} M.,  2002, \mn@doi [\aap]
  {10.1051/0004-6361:20020121}, \href
  {https://ui.adsabs.harvard.edu/abs/2002A&A...385..399B} {385, 399}

\bibitem[\protect\citeauthoryear{{Beelen} et~al.,}{{Beelen}
  et~al.}{2004}]{2004beelen}
{Beelen} A.,  et~al., 2004, \mn@doi [\aap] {10.1051/0004-6361:20040318}, \href
  {https://ui.adsabs.harvard.edu/abs/2004A&A...423..441B} {423, 441}

\bibitem[\protect\citeauthoryear{{Belitsky} et~al.,}{{Belitsky}
  et~al.}{2018a}]{2018bbelitsky}
{Belitsky} V.,  et~al., 2018a, \mn@doi [\aap] {10.1051/0004-6361/201731883},
  \href {https://ui.adsabs.harvard.edu/abs/2018A&A...611A..98B} {611, A98}

\bibitem[\protect\citeauthoryear{{Belitsky} et~al.,}{{Belitsky}
  et~al.}{2018b}]{2018belitsky}
{Belitsky} V.,  et~al., 2018b, \mn@doi [\aap] {10.1051/0004-6361/201731458},
  \href {https://ui.adsabs.harvard.edu/abs/2018A&A...612A..23B} {612, A23}

\bibitem[\protect\citeauthoryear{{Bisbas}, {van Dishoeck}, {Papadopoulos},
  {Sz{\H{u}}cs}, {Bialy}  \& {Zhang}}{{Bisbas} et~al.}{2017}]{2017bisbas}
{Bisbas} T.~G.,  {van Dishoeck} E.~F.,  {Papadopoulos} P.~P.,  {Sz{\H{u}}cs}
  L.,  {Bialy} S.,   {Zhang} Z.-Y.,  2017, \mn@doi [\apj]
  {10.3847/1538-4357/aa696d}, \href
  {https://ui.adsabs.harvard.edu/abs/2017ApJ...839...90B} {839, 90}

\bibitem[\protect\citeauthoryear{{Bischetti} et~al.,}{{Bischetti}
  et~al.}{2021}]{2020bischetti}
{Bischetti} M.,  et~al., 2021, \mn@doi [\aap] {10.1051/0004-6361/202039057},
  \href {https://ui.adsabs.harvard.edu/abs/2021A&A...645A..33B} {645, A33}

\bibitem[\protect\citeauthoryear{{Bolatto}, {Wolfire}  \& {Leroy}}{{Bolatto}
  et~al.}{2013}]{2013bolatto}
{Bolatto} A.~D.,  {Wolfire} M.,   {Leroy} A.~K.,  2013, \mn@doi [\araa]
  {10.1146/annurev-astro-082812-140944}, \href
  {https://ui.adsabs.harvard.edu/abs/2013ARA&A..51..207B} {51, 207}

\bibitem[\protect\citeauthoryear{{Boogaard} et~al.,}{{Boogaard}
  et~al.}{2020}]{2020boogaard}
{Boogaard} L.~A.,  et~al., 2020, \mn@doi [\apj] {10.3847/1538-4357/abb82f},
  \href {https://ui.adsabs.harvard.edu/abs/2020ApJ...902..109B} {902, 109}

\bibitem[\protect\citeauthoryear{{Borisova} et~al.,}{{Borisova}
  et~al.}{2016}]{2016borisova}
{Borisova} E.,  et~al., 2016, \mn@doi [\apj] {10.3847/0004-637X/831/1/39},
  \href {https://ui.adsabs.harvard.edu/abs/2016ApJ...831...39B} {831, 39}

\bibitem[\protect\citeauthoryear{{Bradford} et~al.,}{{Bradford}
  et~al.}{2009}]{2009bradford}
{Bradford} C.~M.,  et~al., 2009, \mn@doi [\apj] {10.1088/0004-637X/705/1/112},
  \href {https://ui.adsabs.harvard.edu/abs/2009ApJ...705..112B} {705, 112}

\bibitem[\protect\citeauthoryear{{Bunker}, {Smith}, {Spinrad}, {Stern}  \&
  {Warren}}{{Bunker} et~al.}{2003}]{2003bunker}
{Bunker} A.,  {Smith} J.,  {Spinrad} H.,  {Stern} D.,   {Warren} S.,  2003,
  \mn@doi [\apss] {10.1023/A:1024038312479}, \href
  {https://ui.adsabs.harvard.edu/abs/2003Ap&SS.284..357B} {284, 357}

\bibitem[\protect\citeauthoryear{{Cai} et~al.,}{{Cai} et~al.}{2017}]{2017cai}
{Cai} Z.,  et~al., 2017, \mn@doi [\apj] {10.3847/1538-4357/aa5d14}, \href
  {https://ui.adsabs.harvard.edu/abs/2017ApJ...837...71C} {837, 71}

\bibitem[\protect\citeauthoryear{{Cai} et~al.,}{{Cai} et~al.}{2019}]{2019cai}
{Cai} Z.,  et~al., 2019, \mn@doi [\apjs] {10.3847/1538-4365/ab4796}, \href
  {https://ui.adsabs.harvard.edu/abs/2019ApJS..245...23C} {245, 23}

\bibitem[\protect\citeauthoryear{{Cantalupo}, {Arrigoni-Battaia}, {Prochaska},
  {Hennawi}  \& {Madau}}{{Cantalupo} et~al.}{2014}]{2014cantalupo}
{Cantalupo} S.,  {Arrigoni-Battaia} F.,  {Prochaska} J.~X.,  {Hennawi} J.~F.,
  {Madau} P.,  2014, \mn@doi [\nat] {10.1038/nature12898}, \href
  {https://ui.adsabs.harvard.edu/abs/2014Natur.506...63C} {506, 63}

\bibitem[\protect\citeauthoryear{{Carilli} \& {Walter}}{{Carilli} \&
  {Walter}}{2013}]{2013carilli}
{Carilli} C.~L.,  {Walter} F.,  2013, \mn@doi [\araa]
  {10.1146/annurev-astro-082812-140953}, \href
  {https://ui.adsabs.harvard.edu/abs/2013ARA&A..51..105C} {51, 105}

\bibitem[\protect\citeauthoryear{{Carilli}, {Lewis}, {Djorgovski}, {Mahabal},
  {Cox}, {Bertoldi}  \& {Omont}}{{Carilli} et~al.}{2003}]{2003carilli}
{Carilli} C.~L.,  {Lewis} G.~F.,  {Djorgovski} S.~G.,  {Mahabal} A.,  {Cox} P.,
   {Bertoldi} F.,   {Omont} A.,  2003, \mn@doi [Science]
  {10.1126/science.1082600}, \href
  {https://ui.adsabs.harvard.edu/abs/2003Sci...300..773C} {300, 773}

\bibitem[\protect\citeauthoryear{{Casey} et~al.,}{{Casey}
  et~al.}{2018}]{2018casey}
{Casey} C.~M.,  et~al., 2018, in {Murphy} E.,  ed.,  Astronomical Society of
  the Pacific Conference Series Vol. 517, Science with a Next Generation Very
  Large Array. p.~629 (\mn@eprint {arXiv} {1810.08258})

\bibitem[\protect\citeauthoryear{{Circosta} et~al.,}{{Circosta}
  et~al.}{2021}]{2021circosta}
{Circosta} C.,  et~al., 2021, \mn@doi [\aap] {10.1051/0004-6361/202039270},
  \href {https://ui.adsabs.harvard.edu/abs/2021A&A...646A..96C} {646, A96}

\bibitem[\protect\citeauthoryear{{Dalgarno} \& {McCray}}{{Dalgarno} \&
  {McCray}}{1972}]{1972dalgarno}
{Dalgarno} A.,  {McCray} R.~A.,  1972, \mn@doi [\araa]
  {10.1146/annurev.aa.10.090172.002111}, \href
  {https://ui.adsabs.harvard.edu/abs/1972ARA&A..10..375D} {10, 375}

\bibitem[\protect\citeauthoryear{{Decarli} et~al.,}{{Decarli}
  et~al.}{2018}]{2018decarli}
{Decarli} R.,  et~al., 2018, \mn@doi [\apj] {10.3847/1538-4357/aaa5aa}, \href
  {https://ui.adsabs.harvard.edu/abs/2018ApJ...854...97D} {854, 97}

\bibitem[\protect\citeauthoryear{{Decarli} et~al.,}{{Decarli}
  et~al.}{2019}]{2019decarli}
{Decarli} R.,  et~al., 2019, \mn@doi [\apj] {10.3847/1538-4357/ab30fe}, \href
  {https://ui.adsabs.harvard.edu/abs/2019ApJ...882..138D} {882, 138}

\bibitem[\protect\citeauthoryear{{Decarli} et~al.,}{{Decarli}
  et~al.}{2020}]{2020decarli}
{Decarli} R.,  et~al., 2020, \mn@doi [\apj] {10.3847/1538-4357/abaa3b}, \href
  {https://ui.adsabs.harvard.edu/abs/2020ApJ...902..110D} {902, 110}

\bibitem[\protect\citeauthoryear{{Decarli}, {Arrigoni-Battaia}, {Hennawi},
  {Walter}, {Prochaska}  \& {Cantalupo}}{{Decarli} et~al.}{2021}]{2021decarli}
{Decarli} R.,  {Arrigoni-Battaia} F.,  {Hennawi} J.~F.,  {Walter} F.,
  {Prochaska} J.~X.,   {Cantalupo} S.,  2021, \mn@doi [\aap]
  {10.1051/0004-6361/202039814}, \href
  {https://ui.adsabs.harvard.edu/abs/2021A&A...645L...3D} {645, L3}

\bibitem[\protect\citeauthoryear{{Dekel} \& {Birnboim}}{{Dekel} \&
  {Birnboim}}{2006}]{2006dekel}
{Dekel} A.,  {Birnboim} Y.,  2006, \mn@doi [\mnras]
  {10.1111/j.1365-2966.2006.10145.x}, \href
  {https://ui.adsabs.harvard.edu/abs/2006MNRAS.368....2D} {368, 2}

\bibitem[\protect\citeauthoryear{{Di Matteo}, {Springel}  \& {Hernquist}}{{Di
  Matteo} et~al.}{2005}]{2005dimatteo}
{Di Matteo} T.,  {Springel} V.,   {Hernquist} L.,  2005, \mn@doi [\nat]
  {10.1038/nature03335}, \href
  {https://ui.adsabs.harvard.edu/abs/2005Natur.433..604D} {433, 604}

\bibitem[\protect\citeauthoryear{{Dijkstra}}{{Dijkstra}}{2017}]{2017dijkstra}
{Dijkstra} M.,  2017, arXiv e-prints, \href
  {https://ui.adsabs.harvard.edu/abs/2017arXiv170403416D} {p. arXiv:1704.03416}

\bibitem[\protect\citeauthoryear{{Downes} \& {Solomon}}{{Downes} \&
  {Solomon}}{1998}]{1998downes}
{Downes} D.,  {Solomon} P.~M.,  1998, \mn@doi [\apj] {10.1086/306339}, \href
  {https://ui.adsabs.harvard.edu/abs/1998ApJ...507..615D} {507, 615}

\bibitem[\protect\citeauthoryear{{Elvis}}{{Elvis}}{2000}]{2000elvis}
{Elvis} M.,  2000, \mn@doi [\apj] {10.1086/317778}, \href
  {https://ui.adsabs.harvard.edu/abs/2000ApJ...545...63E} {545, 63}

\bibitem[\protect\citeauthoryear{{Emonts} et~al.,}{{Emonts}
  et~al.}{2014}]{2014emonts}
{Emonts} B.~H.~C.,  et~al., 2014, \mn@doi [\mnras] {10.1093/mnras/stt2398},
  \href {https://ui.adsabs.harvard.edu/abs/2014MNRAS.438.2898E} {438, 2898}

\bibitem[\protect\citeauthoryear{{Emonts}, {Cai}, {Prochaska}, {Li}  \&
  {Lehnert}}{{Emonts} et~al.}{2019}]{2019emonts}
{Emonts} B. H.~C.,  {Cai} Z.,  {Prochaska} J.~X.,  {Li} Q.,   {Lehnert} M.~D.,
  2019, \mn@doi [\apj] {10.3847/1538-4357/ab45f4}, \href
  {https://ui.adsabs.harvard.edu/abs/2019ApJ...887...86E} {887, 86}

\bibitem[\protect\citeauthoryear{{Farina}, {Falomo}, {Decarli}, {Treves}  \&
  {Kotilainen}}{{Farina} et~al.}{2013}]{2013farina}
{Farina} E.~P.,  {Falomo} R.,  {Decarli} R.,  {Treves} A.,   {Kotilainen}
  J.~K.,  2013, \mn@doi [\mnras] {10.1093/mnras/sts410}, \href
  {https://ui.adsabs.harvard.edu/abs/2013MNRAS.429.1267F} {429, 1267}

\bibitem[\protect\citeauthoryear{{Farina}, {Falomo}, {Scarpa}, {Decarli},
  {Treves}  \& {Kotilainen}}{{Farina} et~al.}{2014}]{2014farina}
{Farina} E.~P.,  {Falomo} R.,  {Scarpa} R.,  {Decarli} R.,  {Treves} A.,
  {Kotilainen} J.~K.,  2014, \mn@doi [\mnras] {10.1093/mnras/stu585}, \href
  {https://ui.adsabs.harvard.edu/abs/2014MNRAS.441..886F} {441, 886}

\bibitem[\protect\citeauthoryear{{Farina} et~al.,}{{Farina}
  et~al.}{2019}]{2019farina}
{Farina} E.~P.,  et~al., 2019, \mn@doi [\apj] {10.3847/1538-4357/ab5847}, \href
  {https://ui.adsabs.harvard.edu/abs/2019ApJ...887..196F} {887, 196}

\bibitem[\protect\citeauthoryear{{Ferrarese} \& {Merritt}}{{Ferrarese} \&
  {Merritt}}{2000}]{2000ferrarese}
{Ferrarese} L.,  {Merritt} D.,  2000, \mn@doi [\apjl] {10.1086/312838}, \href
  {https://ui.adsabs.harvard.edu/abs/2000ApJ...539L...9F} {539, L9}

\bibitem[\protect\citeauthoryear{{Fossati} et~al.,}{{Fossati}
  et~al.}{2021}]{2021fossati}
{Fossati} M.,  et~al., 2021, \mn@doi [\mnras] {10.1093/mnras/stab660}, \href
  {https://ui.adsabs.harvard.edu/abs/2021MNRAS.503.3044F} {503, 3044}

\bibitem[\protect\citeauthoryear{{Fumagalli}, {Cantalupo}, {Dekel}, {Morris},
  {O'Meara}, {Prochaska}  \& {Theuns}}{{Fumagalli}
  et~al.}{2016}]{2016fumagalli}
{Fumagalli} M.,  {Cantalupo} S.,  {Dekel} A.,  {Morris} S.~L.,  {O'Meara}
  J.~M.,  {Prochaska} J.~X.,   {Theuns} T.,  2016, \mn@doi [\mnras]
  {10.1093/mnras/stw1782}, \href
  {https://ui.adsabs.harvard.edu/abs/2016MNRAS.462.1978F} {462, 1978}

\bibitem[\protect\citeauthoryear{{Garc{\'\i}a-Vergara}, {Hennawi}, {Barrientos}
   \& {Rix}}{{Garc{\'\i}a-Vergara} et~al.}{2017}]{2017garciavergara}
{Garc{\'\i}a-Vergara} C.,  {Hennawi} J.~F.,  {Barrientos} L.~F.,   {Rix} H.-W.,
   2017, \mn@doi [\apj] {10.3847/1538-4357/aa8b69}, \href
  {https://ui.adsabs.harvard.edu/abs/2017ApJ...848....7G} {848, 7}

\bibitem[\protect\citeauthoryear{{Garc{\'\i}a-Vergara}, {Hennawi}, {Barrientos}
   \& {Arrigoni Battaia}}{{Garc{\'\i}a-Vergara}
  et~al.}{2019}]{2019garciavergara}
{Garc{\'\i}a-Vergara} C.,  {Hennawi} J.~F.,  {Barrientos} L.~F.,   {Arrigoni
  Battaia} F.,  2019, \mn@doi [\apj] {10.3847/1538-4357/ab4d52}, \href
  {https://ui.adsabs.harvard.edu/abs/2019ApJ...886...79G} {886, 79}

\bibitem[\protect\citeauthoryear{{Goerdt} \& {Ceverino}}{{Goerdt} \&
  {Ceverino}}{2015}]{2015goerdt}
{Goerdt} T.,  {Ceverino} D.,  2015, \mn@doi [\mnras] {10.1093/mnras/stv786},
  \href {https://ui.adsabs.harvard.edu/abs/2015MNRAS.450.3359G} {450, 3359}

\bibitem[\protect\citeauthoryear{{Goldreich} \& {Kwan}}{{Goldreich} \&
  {Kwan}}{1974}]{1974goldreich}
{Goldreich} P.,  {Kwan} J.,  1974, \mn@doi [\apj] {10.1086/152821}, \href
  {https://ui.adsabs.harvard.edu/abs/1974ApJ...189..441G} {189, 441}

\bibitem[\protect\citeauthoryear{{Greve} \& {Sommer-Larsen}}{{Greve} \&
  {Sommer-Larsen}}{2008}]{2008greve}
{Greve} T.~R.,  {Sommer-Larsen} J.,  2008, \mn@doi [\aap]
  {10.1051/0004-6361:20078013}, \href
  {https://ui.adsabs.harvard.edu/abs/2008A&A...480..335G} {480, 335}

\bibitem[\protect\citeauthoryear{{Gronke} \& {Oh}}{{Gronke} \&
  {Oh}}{2018}]{2018gronke}
{Gronke} M.,  {Oh} S.~P.,  2018, \mn@doi [\mnras] {10.1093/mnrasl/sly131},
  \href {https://ui.adsabs.harvard.edu/abs/2018MNRAS.480L.111G} {480, L111}

\bibitem[\protect\citeauthoryear{{Gronke} \& {Oh}}{{Gronke} \&
  {Oh}}{2020}]{2020gronke}
{Gronke} M.,  {Oh} S.~P.,  2020, \mn@doi [\mnras] {10.1093/mnras/stz3332},
  \href {https://ui.adsabs.harvard.edu/abs/2020MNRAS.492.1970G} {492, 1970}

\bibitem[\protect\citeauthoryear{{Gullberg} et~al.,}{{Gullberg}
  et~al.}{2016}]{2016gullberg}
{Gullberg} B.,  et~al., 2016, \mn@doi [\aap] {10.1051/0004-6361/201527647},
  \href {https://ui.adsabs.harvard.edu/abs/2016A&A...591A..73G} {591, A73}

\bibitem[\protect\citeauthoryear{{Haiman} \& {Rees}}{{Haiman} \&
  {Rees}}{2001}]{2001haiman}
{Haiman} Z.,  {Rees} M.~J.,  2001, \mn@doi [\apj] {10.1086/321567}, \href
  {https://ui.adsabs.harvard.edu/abs/2001ApJ...556...87H} {556, 87}

\bibitem[\protect\citeauthoryear{{Harrington} et~al.,}{{Harrington}
  et~al.}{2021}]{2021harrington}
{Harrington} K.~C.,  et~al., 2021, \mn@doi [\apj] {10.3847/1538-4357/abcc01},
  \href {https://ui.adsabs.harvard.edu/abs/2021ApJ...908...95H} {908, 95}

\bibitem[\protect\citeauthoryear{{Heckman}, {Lehnert}, {van Breugel}  \&
  {Miley}}{{Heckman} et~al.}{1991}]{1991heckman}
{Heckman} T.~M.,  {Lehnert} M.~D.,  {van Breugel} W.,   {Miley} G.~K.,  1991,
  \mn@doi [\apj] {10.1086/169794}, \href
  {https://ui.adsabs.harvard.edu/abs/1991ApJ...370...78H} {370, 78}

\bibitem[\protect\citeauthoryear{{Hennawi} \& {Prochaska}}{{Hennawi} \&
  {Prochaska}}{2007}]{2007hennawi}
{Hennawi} J.~F.,  {Prochaska} J.~X.,  2007, \mn@doi [\apj] {10.1086/509770},
  \href {https://ui.adsabs.harvard.edu/abs/2007ApJ...655..735H} {655, 735}

\bibitem[\protect\citeauthoryear{{Hennawi} \& {Prochaska}}{{Hennawi} \&
  {Prochaska}}{2013}]{2013hennawi}
{Hennawi} J.~F.,  {Prochaska} J.~X.,  2013, \mn@doi [\apj]
  {10.1088/0004-637X/766/1/58}, \href
  {https://ui.adsabs.harvard.edu/abs/2013ApJ...766...58H} {766, 58}

\bibitem[\protect\citeauthoryear{{Hennawi} et~al.,}{{Hennawi}
  et~al.}{2006}]{2006hennawi}
{Hennawi} J.~F.,  et~al., 2006, \mn@doi [\apj] {10.1086/507069}, \href
  {https://ui.adsabs.harvard.edu/abs/2006ApJ...651...61H} {651, 61}

\bibitem[\protect\citeauthoryear{{Hennawi}, {Prochaska}, {Cantalupo}  \&
  {Arrigoni-Battaia}}{{Hennawi} et~al.}{2015}]{2015hennawi}
{Hennawi} J.~F.,  {Prochaska} J.~X.,  {Cantalupo} S.,   {Arrigoni-Battaia} F.,
  2015, \mn@doi [Science] {10.1126/science.aaa5397}, \href
  {https://ui.adsabs.harvard.edu/abs/2015Sci...348..779H} {348, 779}

\bibitem[\protect\citeauthoryear{{Hill} et~al.,}{{Hill}
  et~al.}{2019}]{2019hill}
{Hill} R.,  et~al., 2019, \mn@doi [\mnras] {10.1093/mnras/stz429}, \href
  {https://ui.adsabs.harvard.edu/abs/2019MNRAS.485..753H} {485, 753}

\bibitem[\protect\citeauthoryear{{Hummels} et~al.,}{{Hummels}
  et~al.}{2019}]{2019hummels}
{Hummels} C.~B.,  et~al., 2019, \mn@doi [\apj] {10.3847/1538-4357/ab378f},
  \href {https://ui.adsabs.harvard.edu/abs/2019ApJ...882..156H} {882, 156}

\bibitem[\protect\citeauthoryear{{Husband}, {Bremer}, {Stanway}  \&
  {Lehnert}}{{Husband} et~al.}{2015}]{2015husband}
{Husband} K.,  {Bremer} M.~N.,  {Stanway} E.~R.,   {Lehnert} M.~D.,  2015,
  \mn@doi [\mnras] {10.1093/mnras/stv1424}, \href
  {https://ui.adsabs.harvard.edu/abs/2015MNRAS.452.2388H} {452, 2388}

\bibitem[\protect\citeauthoryear{{Husemann}, {Worseck}, {Arrigoni Battaia}  \&
  {Shanks}}{{Husemann} et~al.}{2018}]{2018husemann}
{Husemann} B.,  {Worseck} G.,  {Arrigoni Battaia} F.,   {Shanks} T.,  2018,
  \mn@doi [\aap] {10.1051/0004-6361/201732457}, \href
  {https://ui.adsabs.harvard.edu/abs/2018A&A...610L...7H} {610, L7}

\bibitem[\protect\citeauthoryear{{Ikeda}, {Oka}, {Tatematsu}, {Sekimoto}  \&
  {Yamamoto}}{{Ikeda} et~al.}{2002}]{2002ikeda}
{Ikeda} M.,  {Oka} T.,  {Tatematsu} K.,  {Sekimoto} Y.,   {Yamamoto} S.,  2002,
  \mn@doi [\apjs] {10.1086/338761}, \href
  {https://ui.adsabs.harvard.edu/abs/2002ApJS..139..467I} {139, 467}

\bibitem[\protect\citeauthoryear{{Israel} \& {Baas}}{{Israel} \&
  {Baas}}{2002}]{2002israel}
{Israel} F.~P.,  {Baas} F.,  2002, \mn@doi [\aap] {10.1051/0004-6361:20011736},
  \href {https://ui.adsabs.harvard.edu/abs/2002A&A...383...82I} {383, 82}

\bibitem[\protect\citeauthoryear{{Ivezi{\'c}} et~al.,}{{Ivezi{\'c}}
  et~al.}{2002}]{2002ivezic}
{Ivezi{\'c}} {\v{Z}}.,  et~al., 2002, \mn@doi [\aj] {10.1086/344069}, \href
  {https://ui.adsabs.harvard.edu/abs/2002AJ....124.2364I} {124, 2364}

\bibitem[\protect\citeauthoryear{{Jalan}, {Chand}  \& {Srianand}}{{Jalan}
  et~al.}{2019}]{2019jalan}
{Jalan} P.,  {Chand} H.,   {Srianand} R.,  2019, \mn@doi [\apj]
  {10.3847/1538-4357/ab4191}, \href
  {https://ui.adsabs.harvard.edu/abs/2019ApJ...884..151J} {884, 151}

\bibitem[\protect\citeauthoryear{{Jarvis} et~al.,}{{Jarvis}
  et~al.}{2020}]{2020jarvis}
{Jarvis} M.~E.,  et~al., 2020, \mn@doi [\mnras] {10.1093/mnras/staa2196}, \href
  {https://ui.adsabs.harvard.edu/abs/2020MNRAS.498.1560J} {498, 1560}

\bibitem[\protect\citeauthoryear{{Kanjilal}, {Dutta}  \& {Sharma}}{{Kanjilal}
  et~al.}{2021}]{2021kanjilal}
{Kanjilal} V.,  {Dutta} A.,   {Sharma} P.,  2021, \mn@doi [\mnras]
  {10.1093/mnras/staa3610}, \href
  {https://ui.adsabs.harvard.edu/abs/2021MNRAS.501.1143K} {501, 1143}

\bibitem[\protect\citeauthoryear{{Kauffmann} \& {Haehnelt}}{{Kauffmann} \&
  {Haehnelt}}{2000}]{2000kauffmann}
{Kauffmann} G.,  {Haehnelt} M.,  2000, \mn@doi [\mnras]
  {10.1046/j.1365-8711.2000.03077.x}, \href
  {https://ui.adsabs.harvard.edu/abs/2000MNRAS.311..576K} {311, 576}

\bibitem[\protect\citeauthoryear{{Kere{\v{s}}}, {Katz}, {Weinberg}  \&
  {Dav{\'e}}}{{Kere{\v{s}}} et~al.}{2005}]{2005keres}
{Kere{\v{s}}} D.,  {Katz} N.,  {Weinberg} D.~H.,   {Dav{\'e}} R.,  2005,
  \mn@doi [\mnras] {10.1111/j.1365-2966.2005.09451.x}, \href
  {https://ui.adsabs.harvard.edu/abs/2005MNRAS.363....2K} {363, 2}

\bibitem[\protect\citeauthoryear{{Kim} \& {Croft}}{{Kim} \&
  {Croft}}{2008}]{2008kim}
{Kim} Y.-R.,  {Croft} R. A.~C.,  2008, \mn@doi [\mnras]
  {10.1111/j.1365-2966.2008.13240.x}, \href
  {https://ui.adsabs.harvard.edu/abs/2008MNRAS.387..377K} {387, 377}

\bibitem[\protect\citeauthoryear{{Kormendy} \& {Ho}}{{Kormendy} \&
  {Ho}}{2013}]{2013kormendy}
{Kormendy} J.,  {Ho} L.~C.,  2013, \mn@doi [\araa]
  {10.1146/annurev-astro-082708-101811}, \href
  {https://ui.adsabs.harvard.edu/abs/2013ARA&A..51..511K} {51, 511}

\bibitem[\protect\citeauthoryear{{Lau}, {Prochaska}  \& {Hennawi}}{{Lau}
  et~al.}{2016}]{2016lau}
{Lau} M.~W.,  {Prochaska} J.~X.,   {Hennawi} J.~F.,  2016, \mn@doi [\apjs]
  {10.3847/0067-0049/226/2/25}, \href
  {https://ui.adsabs.harvard.edu/abs/2016ApJS..226...25L} {226, 25}

\bibitem[\protect\citeauthoryear{{Lau}, {Prochaska}  \& {Hennawi}}{{Lau}
  et~al.}{2018}]{2018lau}
{Lau} M.~W.,  {Prochaska} J.~X.,   {Hennawi} J.~F.,  2018, \mn@doi [\apj]
  {10.3847/1538-4357/aab78e}, \href
  {https://ui.adsabs.harvard.edu/abs/2018ApJ...857..126L} {857, 126}

\bibitem[\protect\citeauthoryear{{Laursen}, {Razoumov}  \&
  {Sommer-Larsen}}{{Laursen} et~al.}{2009}]{2009laursen}
{Laursen} P.,  {Razoumov} A.~O.,   {Sommer-Larsen} J.,  2009, \mn@doi [\apj]
  {10.1088/0004-637X/696/1/853}, \href
  {https://ui.adsabs.harvard.edu/abs/2009ApJ...696..853L} {696, 853}

\bibitem[\protect\citeauthoryear{{Law}, {Steidel}, {Erb}, {Larkin}, {Pettini},
  {Shapley}  \& {Wright}}{{Law} et~al.}{2009}]{2009law}
{Law} D.~R.,  {Steidel} C.~C.,  {Erb} D.~K.,  {Larkin} J.~E.,  {Pettini} M.,
  {Shapley} A.~E.,   {Wright} S.~A.,  2009, \mn@doi [\apj]
  {10.1088/0004-637X/697/2/2057}, \href
  {https://ui.adsabs.harvard.edu/abs/2009ApJ...697.2057L} {697, 2057}

\bibitem[\protect\citeauthoryear{{Lu} et~al.,}{{Lu} et~al.}{2018}]{2018lu}
{Lu} N.,  et~al., 2018, \mn@doi [\apj] {10.3847/1538-4357/aad3c9}, \href
  {https://ui.adsabs.harvard.edu/abs/2018ApJ...864...38L} {864, 38}

\bibitem[\protect\citeauthoryear{{Lyke} et~al.,}{{Lyke}
  et~al.}{2020}]{2020lyke}
{Lyke} B.~W.,  et~al., 2020, \mn@doi [\apjs] {10.3847/1538-4365/aba623}, \href
  {https://ui.adsabs.harvard.edu/abs/2020ApJS..250....8L} {250, 8}

\bibitem[\protect\citeauthoryear{{Mackenzie} et~al.,}{{Mackenzie}
  et~al.}{2021}]{2021mackenzie}
{Mackenzie} R.,  et~al., 2021, \mn@doi [\mnras] {10.1093/mnras/staa3277}, \href
  {https://ui.adsabs.harvard.edu/abs/2021MNRAS.502..494M} {502, 494}

\bibitem[\protect\citeauthoryear{{Matuszewski}, {Chang}, {Crabill}, {Martin},
  {Moore}, {Morrissey}  \& {Rahman}}{{Matuszewski} et~al.}{2010}]{2010matu}
{Matuszewski} M.,  {Chang} D.,  {Crabill} R.~M.,  {Martin} D.~C.,  {Moore}
  A.~M.,  {Morrissey} P.,   {Rahman} S.,  2010, in {McLean} I.~S.,  {Ramsay}
  S.~K.,   {Takami} H.,  eds,  Society of Photo-Optical Instrumentation
  Engineers (SPIE) Conference Series Vol. 7735, Ground-based and Airborne
  Instrumentation for Astronomy III. p. 77350P, \mn@doi{10.1117/12.856644}

\bibitem[\protect\citeauthoryear{{McLinden}, {Malhotra}, {Rhoads}, {Hibon},
  {Weijmans}  \& {Tilvi}}{{McLinden} et~al.}{2013}]{2013mclinden}
{McLinden} E.~M.,  {Malhotra} S.,  {Rhoads} J.~E.,  {Hibon} P.,  {Weijmans}
  A.-M.,   {Tilvi} V.,  2013, \mn@doi [\apj] {10.1088/0004-637X/767/1/48},
  \href {https://ui.adsabs.harvard.edu/abs/2013ApJ...767...48M} {767, 48}

\bibitem[\protect\citeauthoryear{{Molina} et~al.,}{{Molina}
  et~al.}{2021}]{2021molina}
{Molina} et~al., 2021, arXiv e-prints, \href
  {https://ui.adsabs.harvard.edu/abs/2021arXiv210100764M} {p. arXiv:2101.00764}

\bibitem[\protect\citeauthoryear{{Morrissey} et~al.,}{{Morrissey}
  et~al.}{2012}]{2012morrissey}
{Morrissey} P.,  et~al., 2012, in {McLean} I.~S.,  {Ramsay} S.~K.,   {Takami}
  H.,  eds,  Society of Photo-Optical Instrumentation Engineers (SPIE)
  Conference Series Vol. 8446, Ground-based and Airborne Instrumentation for
  Astronomy IV. p. 844613, \mn@doi{10.1117/12.924729}

\bibitem[\protect\citeauthoryear{{Moster}, {Naab}  \& {White}}{{Moster}
  et~al.}{2018}]{2018moster}
{Moster} B.~P.,  {Naab} T.,   {White} S. D.~M.,  2018, \mn@doi [\mnras]
  {10.1093/mnras/sty655}, \href
  {https://ui.adsabs.harvard.edu/abs/2018MNRAS.477.1822M} {477, 1822}

\bibitem[\protect\citeauthoryear{{Narayanan} et~al.,}{{Narayanan}
  et~al.}{2006}]{2006narayanan}
{Narayanan} D.,  et~al., 2006, \mn@doi [\apjl] {10.1086/504846}, \href
  {https://ui.adsabs.harvard.edu/abs/2006ApJ...642L.107N} {642, L107}

\bibitem[\protect\citeauthoryear{{Narayanan}, {Cox}, {Hayward}, {Younger}  \&
  {Hernquist}}{{Narayanan} et~al.}{2009}]{2009narayanan}
{Narayanan} D.,  {Cox} T.~J.,  {Hayward} C.~C.,  {Younger} J.~D.,   {Hernquist}
  L.,  2009, \mn@doi [\mnras] {10.1111/j.1365-2966.2009.15581.x}, \href
  {https://ui.adsabs.harvard.edu/abs/2009MNRAS.400.1919N} {400, 1919}

\bibitem[\protect\citeauthoryear{{Neri} et~al.,}{{Neri}
  et~al.}{2003}]{2003neri}
{Neri} R.,  et~al., 2003, \mn@doi [\apjl] {10.1086/379968}, \href
  {https://ui.adsabs.harvard.edu/abs/2003ApJ...597L.113N} {597, L113}

\bibitem[\protect\citeauthoryear{{Neufeld}}{{Neufeld}}{1990}]{1990neufeld}
{Neufeld} D.~A.,  1990, \mn@doi [\apj] {10.1086/168375}, \href
  {https://ui.adsabs.harvard.edu/abs/1990ApJ...350..216N} {350, 216}

\bibitem[\protect\citeauthoryear{{Novak} et~al.,}{{Novak}
  et~al.}{2019}]{2019novak}
{Novak} M.,  et~al., 2019, \mn@doi [\apj] {10.3847/1538-4357/ab2beb}, \href
  {https://ui.adsabs.harvard.edu/abs/2019ApJ...881...63N} {881, 63}

\bibitem[\protect\citeauthoryear{{O'Sullivan}, {Martin}, {Matuszewski},
  {Hoadley}, {Hamden}, {Neill}, {Lin}  \& {Parihar}}{{O'Sullivan}
  et~al.}{2020}]{2020osullivan}
{O'Sullivan} D.~B.,  {Martin} C.,  {Matuszewski} M.,  {Hoadley} K.,  {Hamden}
  E.,  {Neill} J.~D.,  {Lin} Z.,   {Parihar} P.,  2020, \mn@doi [\apj]
  {10.3847/1538-4357/ab838c}, \href
  {https://ui.adsabs.harvard.edu/abs/2020ApJ...894....3O} {894, 3}

\bibitem[\protect\citeauthoryear{{Omont}, {Petitjean}, {Guilloteau}, {McMahon},
  {Solomon}  \& {P{\'e}contal}}{{Omont} et~al.}{1996}]{1996omont}
{Omont} A.,  {Petitjean} P.,  {Guilloteau} S.,  {McMahon} R.~G.,  {Solomon}
  P.~M.,   {P{\'e}contal} E.,  1996, \mn@doi [\nat] {10.1038/382428a0}, \href
  {https://ui.adsabs.harvard.edu/abs/1996Natur.382..428O} {382, 428}

\bibitem[\protect\citeauthoryear{{Ouchi} et~al.,}{{Ouchi}
  et~al.}{2004}]{2004ouchi}
{Ouchi} M.,  et~al., 2004, \mn@doi [\apj] {10.1086/422208}, \href
  {https://ui.adsabs.harvard.edu/abs/2004ApJ...611..685O} {611, 685}

\bibitem[\protect\citeauthoryear{{Papadopoulos} \& {Greve}}{{Papadopoulos} \&
  {Greve}}{2004}]{2004papadopoulos}
{Papadopoulos} P.~P.,  {Greve} T.~R.,  2004, \mn@doi [\apjl] {10.1086/426059},
  \href {https://ui.adsabs.harvard.edu/abs/2004ApJ...615L..29P} {615, L29}

\bibitem[\protect\citeauthoryear{{Peeples} et~al.,}{{Peeples}
  et~al.}{2019}]{2019peeples}
{Peeples} M.~S.,  et~al., 2019, \mn@doi [\apj] {10.3847/1538-4357/ab0654},
  \href {https://ui.adsabs.harvard.edu/abs/2019ApJ...873..129P} {873, 129}

\bibitem[\protect\citeauthoryear{{Perley}, {Chandler}, {Butler}  \&
  {Wrobel}}{{Perley} et~al.}{2011}]{2011perley}
{Perley} R.~A.,  {Chandler} C.~J.,  {Butler} B.~J.,   {Wrobel} J.~M.,  2011,
  \mn@doi [\apjl] {10.1088/2041-8205/739/1/L1}, \href
  {https://ui.adsabs.harvard.edu/abs/2011ApJ...739L...1P} {739, L1}

\bibitem[\protect\citeauthoryear{{Polletta}, {Nesvadba}, {Neri}, {Omont},
  {Berta}  \& {Bergeron}}{{Polletta} et~al.}{2011}]{2011polletta}
{Polletta} M.,  {Nesvadba} N.~P.~H.,  {Neri} R.,  {Omont} A.,  {Berta} S.,
  {Bergeron} J.,  2011, \mn@doi [\aap] {10.1051/0004-6361/201116446}, \href
  {https://ui.adsabs.harvard.edu/abs/2011A&A...533A..20P} {533, A20}

\bibitem[\protect\citeauthoryear{{Porciani}, {Magliocchetti}  \&
  {Norberg}}{{Porciani} et~al.}{2004}]{2004porciani}
{Porciani} C.,  {Magliocchetti} M.,   {Norberg} P.,  2004, \mn@doi [\mnras]
  {10.1111/j.1365-2966.2004.08408.x}, \href
  {https://ui.adsabs.harvard.edu/abs/2004MNRAS.355.1010P} {355, 1010}

\bibitem[\protect\citeauthoryear{{Prescott}, {Martin}  \& {Dey}}{{Prescott}
  et~al.}{2015}]{2015prescott}
{Prescott} M. K.~M.,  {Martin} C.~L.,   {Dey} A.,  2015, \mn@doi [\apj]
  {10.1088/0004-637X/799/1/62}, \href
  {https://ui.adsabs.harvard.edu/abs/2015ApJ...799...62P} {799, 62}

\bibitem[\protect\citeauthoryear{{Prochaska}, {Hennawi}  \&
  {Simcoe}}{{Prochaska} et~al.}{2013}]{2013prochaska}
{Prochaska} J.~X.,  {Hennawi} J.~F.,   {Simcoe} R.~A.,  2013, \mn@doi [\apjl]
  {10.1088/2041-8205/762/2/L19}, \href
  {https://ui.adsabs.harvard.edu/abs/2013ApJ...762L..19P} {762, L19}

\bibitem[\protect\citeauthoryear{{Prochaska}, {Lau}  \& {Hennawi}}{{Prochaska}
  et~al.}{2014}]{2014prochaska}
{Prochaska} J.~X.,  {Lau} M.~W.,   {Hennawi} J.~F.,  2014, \mn@doi [\apj]
  {10.1088/0004-637X/796/2/140}, \href
  {https://ui.adsabs.harvard.edu/abs/2014ApJ...796..140P} {796, 140}

\bibitem[\protect\citeauthoryear{{Rees}}{{Rees}}{1988}]{1988rees}
{Rees} M.~J.,  1988, \mn@doi [\mnras] {10.1093/mnras/231.1.91P}, \href
  {https://ui.adsabs.harvard.edu/abs/1988MNRAS.231P..91R} {231, 91P}

\bibitem[\protect\citeauthoryear{{Richards} et~al.,}{{Richards}
  et~al.}{2006}]{2006richards}
{Richards} G.~T.,  et~al., 2006, \mn@doi [\aj] {10.1086/503559}, \href
  {https://ui.adsabs.harvard.edu/abs/2006AJ....131.2766R} {131, 2766}

\bibitem[\protect\citeauthoryear{{Richstone} et~al.,}{{Richstone}
  et~al.}{1998}]{1998richstone}
{Richstone} D.,  et~al., 1998, \nat, \href
  {https://ui.adsabs.harvard.edu/abs/1998Natur.395A..14R} {385, A14}

\bibitem[\protect\citeauthoryear{{Riechers} et~al.,}{{Riechers}
  et~al.}{2006}]{2006riechers}
{Riechers} D.~A.,  et~al., 2006, \mn@doi [\apj] {10.1086/507014}, \href
  {https://ui.adsabs.harvard.edu/abs/2006ApJ...650..604R} {650, 604}

\bibitem[\protect\citeauthoryear{{Riechers}, {Walter}, {Carilli}  \&
  {Lewis}}{{Riechers} et~al.}{2009}]{2009riechers}
{Riechers} D.~A.,  {Walter} F.,  {Carilli} C.~L.,   {Lewis} G.~F.,  2009,
  \mn@doi [\apj] {10.1088/0004-637X/690/1/463}, \href
  {https://ui.adsabs.harvard.edu/abs/2009ApJ...690..463R} {690, 463}

\bibitem[\protect\citeauthoryear{{Riechers} et~al.,}{{Riechers}
  et~al.}{2011}]{2011riechers}
{Riechers} D.~A.,  et~al., 2011, \mn@doi [\apjl] {10.1088/2041-8205/739/1/L32},
  \href {https://ui.adsabs.harvard.edu/abs/2011ApJ...739L..32R} {739, L32}

\bibitem[\protect\citeauthoryear{{Riechers} et~al.,}{{Riechers}
  et~al.}{2013}]{2013riechers}
{Riechers} D.~A.,  et~al., 2013, \mn@doi [\nat] {10.1038/nature12050}, \href
  {https://ui.adsabs.harvard.edu/abs/2013Natur.496..329R} {496, 329}

\bibitem[\protect\citeauthoryear{{Salak}, {Nakai}, {Seta}  \&
  {Miyamoto}}{{Salak} et~al.}{2019}]{2019salak}
{Salak} D.,  {Nakai} N.,  {Seta} M.,   {Miyamoto} Y.,  2019, \mn@doi [\apj]
  {10.3847/1538-4357/ab55dc}, \href
  {https://ui.adsabs.harvard.edu/abs/2019ApJ...887..143S} {887, 143}

\bibitem[\protect\citeauthoryear{{Schmidt}}{{Schmidt}}{1963}]{1963schmidt}
{Schmidt} M.,  1963, \mn@doi [\nat] {10.1038/1971040a0}, \href
  {https://ui.adsabs.harvard.edu/abs/1963Natur.197.1040S} {197, 1040}

\bibitem[\protect\citeauthoryear{{Schumacher}, {Mart{\'\i}nez-Sansigre},
  {Lacy}, {Rawlings}  \& {Schinnerer}}{{Schumacher}
  et~al.}{2012}]{2012schumacher}
{Schumacher} H.,  {Mart{\'\i}nez-Sansigre} A.,  {Lacy} M.,  {Rawlings} S.,
  {Schinnerer} E.,  2012, \mn@doi [\mnras] {10.1111/j.1365-2966.2012.21024.x},
  \href {https://ui.adsabs.harvard.edu/abs/2012MNRAS.423.2132S} {423, 2132}

\bibitem[\protect\citeauthoryear{{Shangguan}, {Ho}, {Bauer}, {Wang}  \&
  {Treister}}{{Shangguan} et~al.}{2020}]{2020shangguan}
{Shangguan} J.,  {Ho} L.~C.,  {Bauer} F.~E.,  {Wang} R.,   {Treister} E.,
  2020, \mn@doi [\apj] {10.3847/1538-4357/aba8a1}, \href
  {https://ui.adsabs.harvard.edu/abs/2020ApJ...899..112S} {899, 112}

\bibitem[\protect\citeauthoryear{{Shen} et~al.,}{{Shen}
  et~al.}{2007}]{2007shen}
{Shen} Y.,  et~al., 2007, \mn@doi [\aj] {10.1086/513517}, \href
  {https://ui.adsabs.harvard.edu/abs/2007AJ....133.2222S} {133, 2222}

\bibitem[\protect\citeauthoryear{{Shen} et~al.,}{{Shen}
  et~al.}{2016}]{2016shen}
{Shen} Y.,  et~al., 2016, \mn@doi [\apj] {10.3847/0004-637X/831/1/7}, \href
  {https://ui.adsabs.harvard.edu/abs/2016ApJ...831....7S} {831, 7}

\bibitem[\protect\citeauthoryear{{Shen}, {Hopkins}, {Faucher-Gigu{\`e}re},
  {Alexander}, {Richards}, {Ross}  \& {Hickox}}{{Shen} et~al.}{2020}]{2020shen}
{Shen} X.,  {Hopkins} P.~F.,  {Faucher-Gigu{\`e}re} C.-A.,  {Alexander} D.~M.,
  {Richards} G.~T.,  {Ross} N.~P.,   {Hickox} R.~C.,  2020, \mn@doi [\mnras]
  {10.1093/mnras/staa1381}, \href
  {https://ui.adsabs.harvard.edu/abs/2020MNRAS.495.3252S} {495, 3252}

\bibitem[\protect\citeauthoryear{{Silk} \& {Rees}}{{Silk} \&
  {Rees}}{1998}]{1998silk}
{Silk} J.,  {Rees} M.~J.,  1998, \aap, \href
  {https://ui.adsabs.harvard.edu/abs/1998A&A...331L...1S} {331, L1}

\bibitem[\protect\citeauthoryear{{Silva}, {Sajina}, {Lonsdale}  \&
  {Lacy}}{{Silva} et~al.}{2015}]{2015silva}
{Silva} A.,  {Sajina} A.,  {Lonsdale} C.,   {Lacy} M.,  2015, \mn@doi [\apjl]
  {10.1088/2041-8205/806/2/L25}, \href
  {https://ui.adsabs.harvard.edu/abs/2015ApJ...806L..25S} {806, L25}

\bibitem[\protect\citeauthoryear{{Solomon}, {Downes}  \& {Radford}}{{Solomon}
  et~al.}{1992}]{1992solomon}
{Solomon} P.~M.,  {Downes} D.,   {Radford} S.~J.~E.,  1992, \mn@doi [\apjl]
  {10.1086/186569}, \href
  {https://ui.adsabs.harvard.edu/abs/1992ApJ...398L..29S} {398, L29}

\bibitem[\protect\citeauthoryear{{Spingola} et~al.,}{{Spingola}
  et~al.}{2020}]{2020spingola}
{Spingola} C.,  et~al., 2020, \mn@doi [\mnras] {10.1093/mnras/staa1342}, \href
  {https://ui.adsabs.harvard.edu/abs/2020MNRAS.495.2387S} {495, 2387}

\bibitem[\protect\citeauthoryear{{Stacey} et~al.,}{{Stacey}
  et~al.}{2021}]{2021stacey}
{Stacey} H.~R.,  et~al., 2021, \mn@doi [\mnras] {10.1093/mnras/staa3433}, \href
  {https://ui.adsabs.harvard.edu/abs/2021MNRAS.500.3667S} {500, 3667}

\bibitem[\protect\citeauthoryear{{Steinborn}, {Dolag}, {Hirschmann}, {Prieto}
  \& {Remus}}{{Steinborn} et~al.}{2015}]{2015steinborn}
{Steinborn} L.~K.,  {Dolag} K.,  {Hirschmann} M.,  {Prieto} M.~A.,   {Remus}
  R.-S.,  2015, \mn@doi [\mnras] {10.1093/mnras/stv072}, \href
  {https://ui.adsabs.harvard.edu/abs/2015MNRAS.448.1504S} {448, 1504}

\bibitem[\protect\citeauthoryear{{Tacconi} et~al.,}{{Tacconi}
  et~al.}{2018}]{2018tacconi}
{Tacconi} L.~J.,  et~al., 2018, \mn@doi [\apj] {10.3847/1538-4357/aaa4b4},
  \href {https://ui.adsabs.harvard.edu/abs/2018ApJ...853..179T} {853, 179}

\bibitem[\protect\citeauthoryear{{Timlin} et~al.,}{{Timlin}
  et~al.}{2018}]{2018timlin}
{Timlin} J.~D.,  et~al., 2018, \mn@doi [\apj] {10.3847/1538-4357/aab9ac}, \href
  {https://ui.adsabs.harvard.edu/abs/2018ApJ...859...20T} {859, 20}

\bibitem[\protect\citeauthoryear{{Trainor}, {Steidel}, {Strom}  \&
  {Rudie}}{{Trainor} et~al.}{2015}]{2015trainor}
{Trainor} R.~F.,  {Steidel} C.~C.,  {Strom} A.~L.,   {Rudie} G.~C.,  2015,
  \mn@doi [\apj] {10.1088/0004-637X/809/1/89}, \href
  {https://ui.adsabs.harvard.edu/abs/2015ApJ...809...89T} {809, 89}

\bibitem[\protect\citeauthoryear{{Travascio} et~al.,}{{Travascio}
  et~al.}{2020}]{2020travascio}
{Travascio} A.,  et~al., 2020, \mn@doi [\aap] {10.1051/0004-6361/201936197},
  \href {https://ui.adsabs.harvard.edu/abs/2020A&A...635A.157T} {635, A157}

\bibitem[\protect\citeauthoryear{{Tumlinson}, {Peeples}  \& {Werk}}{{Tumlinson}
  et~al.}{2017}]{2017tumlinson}
{Tumlinson} J.,  {Peeples} M.~S.,   {Werk} J.~K.,  2017, \mn@doi [\araa]
  {10.1146/annurev-astro-091916-055240}, \href
  {https://ui.adsabs.harvard.edu/abs/2017ARA&A..55..389T} {55, 389}

\bibitem[\protect\citeauthoryear{{Urry} \& {Padovani}}{{Urry} \&
  {Padovani}}{1995}]{1995urry}
{Urry} C.~M.,  {Padovani} P.,  1995, \mn@doi [\pasp] {10.1086/133630}, \href
  {https://ui.adsabs.harvard.edu/abs/1995PASP..107..803U} {107, 803}

\bibitem[\protect\citeauthoryear{{Valentino} et~al.,}{{Valentino}
  et~al.}{2018}]{2018valentino}
{Valentino} F.,  et~al., 2018, \mn@doi [\apj] {10.3847/1538-4357/aaeb88}, \href
  {https://ui.adsabs.harvard.edu/abs/2018ApJ...869...27V} {869, 27}

\bibitem[\protect\citeauthoryear{{Venemans}, {Walter}, {Zschaechner},
  {Decarli}, {De Rosa}, {Findlay}, {McMahon}  \& {Sutherland}}{{Venemans}
  et~al.}{2016}]{2016venemans}
{Venemans} B.~P.,  {Walter} F.,  {Zschaechner} L.,  {Decarli} R.,  {De Rosa}
  G.,  {Findlay} J.~R.,  {McMahon} R.~G.,   {Sutherland} W.~J.,  2016, \mn@doi
  [\apj] {10.3847/0004-637X/816/1/37}, \href
  {https://ui.adsabs.harvard.edu/abs/2016ApJ...816...37V} {816, 37}

\bibitem[\protect\citeauthoryear{{Venemans} et~al.,}{{Venemans}
  et~al.}{2017}]{2017venemans}
{Venemans} B.~P.,  et~al., 2017, \mn@doi [\apj] {10.3847/1538-4357/aa81cb},
  \href {https://ui.adsabs.harvard.edu/abs/2017ApJ...845..154V} {845, 154}

\bibitem[\protect\citeauthoryear{{Verhamme}, {Schaerer}  \&
  {Maselli}}{{Verhamme} et~al.}{2006}]{2006verhamme}
{Verhamme} A.,  {Schaerer} D.,   {Maselli} A.,  2006, \mn@doi [\aap]
  {10.1051/0004-6361:20065554}, \href
  {https://ui.adsabs.harvard.edu/abs/2006A&A...460..397V} {460, 397}

\bibitem[\protect\citeauthoryear{{Verhamme} et~al.,}{{Verhamme}
  et~al.}{2018}]{2018verhamme}
{Verhamme} A.,  et~al., 2018, \mn@doi [\mnras] {10.1093/mnrasl/sly058}, \href
  {https://ui.adsabs.harvard.edu/abs/2018MNRAS.478L..60V} {478, L60}

\bibitem[\protect\citeauthoryear{{Walter} et~al.,}{{Walter}
  et~al.}{2003}]{2003walter}
{Walter} F.,  et~al., 2003, \mn@doi [\nat] {10.1038/nature01821}, \href
  {https://ui.adsabs.harvard.edu/abs/2003Natur.424..406W} {424, 406}

\bibitem[\protect\citeauthoryear{{Walter}, {Carilli}, {Bertoldi}, {Menten},
  {Cox}, {Lo}, {Fan}  \& {Strauss}}{{Walter} et~al.}{2004}]{2004walter}
{Walter} F.,  {Carilli} C.,  {Bertoldi} F.,  {Menten} K.,  {Cox} P.,  {Lo}
  K.~Y.,  {Fan} X.,   {Strauss} M.~A.,  2004, \mn@doi [\apjl] {10.1086/426017},
  \href {https://ui.adsabs.harvard.edu/abs/2004ApJ...615L..17W} {615, L17}

\bibitem[\protect\citeauthoryear{{Walter}, {Wei{\ss}}, {Downes}, {Decarli}  \&
  {Henkel}}{{Walter} et~al.}{2011}]{2011walter}
{Walter} F.,  {Wei{\ss}} A.,  {Downes} D.,  {Decarli} R.,   {Henkel} C.,  2011,
  \mn@doi [\apj] {10.1088/0004-637X/730/1/18}, \href
  {https://ui.adsabs.harvard.edu/abs/2011ApJ...730...18W} {730, 18}

\bibitem[\protect\citeauthoryear{{Wang} et~al.,}{{Wang}
  et~al.}{2010}]{2010wang}
{Wang} R.,  et~al., 2010, \mn@doi [\apj] {10.1088/0004-637X/714/1/699}, \href
  {https://ui.adsabs.harvard.edu/abs/2010ApJ...714..699W} {714, 699}

\bibitem[\protect\citeauthoryear{{Wang} et~al.,}{{Wang}
  et~al.}{2013}]{2013wang}
{Wang} R.,  et~al., 2013, \mn@doi [\apj] {10.1088/0004-637X/773/1/44}, \href
  {https://ui.adsabs.harvard.edu/abs/2013ApJ...773...44W} {773, 44}

\bibitem[\protect\citeauthoryear{{Wang} et~al.,}{{Wang}
  et~al.}{2016}]{2016wang}
{Wang} R.,  et~al., 2016, \mn@doi [\apj] {10.3847/0004-637X/830/1/53}, \href
  {https://ui.adsabs.harvard.edu/abs/2016ApJ...830...53W} {830, 53}

\bibitem[\protect\citeauthoryear{{Wang}, {Wang}, {Fan}, {Wu}, {Yang}, {Neri}
  \& {Yue}}{{Wang} et~al.}{2019}]{2019wang}
{Wang} F.,  {Wang} R.,  {Fan} X.,  {Wu} X.-B.,  {Yang} J.,  {Neri} R.,   {Yue}
  M.,  2019, \mn@doi [\apj] {10.3847/1538-4357/ab2717}, \href
  {https://ui.adsabs.harvard.edu/abs/2019ApJ...880....2W} {880, 2}

\bibitem[\protect\citeauthoryear{{Wei{\ss}}, {Henkel}, {Downes}  \&
  {Walter}}{{Wei{\ss}} et~al.}{2003}]{2003weiss}
{Wei{\ss}} A.,  {Henkel} C.,  {Downes} D.,   {Walter} F.,  2003, \mn@doi [\aap]
  {10.1051/0004-6361:20031337}, \href
  {https://ui.adsabs.harvard.edu/abs/2003A&A...409L..41W} {409, L41}

\bibitem[\protect\citeauthoryear{{Wei{\ss}}, {Downes}, {Henkel}  \&
  {Walter}}{{Wei{\ss}} et~al.}{2005}]{2005weiss}
{Wei{\ss}} A.,  {Downes} D.,  {Henkel} C.,   {Walter} F.,  2005, \mn@doi [\aap]
  {10.1051/0004-6361:200400085}, \href
  {https://ui.adsabs.harvard.edu/abs/2005A&A...429L..25W} {429, L25}

\bibitem[\protect\citeauthoryear{{Weiss}, {Downes}, {Walter}  \&
  {Henkel}}{{Weiss} et~al.}{2007a}]{2007bweiss}
{Weiss} A.,  {Downes} D.,  {Walter} F.,   {Henkel} C.,  2007a, in {Baker}
  A.~J.,  {Glenn} J.,  {Harris} A.~I.,  {Mangum} J.~G.,   {Yun} M.~S.,  eds,
  Astronomical Society of the Pacific Conference Series Vol. 375, From
  Z-Machines to ALMA: (Sub)Millimeter Spectroscopy of Galaxies. p.~25

\bibitem[\protect\citeauthoryear{{Wei{\ss}}, {Downes}, {Neri}, {Walter},
  {Henkel}, {Wilner}, {Wagg}  \& {Wiklind}}{{Wei{\ss}}
  et~al.}{2007b}]{2007weiss}
{Wei{\ss}} A.,  {Downes} D.,  {Neri} R.,  {Walter} F.,  {Henkel} C.,  {Wilner}
  D.~J.,  {Wagg} J.,   {Wiklind} T.,  2007b, \mn@doi [\aap]
  {10.1051/0004-6361:20066117}, \href
  {https://ui.adsabs.harvard.edu/abs/2007A&A...467..955W} {467, 955}

\bibitem[\protect\citeauthoryear{{White}, {Ellison}, {Claude}, {Dent}  \&
  {Matheson}}{{White} et~al.}{1994}]{1994white}
{White} G.~J.,  {Ellison} B.,  {Claude} S.,  {Dent} W.~R.~F.,   {Matheson}
  D.~N.,  1994, \aap, \href
  {https://ui.adsabs.harvard.edu/abs/1994A&A...284L..23W} {284, L23}

\bibitem[\protect\citeauthoryear{{White} et~al.,}{{White}
  et~al.}{2012}]{2012white}
{White} M.,  et~al., 2012, \mn@doi [\mnras] {10.1111/j.1365-2966.2012.21251.x},
  \href {https://ui.adsabs.harvard.edu/abs/2012MNRAS.424..933W} {424, 933}

\bibitem[\protect\citeauthoryear{{Wootten} \& {Thompson}}{{Wootten} \&
  {Thompson}}{2009}]{2009wootten}
{Wootten} A.,  {Thompson} A.~R.,  2009, \mn@doi [IEEE Proceedings]
  {10.1109/JPROC.2009.2020572}, \href
  {https://ui.adsabs.harvard.edu/abs/2009IEEEP..97.1463W} {97, 1463}

\bibitem[\protect\citeauthoryear{{Yang}, {Stancil}, {Balakrishnan}  \&
  {Forrey}}{{Yang} et~al.}{2010}]{2010yang}
{Yang} B.,  {Stancil} P.~C.,  {Balakrishnan} N.,   {Forrey} R.~C.,  2010,
  \mn@doi [\apj] {10.1088/0004-637X/718/2/1062}, \href
  {https://ui.adsabs.harvard.edu/abs/2010ApJ...718.1062Y} {718, 1062}

\bibitem[\protect\citeauthoryear{{Yang}, {Zabludoff}, {Jahnke}, {Eisenstein},
  {Dav{\'e}}, {Shectman}  \& {Kelson}}{{Yang} et~al.}{2011}]{2011yang}
{Yang} Y.,  {Zabludoff} A.,  {Jahnke} K.,  {Eisenstein} D.,  {Dav{\'e}} R.,
  {Shectman} S.~A.,   {Kelson} D.~D.,  2011, \mn@doi [\apj]
  {10.1088/0004-637X/735/2/87}, \href
  {https://ui.adsabs.harvard.edu/abs/2011ApJ...735...87Y} {735, 87}

\bibitem[\protect\citeauthoryear{{Yang} et~al.,}{{Yang}
  et~al.}{2019}]{2019yang}
{Yang} J.,  et~al., 2019, \mn@doi [\apj] {10.3847/1538-4357/ab2a02}, \href
  {https://ui.adsabs.harvard.edu/abs/2019ApJ...880..153Y} {880, 153}

\bibitem[\protect\citeauthoryear{{Zhang}, {Shi}, {Rieke}, {Xia}, {Wang}, {Sun}
  \& {Wan}}{{Zhang} et~al.}{2016}]{2016zhang}
{Zhang} Z.,  {Shi} Y.,  {Rieke} G.~H.,  {Xia} X.,  {Wang} Y.,  {Sun} B.,
  {Wan} L.,  2016, \mn@doi [\apjl] {10.3847/2041-8205/819/2/L27}, \href
  {https://ui.adsabs.harvard.edu/abs/2016ApJ...819L..27Z} {819, L27}

\bibitem[\protect\citeauthoryear{{van der Tak}, {Black}, {Sch{\"o}ier},
  {Jansen}  \& {van Dishoeck}}{{van der Tak} et~al.}{2007}]{2007vandertak}
{van der Tak} F.~F.~S.,  {Black} J.~H.,  {Sch{\"o}ier} F.~L.,  {Jansen} D.~J.,
   {van Dishoeck} E.~F.,  2007, \mn@doi [\aap] {10.1051/0004-6361:20066820},
  \href {https://ui.adsabs.harvard.edu/abs/2007A&A...468..627V} {468, 627}

\makeatother
\end{thebibliography}

\appendix

\section{Stability of data}
\label{app:stability}
\begin{figure*}
    \centering
    \includegraphics[scale=0.75]{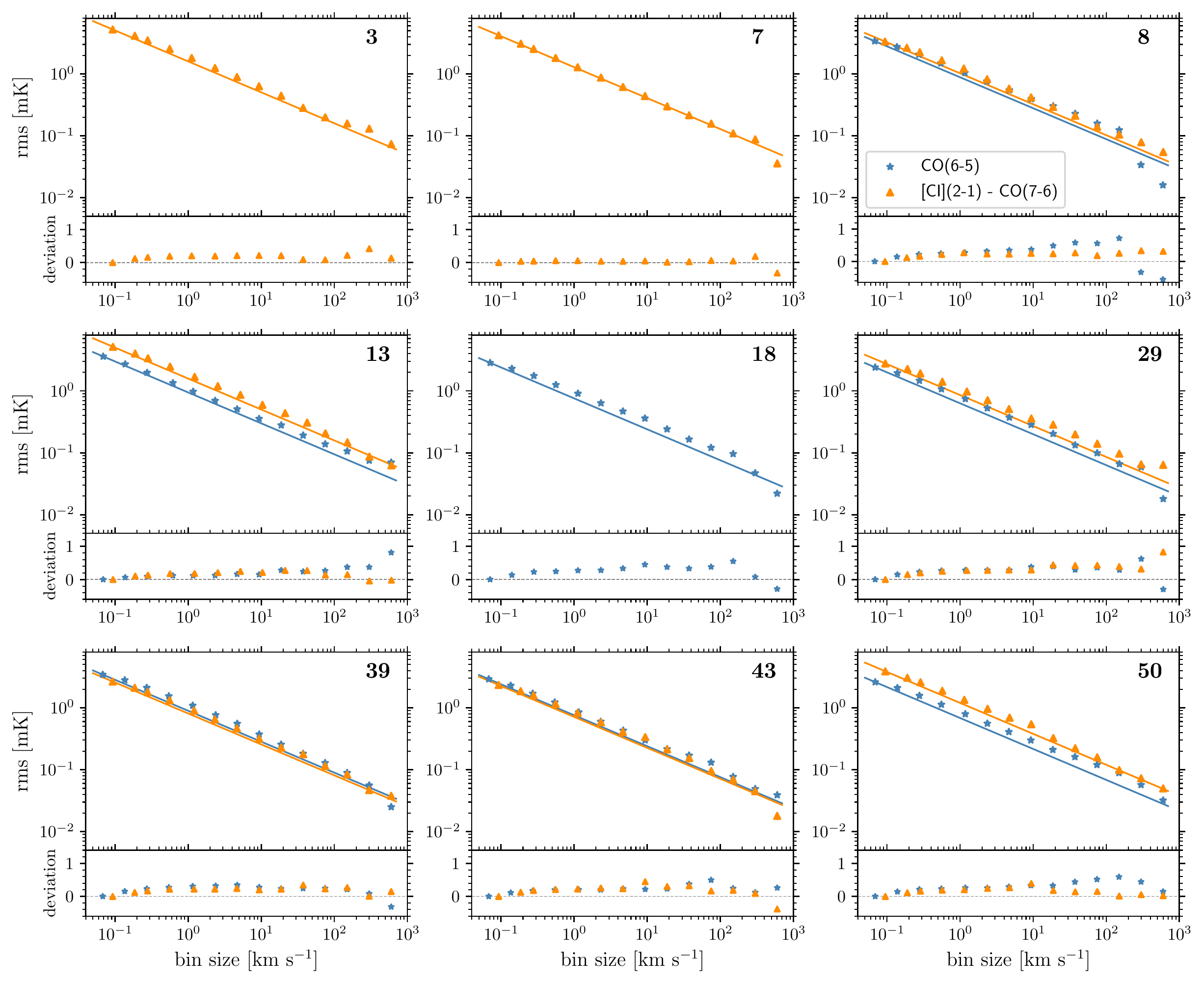}
    \caption{ RMS of the final combined spectra as a function of the bin size for each of the nine sources, for CO(6-5) (blue stars) and [C{\sc i}](2-1) - CO(7-6) (orange triangles) observations. The ID number of each quasar is shown in the top right corner of each panel. The solid lines represent the expected scaling of the rms with increasing bin size (see section~\ref{app:stability}). The lower subpanels show the deviation
    of the observed rms with respect to the predicted value, $\rm (rms_{\rm observed}\ - rms_{\rm expected})/ rms_{\rm expected}$.}
    \label{fig:appendix_rms}
\end{figure*}

We demonstrated the stability of the SEPIA180 receiver by checking if the noise scales down properly when using larger velocity bins. In 
Figure \ref{fig:appendix_rms} we show the rms of the final reduced spectra as a function of the bin size, for each of our sources in the different observed tunings. The smallest bin size corresponds to the original resolution of the spectra (i.e., the unbinned data). The solid lines represent the expected variation of the rms ($\rm rms_{\rm expected}$) with respect to the bin size, which was estimated as:
\begin{equation}
\rm rms_{\rm expected}= rms_{\rm original}\ \times \sqrt{binsize_{\rm original}/binsize_{\rm observed}}      
\end{equation}
where $\rm rms_{\rm original}$ ($\rm binsize_{\rm original}$) is the rms (bin size) at the original resolution, and $\rm binsize_{\rm observed}$ is the bin size used in the final combined spectrum. Each panel has a lower subpanel, which shows the deviation 
of the observed rms from 
the predicted value, estimated as $\rm (rms_{\rm observed}\ - rms_{\rm expected})/ rms_{\rm expected}$, where $\rm rms_{\rm observed}$ is the final rms obtained in the final binned spectrum.

At a bin size of 300 km~s$^{-1}$, we obtained a median 
deviation
of only 12$\%$ and 14$\%$ for CO(6-5) and [C{\sc i}](2-1), respectively. Such a small deviation ensures that we can preform the analysis discussed in the main text. \\

\begin{figure}
    \centering
    \includegraphics[scale=0.55]{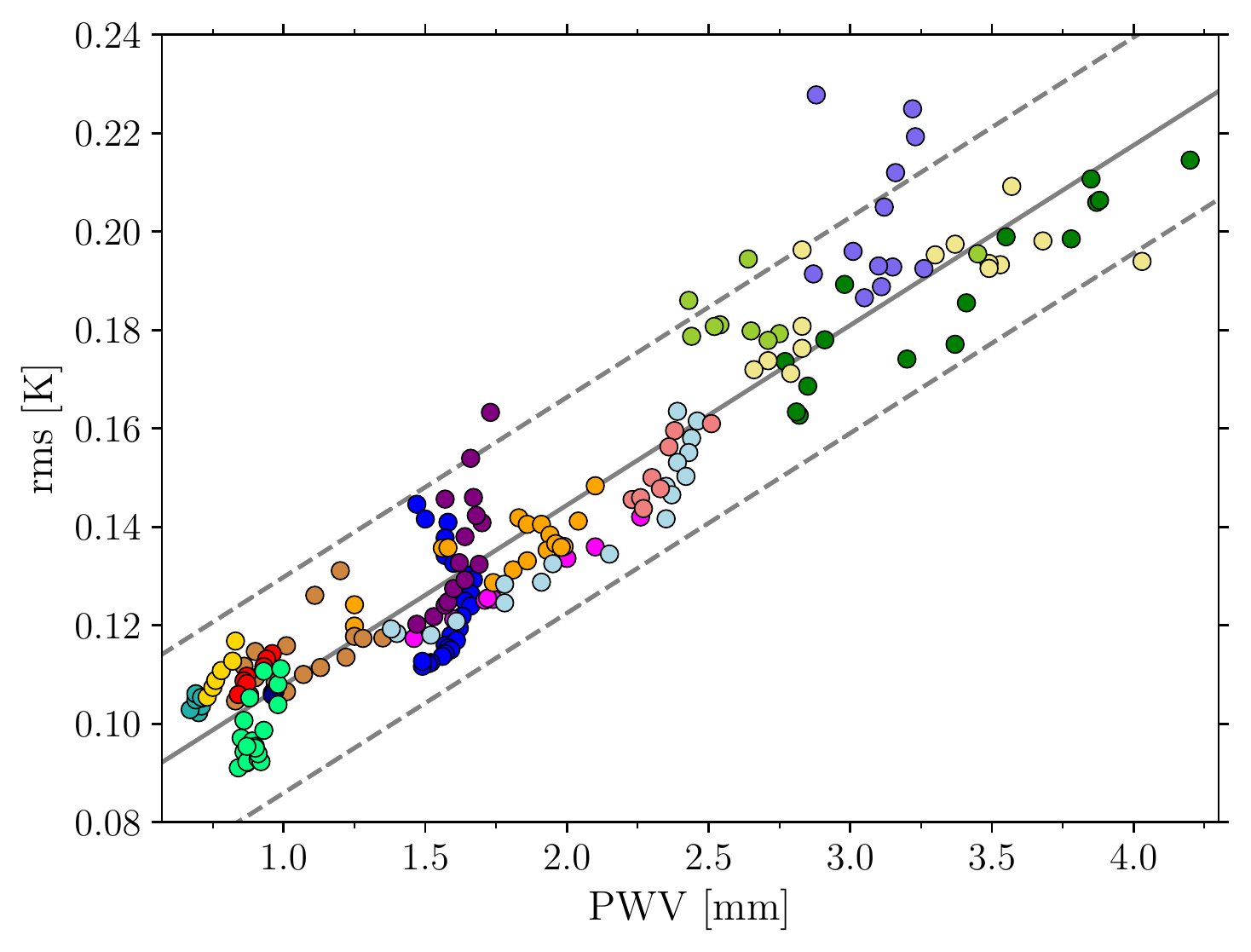}
    \caption{RMS values (for each scan) versus PWV relation for the source SDSS~J0100+2105 (ID 39), corresponding to the CO(6-5) and [C{\sc i}](2-1) - CO(7-6) observations. Each colour indicates a different observation date. The grey solid line represents a linear fit to the points, and the two dashed lines represent the $\pm$2$\sigma$ ($\pm 0.018$ K) deviation from the fit.} 
    \label{fig:pwv_rms}
\end{figure}

Further, as a proof of the stability over time of the calibration of the receiver, in Fig. \ref{fig:pwv_rms} we show the rms for each scan  versus its PWV for the source with ID 39 (J~0100+2105).  Each colour indicates a different observation date, spanning several months of APEX operations. These values correspond to both the CO(6-5) and [C{\sc i}](2-1) - CO(7-6) observations. A linear fit shown by the grey solid line has been applied to the points,  with a resulting squared correlation coefficient of $r^{2} = 0.93$ and a residual standard error of $\sigma = 0.009$ K.  The dashed lines enclose 95 per cent of the measurements (i.e. $\pm$2$\sigma$). Abrupt (vertical) changes of the rms during  one observing date (e.g., blue points) indicate strong changes in the airmass of the source. From this figure, we note that data taken under similar PWV and elevation, but in different dates, are characterized by the same rms.\\

\section{Reliability of line detections}
\label{app:spurious}

We performed two tests to confirm the reliability of the reported lines. First, we performed a negative test that consists in inverting the final combined spectra (multiplying the fluxes by -1) and applying our detection algorithm. Second, we produced jack-knife spectra by inverting each second scan during data reduction. We then apply the detection algorithm to these spectra which should contain only noise.

We stress that in both tests we applied the same detection algorithm used to identify line detections in our spectra (see Section \ref{section:obsresults}). The results of these tests are shown in Table \ref{table:app_observations}.  For completeness, we list in footnotes the presence of features that fulfill only partially our selection criteria (i.e. with peak S/N>2). \\

In summary, we did not find any detections in the negative test. While, in the jack-knife test we found only one line detection (integrated S/N=3.3) for ID8, [C{\sc i}](2-1)-CO(7-6) observations. We note that the final [C{\sc i}](2-1)-CO(7-6) combined spectrum for this source has a feature at 196.1 GHz (or -328 km s$^{-1}$, integrated S/N = 3.1, $I_{\rm line}\sim$3 Jy km s$^{-1}$) that according to our jack-knife test, could be a spurious line.

\begin{table*}   
\centering 
\caption{Numbers of spurious lines in the negative and jack-knife tests.}\label{table:app_observations}
\resizebox{0.9\textwidth}{!}{
\begin{threeparttable}
 \begin{tabular}{lccccc}
\hline\hline 
  & & \multicolumn{2}{c}{Negative test} & \multicolumn{2}{c}{Jackknife test}\\ \cline{3-6}
ID$^{a}$ & Quasar & CO(6-5) tuning       	& [C{\textsc{i}}],CO(7-6) tuning	& CO(6-5) tuning		& [C{\textsc{i}}],CO(7-6) tuning	 \\
\hline 
3  & J 0525-233         & -		& 0			& -		& 0$^{e}$		\\
7  & SDSS J1209+1138    & -		& 0			& -		& 0$^{f}$		\\	
8  & UM683 	            & 0 	& 0       	& 0$^{g}$      & 1$^{h}$ 		\\	
13 & PKS-1017+109       & 0		& 0	        & 0      & 0$^{i}$      \\			
18 & SDSS J1557+1540    & 0$^{b}$ & -	    & 0      & -      \\
29 & Q-0115-30          & 0		& 0	        & 0      & 0$^{j}$      \\	
39 & SDSS J0100+2105    & 0$^{c}$ & 0	    & 0      & 0$^{k}$      \\	
43 & CTSH22.05          & 0		& 0	        & 0       & 0      \\		
50 & SDSS J0819+0823    & 0		& 0$^{d}$ 	& 0       & 0      \\	
\hline
\end{tabular}                                                                                         
\begin{tablenotes}
\footnotesize
\item[\emph{a}]{Identification number taken from the QSO MUSEUM survey \citep{2019arrigoni}.}
\item[\emph{b}]{The negative spectrum has one feature at $+1200$~km~s$^{-1}$ with peak S/N>2, but its integrated S/N is only S/N$_{\rm int}=2.3$.}
\item[\emph{c}]{The negative spectrum has one feature at $+3500$~km~s$^{-1}$ with peak S/N>2, but its integrated S/N is only S/N$_{\rm int}=1.9$.}
\item[\emph{d}]{The negative spectrum has one feature at $+2000$~km~s$^{-1}$ with peak S/N>2, but its integrated S/N is only S/N$_{\rm int}=2.5$.}
\item[\emph{e}]{This jackknife spectrum has one feature with peak S/N>2, but its integrated S/N is only S/N$_{\rm int}=2.8$.}
\item[\emph{f}]{This jackknife spectrum has one feature with peak S/N>2, but its integrated S/N is only S/N$_{\rm int}=1.9$.}
\item[\emph{g}]{This jackknife spectrum has one feature with peak S/N>2, but its integrated S/N is only S/N$_{\rm int}=2.3$.}
\item[\emph{h}]{This jackknife spectrum has one feature with peak S/N=2.8, and its integrated S/N is S/N$_{\rm int}=3.3$.}
\item[\emph{i}]{This jackknife spectrum has one feature with peak S/N>2, but its integrated S/N is only S/N$_{\rm int}=2.1$.}
\item[\emph{j}]{This jackknife spectrum has two features with peak S/N>2, but their integrated S/N is only S/N$_{\rm int}=2.1$ and $2.5$.}
\item[\emph{k}]{This jackknife spectrum has one feature with peak S/N>2, but its integrated S/N is only S/N$_{\rm int}=1.9$.}

\end{tablenotes}                                  
\end{threeparttable}
}
\end{table*}



\bsp	
\label{lastpage}
\end{document}